\documentclass[%
 reprint,
preprintnumbers,
nofootinbib,
amsmath,amssymb,
aps,
prd,
floatfix,
showkeys,
]{revtex4-2}

\usepackage{graphicx}
\usepackage{dcolumn}
\usepackage{bm}
\usepackage{hyperref}
\usepackage{xcolor}
\usepackage{comment}
\usepackage{multirow}
\usepackage{listings}

\usepackage{xcolor}

\definecolor{codeorange}{rgb}{0.984,0.0.521,0}
\definecolor{codegray}{rgb}{0.5,0.5,0.5}
\definecolor{codepurple}{rgb}{0.0196,0.188,0.278}
\definecolor{codeblue}{rgb}{0.1333,0.619,0.737}

\lstdefinestyle{mystyle}{ 
    commentstyle=\color{codeorange},
    keywordstyle=\color{codeblue},
    numberstyle=\tiny\color{codegray},
    stringstyle=\color{codepurple},
    basicstyle=\ttfamily\footnotesize,
    breakatwhitespace=false,         
    breaklines=true,                 
    captionpos=b,                    
    keepspaces=true,                 
    numbers=left,                    
    numbersep=5pt,                  
    showspaces=false,                
    showstringspaces=false,
    showtabs=false,                  
    tabsize=2
}

\usepackage{fontawesome}
\usepackage{mathtools}
\usepackage{pifont}
\usepackage{orcidlink}

\newcommand{\cmark}{\ding{51}}%
\newcommand{\xmark}{\ding{55}}%

\DeclarePairedDelimiter{\norm}{\lVert}{\rVert}

\begin{document}

\preprint{ULB-TH/25-02}

\title{Neural network emulation of reionization \\
to constrain new physics with early- and late-time probes}

\author{Ga\'{e}tan Facchinetti \orcidlink{0000-0002-5815-7540}}
\email{gaetan.facchinetti@ulb.be}
\affiliation{Service de Physique Th\'{e}orique, \\ 
Brussels Laboratory of the Universe BLU-ULB, \\
Universit\'{e} Libre de Bruxelles, Boulevard du Triomphe,\\
C.P. 225, B-1050 Brussels, Belgium}

\date{\today}

\begin{abstract}

The optical depth to reionization, a key parameter of the $\Lambda$CDM model, can be computed within astrophysical frameworks for star formation by modeling the evolution of the intergalactic medium. Accurate evaluation of this parameter is thus crucial for joint statistical analyses of CMB data and late-time probes such as the 21 cm power spectrum, requiring consistent integration into cosmological solvers. However, modeling the optical depth with sufficient precision in a computationally feasible manner for MCMC analyses is challenging due to the complexities of the nonlinear astrophysics. We introduce {\tt NNERO} ({\tt N}eural {\tt N}etwork {\tt E}mulator for {\tt R}eionization and {\tt O}ptical depth), a framework that leverages neural networks to emulate the evolution of the free-electron fraction during cosmic dawn and reionization. We demonstrate its effectiveness by simultaneously constraining cosmological and astrophysical parameters in both standard cold dark matter and non-cold dark matter scenarios, including models with massive neutrinos and warm dark matter, showcasing its potential for efficient and accurate parameter inference.
\href{https://github.com/gaetanfacchinetti/NNERO}{\faGithub}

\end{abstract}

\keywords{Dark Matter, Evolution of the Universe, Machine Learning, Cosmological Parameters}
\maketitle

\section{Introduction}

The evolution of the Universe provides a wealth of information about dark particle properties, including exotic dark matter species and neutrinos. Scenarios that affect the distribution of matter or the evolution of the intergalactic medium (IGM) -- its temperature and ionization fractions -- leave distinctive imprints on various cosmological observables. For over two decades, cosmic microwave background (CMB) anisotropies have been a cornerstone in probing these phenomena. The CMB is highly sensitive to both the optical depth to reionization and the linear matter power spectrum, and its behavior can be modeled with remarkable precision. Consequently, data from WMAP \cite{WMAP:2010qai} and Planck \cite{Planck:2018vyg} have been extensively analyzed for signs of dark matter, leading to robust constraints on its properties \cite{Zhang:2006fr, Zhang:2007zzh, Mapelli:2006ej, Padmanabhan:2005es, Galli:2009zc, Galli:2011rz, Finkbeiner:2011dx, Galli:2013dna, Slatyer:2015jla, Slatyer:2015kla, Poulin:2015pna, Slatyer:2016qyl, Liu:2016cnk, Slatyer:2012yq, Lopez-Honorez:2013cua, Capozzi:2023xie}.

Late-time probes, such as the Lyman-$\alpha$ forest from Quasi-Stellar Objects and the 21 cm line of neutral hydrogen, provide complementary insights into the evolution of the IGM during cosmic dawn and reionization. They are particularly sensitive to the formation history of the dark-matter halos. The latter are indeed the fundamental building blocks of galaxy formation, and they consequently impact the stellar emission of ionizing UV and heating X-ray light together with the distribution of baryons. Even though the Universe is more difficult to model in these periods due to the nonlinear nature of the astrophysical processes at play, thanks to a combination of improvements in simulations and instrument capabilities, the available data have become sufficiently precise to offer a competitive look at dark matter.

For example, warm dark matter or massive neutrinos can suppress the matter power spectrum at small scales, compared to the vanilla $\Lambda$CDM model. Hence, they deplete both the amount of small neutral hydrogen clouds seen in the Lyman-$\alpha$ forest and the density of the lightest star-hosting halos, thus the amount of UV sources. The Lyman-$\alpha$ forest observations, in combination with the CMB data, have already placed strong constraints on such scenarios \cite{Narayanan:2000tp, Viel:2005qj, Boyarsky:2008xj, Ballesteros:2020adh, Viel:2013fqw, Garzilli:2015iwa, Irsic:2017ixq, Garzilli:2019qki, Hooper:2022byl, Villasenor:2022aiy, Rossi:2014wsa, Palanque-Delabrouille:2015pga, Palanque-Delabrouille:2019iyz}. Today, increasing attention is being drawn to the power spectrum of the 21cm signal, which could be detected in the near future with the Hydrogen Epoch of Reionization Array (HERA) \cite{DeBoer:2016tnn, HERA:2021noe}. The sensitivity of HERA and the Square Kilometer Array (SKA) \cite{Carilli:2004nx} to many exotic dark matter models has thus been assessed in several analysis \cite{Munoz:2019hjh, Cole:2019zhu, Hotinli:2021vxg, Flitter:2022pzf, Facchinetti:2023slb, Sun:2023acy, Gessey-Jones:2023amq, Cruz:2023rmo, Plombat:2024kla, Decant:2024bpg}. 

Combining CMB and 21cm data offers a powerful way to constrain the entire history of the Universe, from recombination to reionization. A key technical challenge in this combination is that the optical depth to reionization, $\tau$, is treated as a free parameter in $\Lambda$CDM analyses of the CMB, whereas it is directly predicted by astrophysical models used to simulate the 21cm signal. Using the semi-analytical simulation code {\tt 21cmFAST} \cite{Mesinger:2010ne, Murray:2020trn}, Refs.~\cite{Liu:2015txa, Shmueli:2023box} addressed this issue by self-consistently computing $\tau$ and incorporating it into the CMB covariance matrix, removing it as a free parameter. They achieved this by computing its variations around a fixed value alongside all the other parameters. Their results demonstrated that this approach could enhance the sensitivity to key parameters, such as reducing the uncertainty on neutrino masses by approximately 40\%. While their treatment was consistent, it was limited to Fisher forecasts, relying on linear approximations around a fixed fiducial model. Alternatively, Ref.~\cite{Qin:2020xrg} investigated the joint constraining power of reionization data and CMB anisotropies by employing a Monte Carlo Markov Chain (MCMC) approach. While their analysis offered a different perspective, it was confined to the $\Lambda$CDM framework and likely required months to converge, reflecting the significant computational challenges of such studies.

In this work, we take a additional step forward by introducing an emulator trained on {\tt 21cmFAST} simulations \cite{Mesinger:2010ne}. Emulators offer numerous advantages and have gained traction in recent years for studying the late-time Universe \cite{Jennings:2018eko, Kern:2017ccn, Schmit:2017pho, LaPlante:2018pst, Breitman:2023pcj}. First, unlike MCMC, the simulations used to train the emulator can be run in parallel. This approach allowed us to obtain approximately 130,000 data samples in just a few days, compared to the several months typically required for an MCMC analysis. Second, once trained, the emulator can be combined not only with CMB data but also with other relevant observables, such as Lyman-$\alpha$ forest data or the 21cm power spectrum. Our new emulator is both lightweight and accurate, specifically designed to emulate the reionization history -- see Ref.~\cite{Montero-Camacho:2024dzs} for a complementary approach using symbolic regression techniques to provide a fit to the reionization in a $\Lambda$CDM scenario. It is trained on data produced with {\tt 21cmFAST v3}, enhanced to account for exotic cosmologies with non-standard matter power spectra through seamless integration with {\tt CLASS} \cite{Blas:2011rf, Lesgourgues:2011rh}. This newly developed version of {\tt 21cmFAST} is referred to as {\tt 21cmCLAST}\footnote{\href{https://github.com/gaetanfacchinetti/21cmCLAST/tree/ncdm}{\tt \faGithub /gaetanfacchinetti/21cmCLAST/ncdm}} and was simultaneously developed and utilized in Ref.~\cite{Dandoy:2025}. More specifically, we apply these new tools to scenarios involving either massive neutrinos or warm dark matter. The emulator is built using the {\tt PyTorch} library and is available as a Python package named {\tt NNERO}\footnote{\href{https://github.com/gaetanfacchinetti/NNERO}{\tt \faGithub /gaetanfacchinetti/NNERO}} ({\tt N}eural {\tt N}etwork {\tt E}mulator of {\tt R}eionization and {\tt O}ptical depth).

The structure of the paper is organized as follows. In Section~\ref{sec:reio_history}, we define the ionization fraction and the optical depth to reionization while introducing the effective astrophysical model implemented in {\tt 21cmFAST}. Section~\ref{sec:method} discusses the machine learning model utilized to develop the emulator. In Section~\ref{sec:inference}, we apply the emulator to derive combined constraints on astrophysical and cosmological parameters within a non-cold dark matter framework, using both CMB and reionization history data. Finally, we conclude in Section~\ref{sec:conclusion}. Throughout the paper, we adopt the normal unit convention $\hbar = c = k_{\rm B} = 1$.

\section{Reionization history}
\label{sec:reio_history}

Let us introduce here some notations and conventions related to the reionization history, as well as the model behind {\tt 21cmFAST} and the impact of non-standard dark matter on the observables.

\subsection{Free-electron fraction and optical depth}

Reionization of neutral hydrogen and helium is the most recent major phase transition in the history of the Universe. The exact evolution of the average free-electron fraction $x_e \equiv \overline{n_e}/\overline{n_{\rm H}}$, is challenging to probe during the EoR. Nonetheless, the CMB anisotropies are an indirect but powerful probe through the optical depth, $\tau$. As reionization progresses, the Universe becomes thicker to CMB photons, as they scatter on more free electrons; hence, the observed amplitude of the observed primordial anisotropies decreases as $\sim \exp(-\tau)$. The optical depth to reionization is more particularly defined as,
\begin{equation}
\begin{split}
    & \tau[x_e]  = \overline{n_{\rm H, 0}}\sigma_{\rm T} \frac{h}{H_0}\int_0^{z_\star} \frac{ (1+z)^2 x_e(z)}{\sqrt{\omega_\Lambda + \omega_{\rm m}(1+z)^3}}{\rm d} z \, .
    \end{split}
\end{equation}
Here, $\sigma_{\rm T}$ represents the Thomson scattering cross section, and $h$ is the reduced Hubble constant (such that $H_0 = 100 h~{\rm km/s/Mpc}$). The upper bound of the integral corresponds to the onset of reionization, but it has no precise definition. For instance, in the {\tt CLASS} code, $z_\star = {\rm argmin}(x_e)$ to accommodate for scenarios with exotic energy injection leading to non-standard evolutions of $x_e$ with an early start of the reionization. For consistency, we follow the same convention. In addition, $\omega_{\rm i} \equiv \Omega_{i} h^2$ for $i={\rm m, \Lambda}$ are the reduced abundances of matter and dark energy respectively. Finally, the mean Hydrogen density today is given by
\begin{equation}
\begin{split}
    \overline{n_{\rm H, 0}} & = \frac{\rho_{\rm c, 0}\omega_{\rm b}}{h^2 m_{\rm H}} (1-Y_{\rm He}) \, ,
    \end{split}
\end{equation}
with $\omega_{\rm b}$ the reduced abundance of baryons, $m_{\rm He}$ and $m_{\rm H}$ the proton and Helium masses, $\rho_{\rm c, 0}$ the critical density of the Universe and $Y_{\rm He} \equiv \overline{\rho_{\rm He}}/\overline{\rho_{\rm b}}$ the ratio of the Helium energy density over the total baryon energy density. Setting $Y_{\rm He}= 0.245$ as the benchmark value used throughout this analysis, $\tau$ can be estimated as
\begin{equation}
\begin{split}
    \tau[x_e] \simeq 0.0012 & \left(\frac{\omega_{\rm b}}{0.0224}\right) \\
    & \times \int_0^{z_\star} \frac{ (1+z)^2 x_e(z)}{\sqrt{\omega_\Lambda + \omega_{\rm m}(1+z)^3}}{\rm d} z  \, .
    \end{split}
\end{equation}

Following \cite{Liu:2015txa, Shmueli:2023box}, assuming that Helium is singly ionized simultaneously to Hydrogen \cite{Trac:2006vr}, the free electron fraction is 
\begin{equation}
\begin{split}
    x_e & = \frac{1}{\overline{n_{\rm H}}} \left\{\overline{x_{\rm HII} n_{\rm H}} + \overline{x_{\rm HeII} n_{\rm He}} + \overline{x_{\rm HeIII} n_{\rm He}}  \right\}\\
     & = \frac{\overline{n_{\rm b}}}{\overline{n_{\rm H}}} \left\{ \overline{x_{\rm HII}[1+\delta_{\rm b}]} + \frac{f_{\rm He}}{1+f_{\rm He}}\overline{x_{\rm HeIII}[1+\delta_{\rm b}]}\right\}\, ,
    \end{split}
\end{equation}
with $x_{\rm HII} = \overline{n_{\rm HII}} / \overline{n_{\rm H}}$ the fraction of ionized Hydrogen, $x_{\rm HeII} = (\overline{n_{\rm HeII}} + \overline{n_{\rm HeIII}})/\overline{n_{\rm He}}$ the fraction of Helium that is, at least one time ionized, and $x_{\rm HeIII} = \overline{n_{\rm HeIII}}/\overline{n_{\rm He}}$ the fraction of doubly ionized Helium. In addition, $f_{\rm He}\equiv n_{\rm He}/n_{\rm H} = m_{\rm H} Y_{\rm He}/(1-Y_{\rm He})/m_{\rm He}$ and $\delta_{\rm b}$ is the baryon density contrast. In the following we introduce
\begin{equation}
    X_e \equiv \overline{x_{\rm HII}[1+\delta_{\rm b}]} \, ,
\end{equation}
which we can evaluate with {\tt 21cmCLAST}. Moreover, we consider that second ionization of Helium happens instantaneously at redshift 3, such that
\begin{equation}
    \overline{x_{\rm HeIII}[1+\delta_{\rm b}]} \simeq \Theta (3-z) \, ,
\end{equation}
with $\Theta$ the Heaviside distribution. In the standard CMB anisotropies analysis with Planck data \cite{Planck:2018vyg}, the evolution of $X_e$ during reionization is modeled with a simple hyperbolic tangent,
\begin{equation}
    X_e^{\rm t} \equiv  \frac{1}{2} \left\{ 1+\tanh\left(  \frac{1+z_{\rm r}}{1.5 \delta_z}\left[1-\left(\frac{1+z}{1+z_{\rm r}}\right)^{1.5}\right]   \right) \right\} \, ,
\end{equation}
with $\delta_z = 0.5$ and $z_{\rm r}$ the reionization redshift (in one-to-one relationship with $\tau$). The precise choice of the shape does not actually really impact the inference of the cosmological parameters as the CMB data is mainly sensitive to $\tau$ (or $z_{\rm r}$), which is considered a free parameter of the model. However, as we discuss in the next subsection, $\tau$ and $X_e$ can be computed by solving the evolution of the IGM and introducing a model for star formation.

\subsection{The astrophysical model}

Determining the evolution of the IGM during the EoR, particularly the gas temperature, is a challenging task \cite{Puchwein:2018arm}. In this work, we rely on {\tt 21cmFAST}, which provides a robust framework for tracking the free-electron fraction. More precisely, fully ionized regions are identified using an excursion set formalism, where a region is flagged as ionized once the cumulative number of ionizing photons, $n_{\rm ion}$, exceeds a recombination threshold. This approach, however, underestimates the true number of ionizing photons, not accounting for leakage in adjacent regions. To address this, the average volume-filling factor of ionized hydrogen is evaluated analytically under simplified conditions and compared to the averaged result of the corresponding simplified simulation. This comparison enables a correction for leaked photons, ensuring accurate results when the simulation is run with full settings. See Ref.~\cite{Park:2021eux} for more details on the conservation of the number of photons and Ref.~\cite{Mesinger:2010ne} for a description of the excursion set approach. In the Park model \cite{Park:2018ljd} as implemented in {\tt 21cmFAST v3}, $n_{\rm ion}$ is evaluated as
\begin{equation}
    n_{\rm ion} \equiv \frac{1}{\overline{\rho_{\rm b}}} \int {\rm d} M \, \frac{\partial n(M, z \, | \, R, \delta_R)}{\partial M} f_{\rm duty} M_\star f_{\rm esc} N_{\gamma/{\rm b}} \, .
    \label{eq:nion}
\end{equation}
with $\partial n / \partial M$ the conditional halo mass function in region of size $R$ and overdensity $\delta_R$. Therefore, any modification of the matter power spectrum directly impacts the halo mass function and thus $n_{\rm ion}$. The stellar mass in a halo of mass $M$ is modeled with two parameters $f_{\star, 10}$ and $\alpha_\star$, according to \cite{Qin:2020xyh, Stefanon_2021, Shuntov:2022qwu}
\begin{equation}
    M_\star = M \left(\frac{\omega_{\rm b}}{\omega_{\rm m}}\right) {\rm min}\left\{1, f_{\star, 10} \left(\frac{M}{10^{10} \, {\rm M_\odot}}\right)^{\alpha_\star}\right\} \, .
    \label{eq:mstar}
\end{equation}
The fraction of UV photon escaping the galaxy is similarly given in terms of two other parameters, $f_{{\rm esc}, 10}$ and $\alpha_{\rm esc}$,
\begin{equation}
    f_{\rm esc} = {\rm min}\left\{1, f_{{\rm esc}, 10} \left(\frac{M}{10^{10} \, {\rm M_\odot}}\right)^{\alpha_{\rm esc}}\right\}\, .
\end{equation}
In addition, $N_{\gamma/b}$ is the number of ionizing photons per stellar baryons, fixed to 5000 (otherwise degenerate with $f_{\star, 10}$). Finally, the duty cycle encodes all processes that suppresses the formation of small galaxies, with a parameter called $M_{\rm turn}$ such that \cite{Qin:2020xyh}
\begin{equation}
    f_{\rm duty} = \exp\left(-M_{\rm turn}\frac{1\, {\rm M_\odot}}{M}\right) \, .
\end{equation}
This choice of model can only account for a single population of stars, formed in atomic-cooling galaxies. Even if uncertain, the existence of molecular cooling galaxies is of particular interest in light of the recent discoveries of high redshift galaxies with the James Webb Space Telescope \cite{Zackrisson:2011ct, Rydberg:2012ez, Riaz:2022prd, Larkin:2022asx}. A two-population model for star formation is, however, more complex and adds several free parameters. Furthermore, the analytical patch for the leakage of ionizing photons is currently not available when including molecular cooling galaxies in {\tt 21cmFAST}, and its implementation is out of the scope of this analysis. Finally, most parameter inferences available in the literature have been performed with a single population of stars. Using the same set of parameters allows for better comparisons.

For precision, we also need to account for the ionization of the mostly neutral IGM from X-rays, which happens prior to reionization, to the total ionization fraction. X-ray injection from the stars is described with two extra parameters, respectively denoted $L_{\rm X}$ and $E_0$, controlling the total amplitude of the emission and the minimal energy (in eV) needed to escape galaxies and enter the IGM \cite{Das:2017fys}. Eventually, to have a self-consistent model, the star formation rate is given by $\dot M_\star \equiv M_\star H/ t_{\star}$ where $t_\star$ is another free parameter representing the typical time scale between 0 and 1. This parameter is used, in particular, in Appendix~\ref{app:UVLF} with the discussion of UV-luminosity functions. The set of all astrophysical parameters considered in this analysis is summarized in the right part of table~\ref{tab:input_parameters}.

\begin{table}[ht]
    \centering
    \begin{tabular}{c | c || c | c  } 
        Parameter & Range & Parameter & Range \\
        \hline
        $\omega_{\rm b}$  & $[0.1, 0.14]$ & $\log_{10} f_{\star, 10}$ & $[-2.5,0.0]$\\ 
        $\omega_{\rm dm}$ & $[0.0213, 0.0235]$ & $\log_{10} f_{\rm esc, 10}$ & $[-2.5,0.0]$ \\ 
        $h$ & $[0.6, 0.75]$ & $\alpha_\star$ & $[-0.1, 1.0]$\\ 
        $\ln(10^{10}A_{\rm s})$ & $[2.98,3.1]$ & $\alpha_{\rm esc}$ & $[-1.0, 0.5]$ \\ 
        $n_s$ & $[0.940,0.985]$ & $\log_{10} M_{\rm turn}$ & $[6.0,10.0]$\\
        $m_{\nu, 1}~{\rm [eV]}$ & $[0,0.1]$ & $t_\star$ & $[0, 1.0]$\\
        $f_{\rm WDM}$ & $[0,1.0]$ & $\log_{10}L_{\rm X}$ & $[38, 42]$\\
        $\mu_{\rm WDM}$ & $[0,4.0]$ & $E_0$ & $[100, 1500]$\\
    \end{tabular}

    \caption{Free parameters of the model (cosmology on the left, astrophysics on the right) and their prior range.}
    \label{tab:input_parameters}
\end{table}

\subsection{Warm dark matter and massive neutrinos}

As a representative example of exotic scenarios that can impact reionization, we examine massive neutrinos and warm dark matter. Both lead to a suppression of the matter power spectrum on small scales, albeit in different ways. Massive neutrinos cause a mild suppression over a broad range of scales, whereas warm dark matter sharply suppresses the power spectrum below a characteristic scale. By reducing the abundance of dark matter halos that can form at a given redshift, both scenarios diminish the halo mass function. Consequently, as described by Eq.~(\ref{eq:nion}), this leads to delayed reionization and a lower optical depth to reionization.

Although the matter power spectrum is primarily sensitive to the sum of neutrino masses, we adopt a physically motivated framework by assuming a normal mass hierarchy\footnote{The inverse mass hierarchy is under more observational pressure than the normal hierarchy as it predicts higher mass sums.} \cite{deSalas:2020pgw}, with mass splittings $\Delta m_{21}^2 = 7.5 \times 10^{-5}$ eV$^2$ and $\Delta m_{31}^2 = 2.55 \times 10^{-3}$ eV$^2$. Additionally, we consider scenarios where only a fraction of the dark matter is warm. To describe these two cases, we introduce three parameters: the mass of the lightest neutrino, $m_{\nu 1}$; the mass of the warm dark matter particle, $m_{\rm WDM}$; and the fraction of dark matter that is warm, $f_{\rm WDM} \in [0, 1]$. For convenience, we redefine the warm dark matter mass in terms of its inverse, normalized to 1 keV, as $\mu_{\rm WDM} \equiv (1~{\rm keV})/m_{\rm WDM}$. From these definitions, the total dark matter abundance is thus 
\begin{equation}
    \omega_{\rm dm} \equiv \frac{1}{(1-f_{\rm wdm})}\omega_{\rm cdm} + \omega_{\nu}\, ,
\end{equation}
with the neutrino abundance being related to the mass of the lightest neutrino following
\begin{equation}
    \omega_\nu = \frac{m_{\nu 1}}{93.14 ~ {\rm eV}} \left\{1+\sqrt{1+\frac{\Delta m_{21}^2}{m_{\nu1}^2}} + \sqrt{1+\frac{\Delta m_{31}^2}{m_{\nu1}^2}}\right\} \, .
\end{equation}
The cosmological parameters used in this analysis are summarized in the left part of table~\ref{tab:input_parameters}. 

In {\tt 21cmCLAST}, the halo mass function computation has been updated to use a sharp-k window function with parameters calibrated from warm dark matter simulations, following Ref.~\cite{Schneider:2014rda}. This modification replaces the default, less accurate warm dark matter implementation in {\tt 21cmFAST}. However, this choice introduces a slight deviation from accuracy when $f_{\rm WDM}$ or $\mu_{\rm WDM}$ approaches 0, effectively reverting to the cold dark matter (CDM) scenario. Given that the choice of halo mass function can influence parameter inference, as shown in Ref.~\cite{Greig:2024zso}, the results should be interpreted with caution. Nonetheless, the adopted approach represents the most reliable analytical description of warm / non-cold dark matter currently available for semi-analytical codes such as {\tt 21cmFAST}. Any resulting biases are expected to be minor and unlikely to affect our conclusions.

In section \ref{sec:inference}, we will investigate the impact of the neutrino mass and of a warm dark matter component on the reionization history and infer constraints using a combination of likelihoods. Before, in the next section, we detail how we can emulate reionization with neural networks.

\section{Emulation with Neural Networks}
\label{sec:method}

Emulating the ionization fraction with high precision is complex due to the sharp transition between the quasi-neutral regime and the fully ionized regime. In addition, for some parameter choices, reionization does not complete before the end of the {\tt 21cmCLAST} simulation. The strategy adopted in this work is to divide the problem into two simpler tasks. Firstly, we introduce a classifier that is trained to identify if, given a set of parameters $\boldsymbol\theta$, reionization happens early enough. In practice, the limit is set from the 5$\sigma$ bound of Ref.~\cite{McGreer:2014qwa}, the classifier identifies a set of parameters as valid only if $X_{\rm e}(z=5.9) > 0.69$. Secondly, a regressor is trained to emulate the curves that pass the aforementioned \emph{selection cut}. Both the classifier and regressor are implemented using rather shallow neural networks. A schematic representation of the entire process is given in fig.~\ref{fig:schema}. A quick instruction manual on how to use {\tt NNERO} is given in appendix~\ref{app:NNERO_manual}.

\begin{figure}[ht]
    \centering
    \includegraphics[width=1\linewidth]{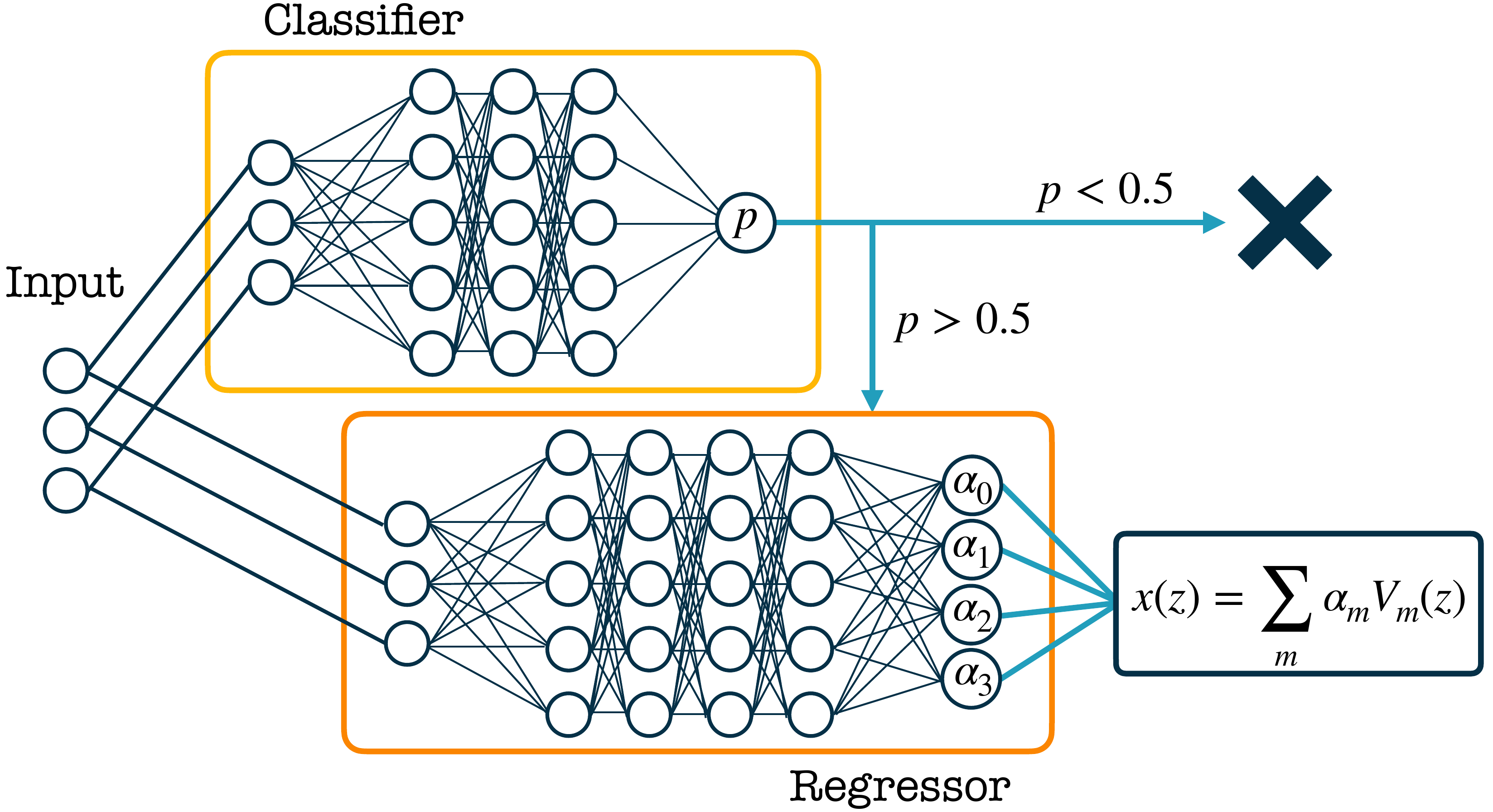}
    \caption{Schematic view of the structure of the code, divided into a two neural networks, one for classification and the other one for regression.}
    \label{fig:schema}
\end{figure}

\subsection{Sampling and simulations}

In order to generate data to feed the neural networks, {\tt 21cmCLAST} simulations have been run varying the set of input parameters $\boldsymbol\theta$. This set contains all cosmological and astrophysical parameters given in Tab.~\ref{tab:input_parameters}. For each simulation, all parameters are randomly drawn from a uniform distribution within a prior range reported in the second column of the same table. In addition, the seed of the simulation, that is used to set the initial conditions, is also chosen randomly. Therefore, for a given $\boldsymbol\theta$, $x_e$ is not exactly determined. In practice, we find that the initial conditions have a negligible impact on $x_e$ and one can safely associate one free electron fraction curve $x_e$, to one set of input parameters $\boldsymbol\theta$.

\begin{figure}[htbp]
    \centering
    \includegraphics[width=\linewidth]{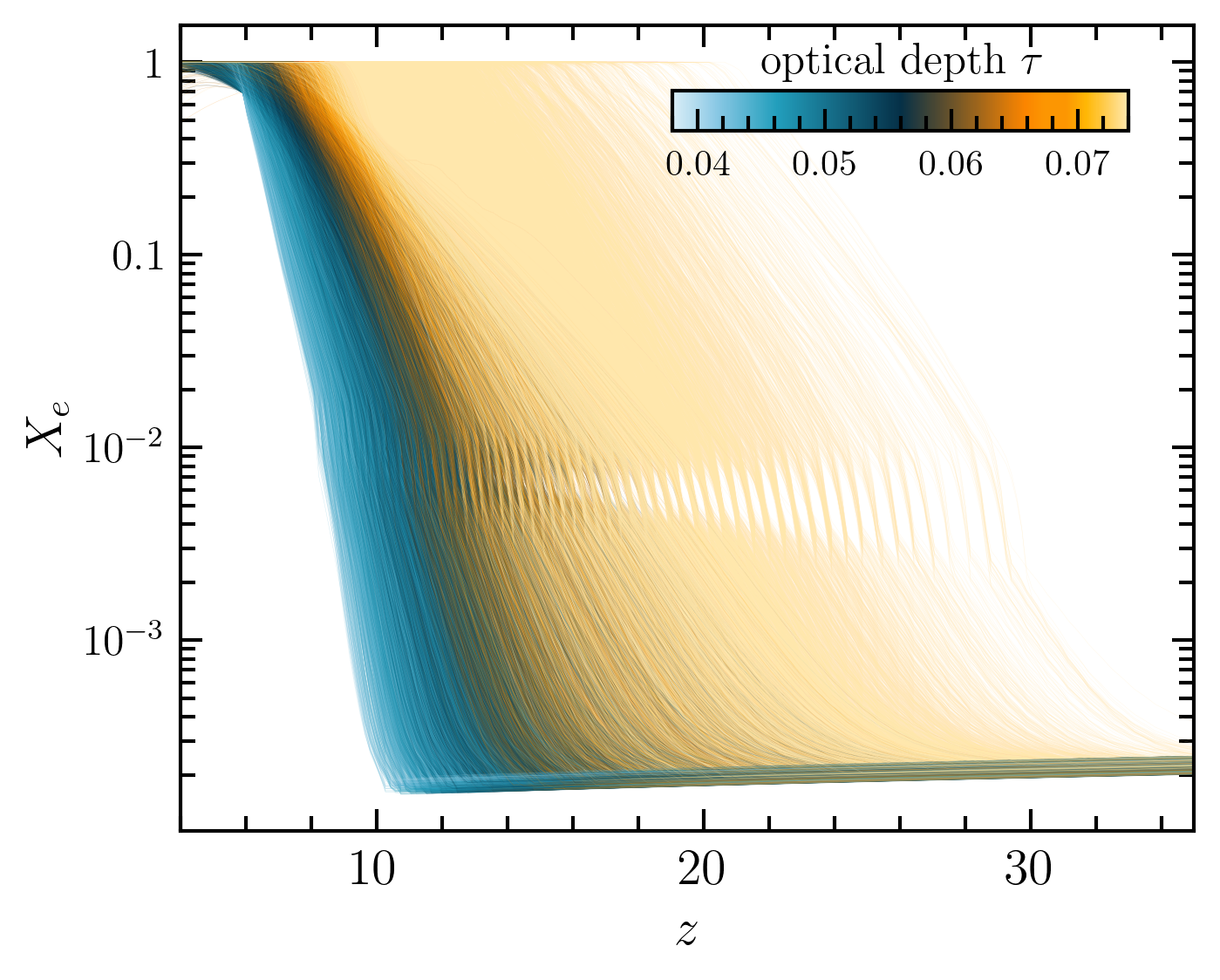}
    \caption{Simulated evolution of the free-electron fraction $X_e$ vs the redshift $z$ for various $43641$ different input parameters. The corresponding optical depth to reionization $\tau$ is shown in color, the central, darkest color being equal to the Planck constraint $\tau = 0.056$ \cite{Planck:2018vyg}.}
    \label{fig:xe_vs_z}
\end{figure}

More precisely, in this analysis, a total of 127401 simulations have been performed. For the classifier, this entire set was divided into a training dataset of 101921 simulations, as well as a validation and a test dataset, each composed of 12740 simulations. In total, with our choice of prior ranges, 43641 simulations passed the selection cut. Out of these, 34913 were used for training, 4364 were used for validation, and similarly for testing. In Fig.~\ref{fig:xe_vs_z} we show all curves passing the selection cut, colored according to the corresponding optical depth to reionization. The samples in the palest shade all predict an optical depth that is outside the Planck constraints at the 68\% confidence level (CL), $\tau \sim 0.054 \pm 0.008$.

In the following subsections, we give more details about the technical implementation of the neural networks as well as the training procedure and cost functions.

\subsection{The classifier}

The classifier implements the \emph{selection cut} to distinguish between late, excluded reionization histories (excluded at more than 5$\sigma$ by Ref.~\cite{McGreer:2014qwa}) and valid ones. Numerically, we assign a label of $y=0$ to the excluded histories and $y=1$ to the valid ones. The classifier is constructed using a neural network, which can be described as a function  $p : \boldsymbol\theta \mapsto (\boldsymbol\theta, \boldsymbol\phi)$  where $\boldsymbol\theta$ represents the input parameters and  $\boldsymbol\phi$ the network's weights. In this setup, the goal is for $p$ to approximate the probability distribution of $y(\boldsymbol\theta)$.  A common and effective approach is to train the neural network by minimizing the binary cross-entropy loss function,
\begin{equation}
\begin{split}
    \mathcal{C}_{\rm C}(\boldsymbol\phi) & = \int {\rm d} \boldsymbol\theta \,  \pi(\boldsymbol\theta)  \ln  \left\{  \frac{[p(\boldsymbol\theta, \boldsymbol\phi)]^{y(\boldsymbol\theta)}}{[1-p( \boldsymbol\theta_i, \boldsymbol\phi)]^{y(\boldsymbol\theta)-1}}\right\} \\
    & \simeq -\frac{1}{N}\sum_{i=1}^{N}   \ln \left\{  \frac{[p(\boldsymbol\theta_i, \boldsymbol\phi)]^{y(\boldsymbol\theta_i)}}{[1-p( \boldsymbol\theta_i, \boldsymbol\phi)]^{y(\boldsymbol\theta_i)-1}}\right\} \, ,
    \end{split}
\end{equation}
with respect to $\boldsymbol\phi$. We denote by $\pi(\boldsymbol\theta)$ the prior distribution. Intuitively, if $y(\boldsymbol\theta_i) = 1$ and $p(\boldsymbol\theta_i, \boldsymbol\phi) \sim 0$ then the cost function increases due to the numerators inside the log. On the other hand, if $y(\boldsymbol\theta_i) = 0$ and $p(\boldsymbol\theta_i, \boldsymbol\phi) \sim 1$, the cost increases due to the denominator.

\begin{figure*}
    \centering
    \includegraphics[width=0.9\linewidth]{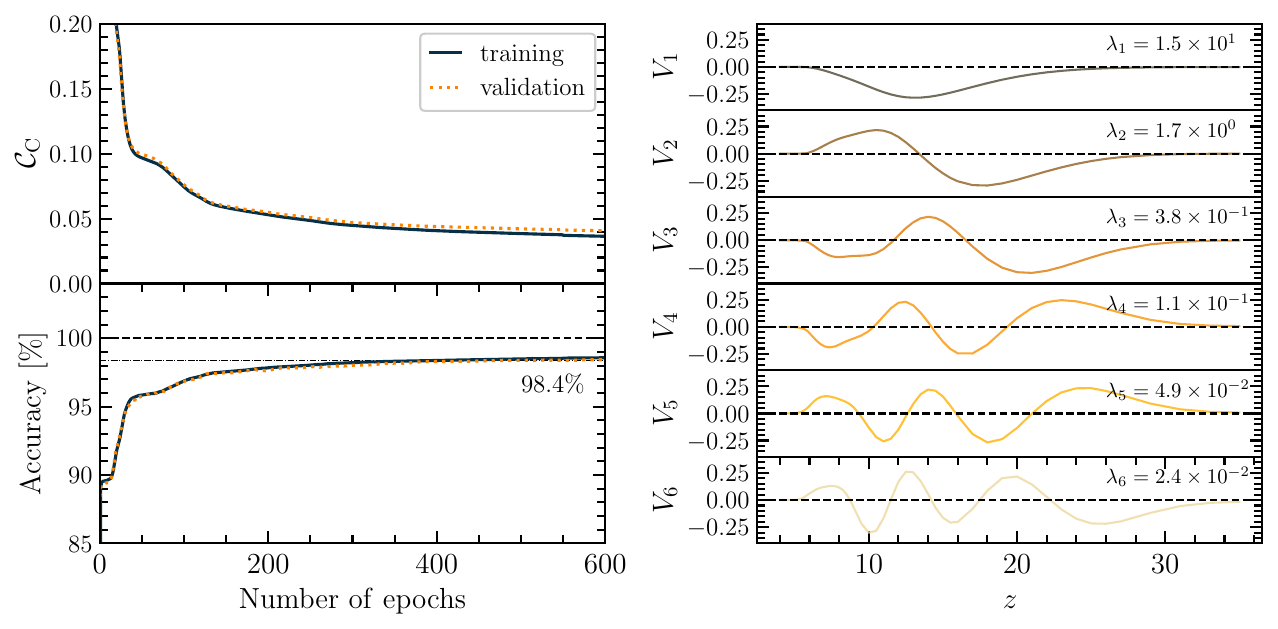}
    \caption{{\bf Left panel.} Evolution of the cost function and accuracy, evaluated on the training (solid black) and validation (dashed orange) datasets, with the number of epochs. The dash dotted black line on the lower panel shows the accuracy evaluated on the test dataset. {\bf Right panel.} First 6 eigenvectors of the principal component analysis with their associated eigenvalues.}
    \label{fig:classifier_perf_and_eigv}
\end{figure*}

Given the simplicity of the task, a 2-layer neural network with 30 hidden neurons per layer and ReLU activation functions is sufficient to achieve the desired accuracy. The left panel of Fig.~\ref{fig:classifier_perf_and_eigv} shows the evolution of the cost function during training and validation. After 600 epochs, the network achieved an accuracy exceeding $98\%$, with a precise accuracy of $98.4\%$ on the test dataset. Furthermore, we verified that the majority of misclassifications occur for points near the boundary of the selection cut. Such misclassifications are not problematic when performing MCMCs. Indeed, false positives are almost systematically rejected at $\sim 5\sigma$ by the likelihood constraint from Ref.~\cite{McGreer:2014qwa} (see Section~\ref{sec:inference}). Similarly, false negatives would have been excluded regardless.

\subsection{The regressor}

The goal of the regressor is to approximate $X_e(z \, | \, \boldsymbol\theta)$ when it passes the selection cut. For this purpose, we use a neural network that serves as a universal interpolator, and which can be described as a function $\xi(z \, | \, \boldsymbol\theta, \boldsymbol\phi)$ of weights $\boldsymbol\phi$. These weights are adjusted to optimize the cost function, forcing $\xi$ to \emph{converge} toward $X_e$.

\subsubsection{Cost function}

Let $\hat{\boldsymbol\phi}$ represent the value of $\boldsymbol\phi$ that optimizes the cost function. Up to a certain level of precision, the neural network needs to fulfill $X_e(\cdot \, | \, \boldsymbol\theta) \simeq \xi(\cdot \, | \, \boldsymbol\theta, \hat{\boldsymbol\phi})$. To that end, it can \emph{learn} by minimizing the following weighted $L_1$-norm of the relative error
\begin{equation}
    \norm{\mathcal{E}_X(\, \boldsymbol\theta, \boldsymbol\phi)} \equiv  \frac{1}{\Delta z}\int \left|  \mathcal{E}_X(z \, |  \, \boldsymbol\theta, \boldsymbol\phi )  \right| w(z) {\rm d} z \, ,
\end{equation}
where $\Delta z = |z_{\rm max} - z_{\rm min}|$ is the total redshift range and the relative error function is defined as 
\begin{equation}
   \mathcal{E}_X(z \, |  \, \boldsymbol\theta, \boldsymbol\phi )  \equiv 1-\frac{\xi(z\, | \, \boldsymbol\theta, \boldsymbol\phi )}{X_e(z \, | \, \boldsymbol\theta)}\, .
\end{equation}
Here $w$ is a function normalized such that
\begin{equation}
    \frac{1}{\Delta z}\int w(z){\rm d} z = 1
\end{equation}
and whose main purpose is to put more weight at some redshifts where, for instance, the variations of the ionization fraction are more important. In practice, the free-electron fractions are stored on $n=50$ redshift steps. From the {\tt 21cmFAST} output, the array of redshifts spans the range $[4, 35]$. However, the free-electron fractions are all extremely smooth around $z\sim 30$ (see Fig.~\ref{fig:xe_vs_z}) as, at these redshifts, $X_e$ mostly follows the cosmological background evolution before the first stars ignite reionization. This is the reason why the redshift steps are not chosen of equal length and the density of points is higher in the range $ 4 \le z \le 15$.

In numerous applications requiring the computation of the free-electron fraction, while the evolution with redshift is of interest, it is predominantly essential to precisely determine the associated optical depth to reionization (e.g. for CMB analysis). Consequently, for each configuration of input parameters $\boldsymbol\theta$, we seek to minimize the relative error between the optical depths to reionization assessed at $x_e$ and $\xi$,
\begin{equation}
    \mathcal{E}_\tau(\, \boldsymbol\theta, \boldsymbol\phi) \equiv  1-  \frac{\tau\left[ \xi\, | \, \boldsymbol\theta, \boldsymbol\phi \right]}{\tau \left[x_e \, | \, \boldsymbol\theta \right]} \, .
\end{equation}
The total cost function for the regressor is thus,
\begin{equation}
\begin{split}
    \mathcal{C}_{\rm R}(\boldsymbol\phi) & \equiv \int {\rm d}\boldsymbol\theta \, \pi_{\rm R}(\boldsymbol\theta) \frac{1}{2} \left\{\norm{\mathcal{E}_X(\boldsymbol\theta, \boldsymbol\phi)} + \mathcal{E}_\tau( \boldsymbol\theta, \boldsymbol\phi) \right\}\\
    & \simeq \frac{1}{N_{\rm R}} \sum_{i=1}^{N_{\rm R}}  \frac{1}{2} \left\{\norm{\mathcal{E}_X(\boldsymbol\theta_i, \boldsymbol\phi)} + \mathcal{E}_\tau( \boldsymbol\theta_i, \boldsymbol\phi)\right\}
    \end{split}
\end{equation}
where $\pi_{\rm R}(\theta)$ is the truncated prior distribution of parameters passing the selection cut. Therefore, the sum runs over the $N_{\rm R}$ drawn samples of input parameters that pass the selection cut. Including $\mathcal{E}_\tau$ can actually seem superfluous as approaching the minimum of $\mathcal{E}_X$ should automatically drive $\boldsymbol\phi$ towards the minimum of $\mathcal{E}_\tau$. In practice, adding the constraint on the optical depth to reionization helps the neural networks converge and ensures that the latter is recovered with the desired precision.

\begin{figure*}[htbp]
    \centering
    \includegraphics[width=0.85\linewidth]{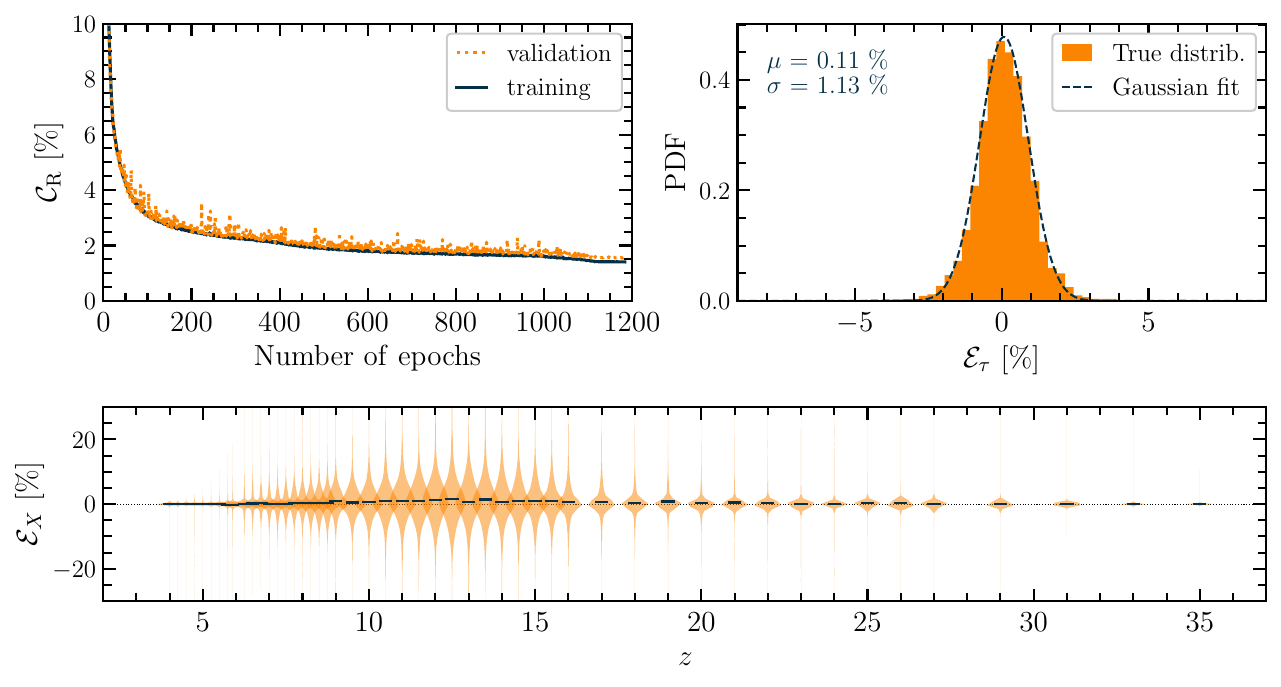}
    \caption{Performance of the regressor. {\bf Upper left panel.} Evolution of the training (solid black) and validation (dashed orange) cost function with the number of trainign epochs. {\bf Upper right panel.} Distribution of relative error on $\tau$ evaluated on the test dataset. {\bf Lower panel}. Distributions of relative error on $X_e$ at each redshift (orange violins). The black lines represent the average of the distributions.  }
    \label{fig:regressor_perf}
\end{figure*}

\begin{table*}[ht]
    \centering
    \begin{tabular}{c || c | c | c | c | c | c} 
        
        \multirow{3}{*}{scenario} & \multirow{3}{*}{parameters} & \multicolumn{2}{|c|}{\bf Planck18} & \multirow{2}{*}{\bf Reio.} & \multirow{2}{*}{\bf UV-LFs} &  \multirow{2}{*}{\bf HERA} \\
        & & \phantom{a} {\small TTTEEE} \phantom{a} & \phantom{a} \multirow{1}{*}{{\small lensing}} \phantom{a} & & \\  
        & & {\small + lowE} & {\small + BAO} & & \\
        \hline
        \multirow{3}{*}{\bf cold dark matter} & $\boldsymbol\theta_{\rm \Lambda CDM} + \tau$ & \cmark & \cmark & \xmark & \xmark & \xmark \\
        & $\boldsymbol\theta_{\rm \Lambda CDM} + \boldsymbol\theta_{\rm astro}$ & \cmark & \cmark & \cmark & \cmark & \xmark\\
        & $\boldsymbol\theta_{\rm astro}$ ($\boldsymbol\theta_{\rm \Lambda CDM}$ fixed)  & \xmark & \xmark & \cmark & \cmark & \xmark\\
        \hline
         \multirow{3}{*}{\bf massive neutrinos} & $\boldsymbol\theta_{\rm \Lambda CDM} + \tau + m_{\nu 1}$ & \cmark & \cmark & \xmark & \xmark & \xmark \\
        & $\boldsymbol\theta_{\rm \Lambda CDM} + \boldsymbol\theta_{\rm astro} + m_{\nu 1}$ & \cmark & \cmark & \cmark & \cmark & \xmark\\
        & $\boldsymbol\theta_{\rm \Lambda CDM} + \boldsymbol\theta_{\rm astro} + m_{\nu 1}$ & \cmark & \cmark & \cmark & \cmark & \cmark\\
        \hline
        \multirow{2}{*}{\bf warm dark matter} & $\boldsymbol\theta_{\rm \Lambda CDM} +  \boldsymbol\theta_{\rm astro} + \mu_{\rm WDM} \, (f_{\rm WDM} = 1)$ & \cmark & \xmark & \cmark & \cmark & \xmark \\
         & $\boldsymbol\theta_{\rm \Lambda CDM} +  \boldsymbol\theta_{\rm astro} + \mu_{\rm WDM} + f_{\rm WDM}$ & \cmark & \xmark & \cmark & \cmark & \xmark 
    \end{tabular}

    \caption{Overview of the studied scenarios, including the varied parameters and the likelihoods used in the MCMC analysis. A \cmark{} indicates that the likelihood is included, while a \xmark{} signifies that it is not.}
    \label{tab:mcmcs}
\end{table*}

\subsubsection{Principal component decomposition}

A first approach to build a regressor would be to implement a neural network with a number of input nodes equal to the number of parameters in $\boldsymbol\theta$ and $n$ output nodes giving the vector
\begin{equation}
    {\bf X}(\boldsymbol\theta) \equiv \left(X_e(z_0  \, | \, \boldsymbol\theta), X_e(z_1 \, | \, \boldsymbol\theta), \dots, X_e(z_n \, | \, \boldsymbol\theta) \right)\, .
\end{equation}
However, as shown in Fig.~\ref{fig:xe_vs_z}, the evolution of $X_e$ with redshift $z$ is not entirely random -- these curves exhibit clear similarities. Therefore, as is common in regression tasks, we can capture this structured evolution by performing a principal component analysis (PCA) \cite{Habib:2007ca, HIGDON20082431, Kern:2017ccn}. This allows us to express the final result in a more convenient basis. Furthermore, since the free-electron fraction spans several orders of magnitude, we apply the principal component decomposition to its logarithm to better capture its variations,
\begin{equation}
    {\bf Y}(\boldsymbol\theta) \equiv \log_{10}{\bf X}(\boldsymbol\theta)  \, .
\end{equation}
The mean and covariance matrix associated to this function are respectively denoted
\begin{align}
    \boldsymbol \mu_{\bf Y} & \equiv \left< {\bf Y}(\boldsymbol\theta)\right>_{\boldsymbol\theta} \, , \\ 
    {\bf C}_{\bf Y} & \equiv \left< \left({\bf Y}(\boldsymbol\theta) -  \boldsymbol \mu  \right) \left({\bf Y}(\boldsymbol\theta) - \boldsymbol\mu \right)^{\rm T}\right>_{\boldsymbol\theta} \, .
\end{align}
The covariance is real symmetric, hence it can be diagonalized. If the free-electron fraction vectors were independent of the parameters $\boldsymbol\theta$, then ${\bf C}_{\bf Y}$ would have a single trivial non-zero eigenvalue associated with the eigenvector ${\bf Y}(\boldsymbol\theta) -  \boldsymbol \mu_{\bf Y}$. In practice, ${\bf C}_{\bf Y}$ has a finite spectrum of eigenvalues $\lambda_k$ associated with orthonormal eigenvectors ${\bf V}_k$. One can always write the free-electron fraction vector in the eigenbasis using the coefficients $a_k(\boldsymbol\theta) \equiv {\bf V}_k^{\rm T} \cdot  \left({\bf Y}(\boldsymbol \theta) - \boldsymbol\mu_{\bf Y} \right)$ such that
\begin{equation}
    {\bf Y}(\boldsymbol \theta) = \boldsymbol\mu_{\bf Y} +  \sum_{k=1}^{n} a_k(\boldsymbol\theta) {\bf V}_k \, .
    \label{eq:PCA}
\end{equation}
These coefficients straightforwardly satisfy $\left< a_k(\boldsymbol \theta)\right>_{\boldsymbol \theta} = 0$ as well as
\begin{equation}
    \left< a_k(\boldsymbol \theta) a_j(\boldsymbol \theta)\right>_{\boldsymbol \theta} = \lambda_k \delta_{kj} \, ,
\end{equation}
for all $k$ and $j$, with $\delta_{kj}$ the Kronecker symbol. As a result, only eigenvectors associated with large eigenvalues contribute significantly to Eq.~(\ref{eq:PCA}). To reduce the dimensionality of the neural network output, the sum over k  can be truncated. The right panel of Fig.~\ref{fig:classifier_perf_and_eigv} displays the first six eigenvectors along with their corresponding eigenvalues. Lower eigenvalues correspond to eigenvectors that capture finer details of the evolution. Notably, the 6$^{\rm th}$  eigenvalue is already about three orders of magnitude smaller than the first. In practice, with  $n = 50$  redshift steps, we find that retaining only the first  $m \simeq 30$  eigenvectors is sufficient to reconstruct  ${\bf Y}(\boldsymbol \theta)$  with percent-level accuracy.  Consequently, the neural network outputs are the $m$ first coefficients $\alpha_k(\boldsymbol\theta \, | \, \hat{\boldsymbol \phi})$ of the decomposition in the eigenbasis and
\begin{equation}
    {\bf Y}(\boldsymbol\theta) \sim \sum_{i=1}^{m} \alpha_i(\boldsymbol\theta \, | \, \hat{\boldsymbol \phi}) {\bf V}_i \, .
\end{equation}
More particularly, with this technique, the accuracy  -- discussed in more details the following subsection -- is obtained with a network of 6 hidden layers, each with 80 hidden features.

\subsubsection{Performance}

The performance of the regressor is assessed based on its precision in reconstructing the free-electron fraction curve, $X_e$, and the optical depth to reionization, $\tau$. Results are summarized in Fig.~\ref{fig:regressor_perf}. The upper-left panel shows that validation and training losses closely track each other, a good indication that the network is not overfitting the training data. The upper right panel illustrates the distribution of relative errors in $\tau$ from the test dataset (orange histogram), which is well fitted by a Gaussian (black dashed curve), with mean and standard deviation of $0.11\%$ and $1.13\%$ respectively. Therefore, the total relative error on $\tau$ is below 1.2\% at $1\sigma$, well within the  $1\sigma$ error from CMB measurements (corresponding to a $\sim 10\%$ relative error), demonstrating the high precision of the regressor.

The bottom panel shows violin plots of the relative error on $X_{\rm e}$ in the test dataset across redshifts. Errors are largest between redshifts 8 and 17 due to the sharp transition in $X_{\rm e}$ at those redshifts but remain below 10\% at $1\sigma$. This precision satisfies the accuracy required for modeling $X_{\rm e}$.

\section{Statistical inferences}
\label{sec:inference}

Once the model is fully trained, we can predict the reionization history in a fraction of a second for any value of the input parameters. A straightforward application is to make a joint CMB and reionization analysis for exotic dark matter models. We outline the method below before showing and discussing the results of MCMC analyses performed in various scenarios.

\subsection{Likelihoods and models}

The full set of parameters, on which the neural networks were trained, are divided into 5 $\Lambda$CDM parameters,
\begin{equation*}
    \boldsymbol\theta_{\rm \Lambda CDM} = \left\{\omega_{\rm b}, \omega_{\rm dm}, h, \ln\left(10^{10} A_{\rm s}\right), n_{\rm s}\right\}
\end{equation*}
and 9 astrophysical parameters, 
\begin{equation*}
\begin{split}
    \boldsymbol\theta_{\rm astro}  = & \left\{\log_{10}  f_{\star, 10}, \log_{10} f_{\rm esc, 10}, \right.\\
      & \qquad \left. \alpha_\star, \alpha_{\rm esc}, M_{\rm turn}, t_\star, \log_{10}L_{\rm X}, E_{\rm 0} \right\} \, .
\end{split}
\end{equation*}
Note that, in this work, the $\Lambda$CDM parameters are, by default, reduced to 5 because the optical depth to reionization can be computed from other parameters. In order to test how much reionization data can constrain exotic parameters, we consider three scenarios referred to as {\bf cold dark matter}, {\bf massive neutrinos}, and {\bf warm dark matter}. For each of them, we run several MCMCs, changing either the likelihoods or the set of parameters that are varied. By default, (that is, if not varied and unless specified otherwise) $m_{\nu,1} = 0$ eV, which corresponds to a sum of the neutrino masses equal to $\sim 0.06$ eV and the fraction of warm dark matter is set to 0.

Using a custom-made, slightly modified version of {\tt MontePython}\footnote{ \href{https://github.com/gaetanfacchinetti/montepython_public}{\tt \faGithub /gaetanfacchinetti/montepython\_public}} \cite{Audren:2012wb}, we can perform MCMCs that combine cosmological and astrophysical likelihoods. While the former ones only constrain the cosmological parameters, the latter ones depend both on the cosmology and on the astrophysics model. In particular, we use the following likelihoods.
\begin{itemize}
\item {\bf Planck18}. This likelihood corresponds to the likelihood combination associated with the Planck 2018  TTTEEE+lowE+lensing data \cite{Planck:2018vyg} as well as BAO data from BOSS \cite{BOSS:2016wmc, Beutler:2011hx, Ross:2014qpa}.
\item {\bf Reionization}. This likelihood corresponds to the reionization data from Ref.~\cite{McGreer:2014qwa} -- also referred to as McGreer (2014) in the following. It is calculated from a truncated Gaussian distribution requiring $X_{\rm e}(z=5.6) > 0.96$ $(- 0.05)$ and $X_{\rm e}(z=5.9)  > 0.94$ $(- 0.05)$ at 1$\sigma$.
\item {\bf UV-LFs}. This likelihood refers to UV-luminosity functions. Following Ref.~\cite{HERA:2021noe} it is taken as a split Gaussian, evaluated using Ref.~\cite{Bouwens_2017, Bouwens_2021} data, truncated to a magnitude $M_{\rm UV} > -20$. See appendix~\ref{app:UVLF} for more details.
\end{itemize}
In addition, we have integrated an MCMC toolbox into {\tt NNERO}, enabling efficient MCMC runs with the {\tt emcee} package when $\boldsymbol\theta_{\rm \Lambda CDM}$ is held constant. This functionality is used to cross-check the results from {\tt MontePython} and to estimate potential posterior biases under a fixed cosmological setup.

Since the UV-LFs and Reionization likelihoods are derived assuming fixed cosmologies -- ($\Omega_{\rm m} = 0.3$, $h = 0.7$) for the former and ($\Omega_{\rm m} = 0.28$, $h = 0.7$) for the latter -- there is a minor inconsistency due to the fact that these parameters vary in the MCMC. However, since these fixed values are close to the best-fit parameters from the Planck18 likelihood, we expect that making both likelihoods explicitly cosmology-dependent would not significantly alter our conclusions. Moreover, as noted in \cite{McGreer:2014qwa}, the choice of cosmology has a negligible impact on their constraints, as it only affects the conversion of redshifts to comoving distances.

Finally, MCMC runs that jointly constrain $\boldsymbol\theta_{\rm \Lambda CDM}$ and $\boldsymbol\theta_{\rm astro}$ can be seamlessly extended to include 21cm data, as $\tau$ is computed consistently from the same astrophysics model that is used in {\tt 21cmFAST} to compute the 21cm signal. Specifically, we perform Fisher forecasts for the 21cm power spectrum, assuming HERA sensitivity with 1000 hours of observations, using {\tt 21cmCLAST} and {\tt 21cmCAST}\footnote{\href{https://github.com/gaetanfacchinetti/21cmCAST}{\tt \faGithub /gaetanfacchinetti/21cmCAST}} following the method detailed in Ref.~\cite{Facchinetti:2023slb}. Unless specified otherwise, the fiducial model is selected to match the median values of the MCMC posteriors. Denoting the Fisher matrix for HERA as ${\bf F_{\rm HERA}}$  the combined sensitivity can be estimated from the following covariance matrix:
\begin{equation}
    {\bf C} \equiv \left[{\bf F}_{\rm HERA} + {\bf C}^{-1}_{\rm MCMC}\right]^{-1} \, ,
\end{equation}
where ${\bf C}_{\rm MCMC}$ represents the covariance matrix obtained from the MCMC analysis.

The list of parameters and likelihoods included in each inference run or forecast is reported in table~\ref{tab:mcmcs}. We consider MCMCs performed with {\tt MontePython} to have converged when they meet the Gelman-Rubin criterion of $|R - 1| < 0.1$ for all parameters, as defined in {\tt MontePython} for chains of varying lengths. We consider MCMC performed with {\tt emcee} to have converged when the total length of the chains is higher than 50 times the autocorrelation length for all parameters.

\subsection{Results and discussion}

\begin{figure*}[htbp]
    \centering
    \includegraphics[width=0.99\linewidth]{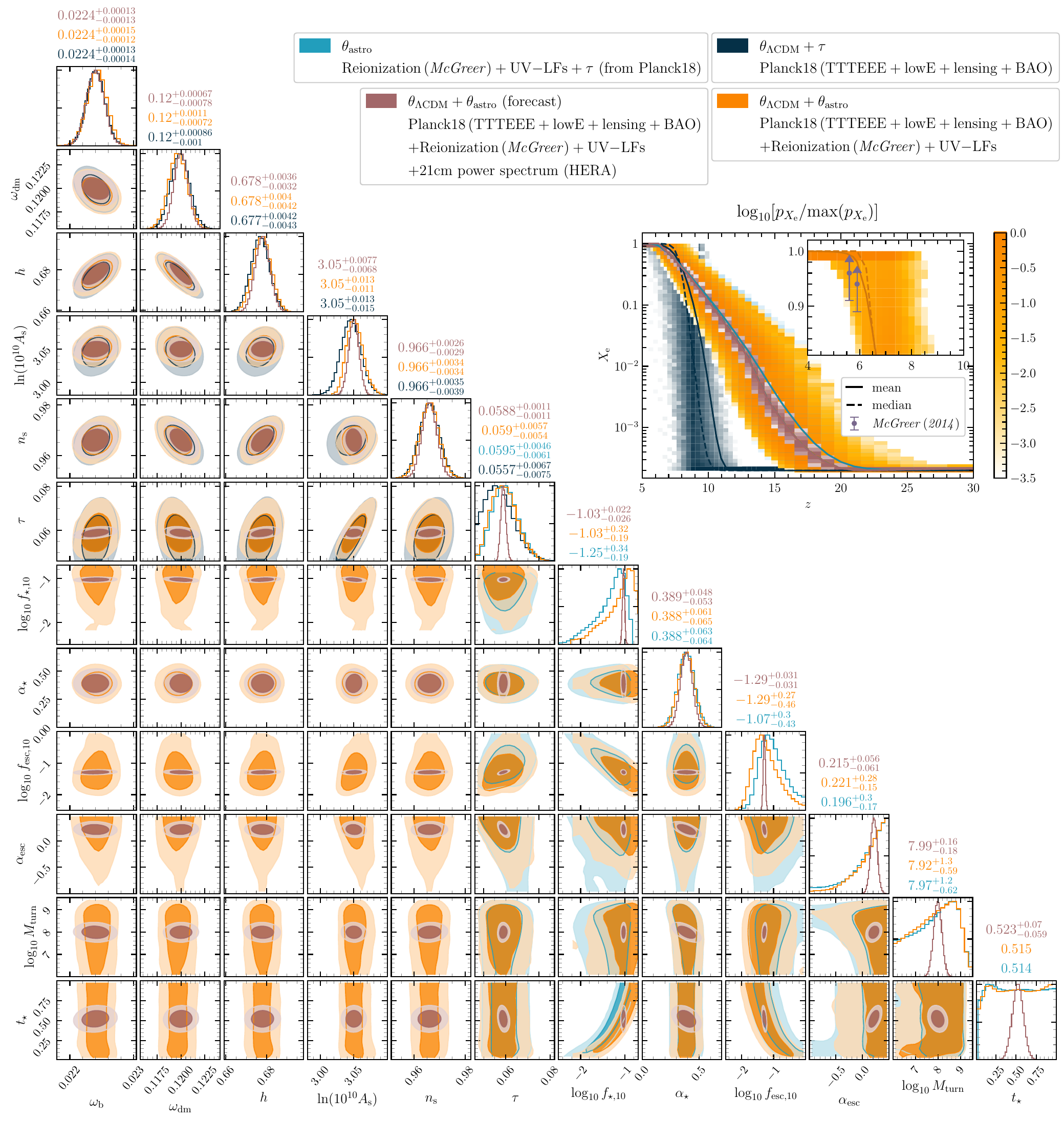}
    \caption{Posteriors for the combined inference on cosmological and astrophysical parameters in the cold dark matter scenario. The dark-shaded contours show the MCMC results for $\boldsymbol\theta_{\rm \Lambda CDM} + \tau$, where $\tau$ is treated as a free parameter using Planck18 data alone. The orange-shaded contours represent the MCMC results when including both $\boldsymbol\theta_{\rm \Lambda CDM}$ and astrophysical parameters $\boldsymbol\theta_{\rm astro}$, incorporating Planck18, reionization, and UV-LF data. The blue-shaded contours correspond to an MCMC analysis on $\boldsymbol\theta_{\rm astro}$ alone, fixing $\boldsymbol\theta_{\rm \Lambda CDM}$ and using only the reionization and UV-LF data. Finally, the taupe contours combine the covariance matrix from the orange posterior with a Fisher forecast for the 21cm power spectrum. The upper right panel shows the evolution of the free-electron fraction $X_e$ with redshift, where the Planck18-only run uses the hyperbolic tangent parametrization, while all other cases use {\tt NNERO}. The inset zooms in on redshift $z \sim 6$, highlighting reionization constraints at the 1$\sigma$ level, marked by purple arrows. For clarity, we omit the posteriors on $\log_{10} L_X$ and $E_0$ as these parameters remain largely unconstrained. Median values and 68\% confidence intervals are displayed at the top of each column.}
    \label{fig:mcmc_cdm}
\end{figure*}

\begin{figure*}[htbp]
    \centering
    \includegraphics[width=0.99\linewidth]{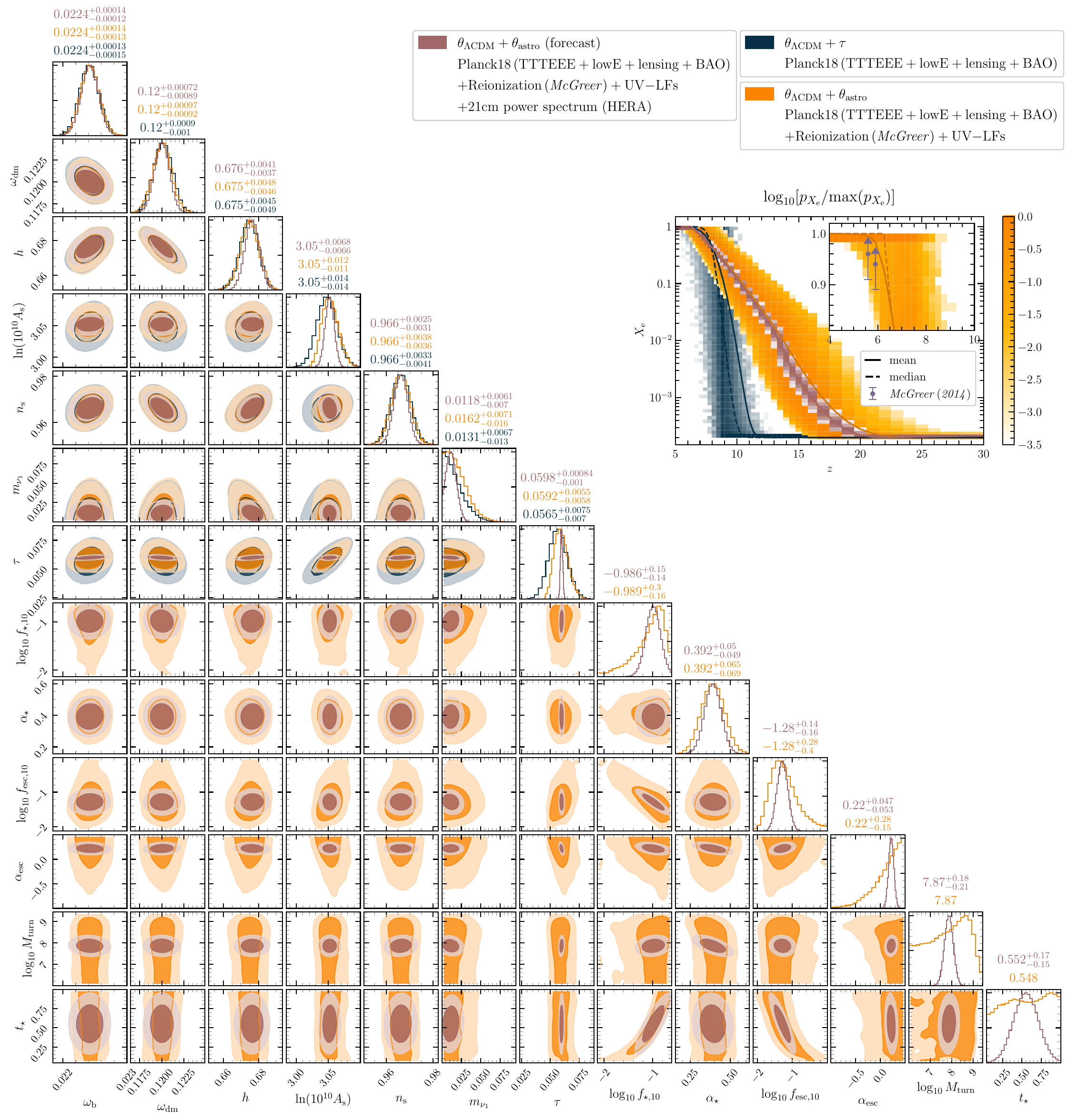}
    \caption{Posteriors for the combined inference on cosmological and astrophysical parameters in the massive neutrino scenario. The dark-shaded contours show the MCMC results for $\boldsymbol\theta_{\rm \Lambda CDM} + \tau  + m_{\rm \nu,1}$, where $\tau$ is treated as a free parameter using Planck18 data alone. The orange-shaded contours represent the MCMC results when including both $\boldsymbol\theta_{\rm \Lambda CDM} + m_{\rm \nu,1}$ and astrophysical parameters $\boldsymbol\theta_{\rm astro}$, incorporating Planck18, reionization, and UV-LF data. The taupe contours combine the covariance matrix from the orange posterior with a Fisher forecast for the 21cm power spectrum. The upper right panel shows the evolution of the free-electron fraction $X_e$ with redshift, where the Planck18-only run adopts the hyperbolic tangent parametrization, while all other cases use {\tt NNERO}. The inset zooms in on redshift $z \sim 6$, highlighting reionization constraints at the 1$\sigma$ level, marked by purple arrows. For clarity, we omit the posteriors on $\log_{10} L_X$ and $E_0$ as these parameters remain largely unconstrained. Median values and 68\% confidence intervals are displayed at the top of each column.}
    \label{fig:mcmc_mnu}
\end{figure*}

\begin{figure*}[htbp]
    \centering
    \includegraphics[width=0.99\linewidth]{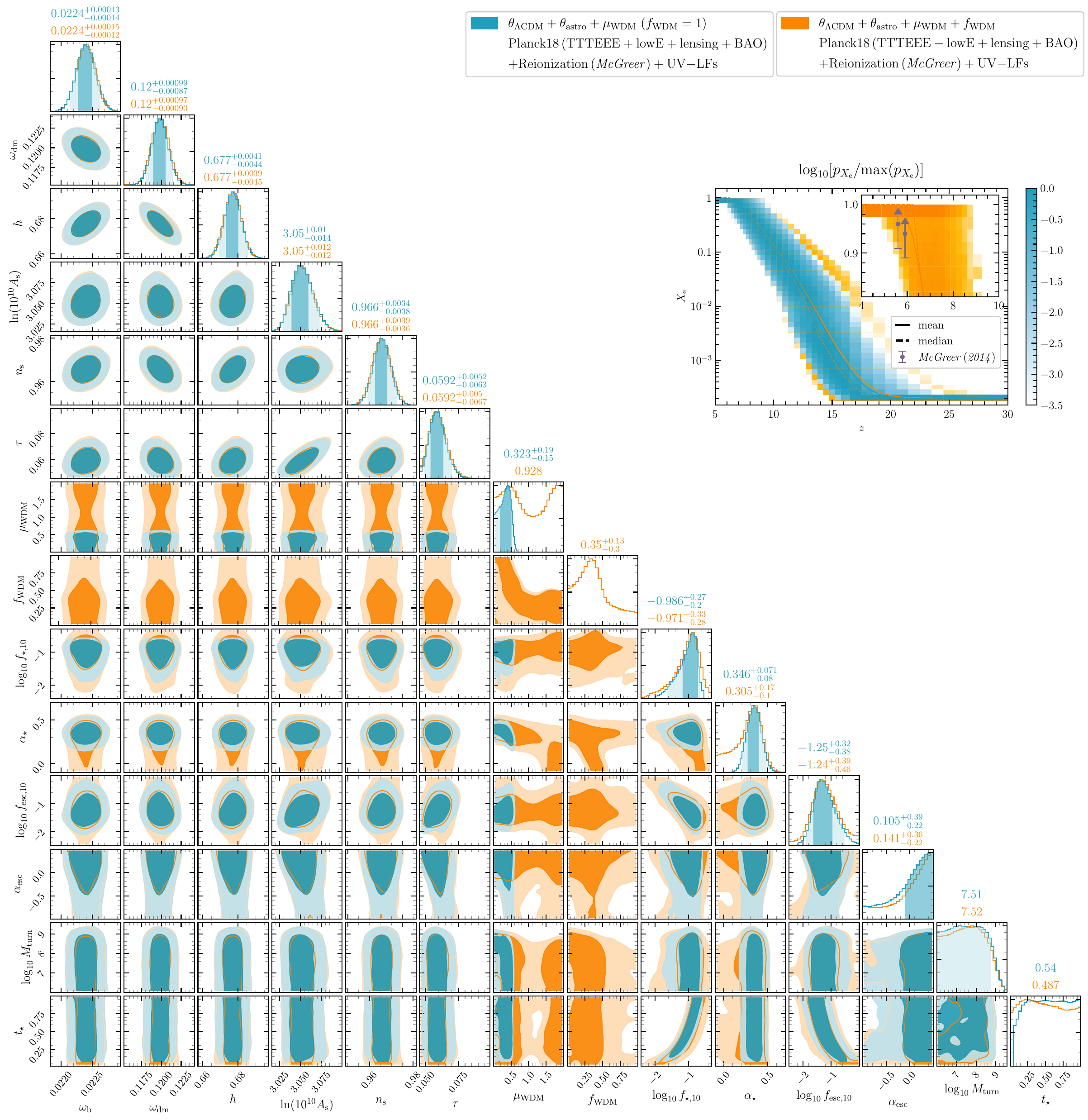}
    \caption{Posteriors for the combined inference on cosmological and astrophysical parameters in the warm dark matter scenario. The blue-shaded contours show MCMC results for $\boldsymbol\theta_{\rm \Lambda CDM}  + \boldsymbol\theta_{\rm astro} + \mu_{\rm WDM}$, assuming a fixed $f_{\rm WDM} = 1$, while the orange-shaded contours allow $f_{\rm WDM}$ to vary freely. These analyses incorporate Planck18, reionization, and UV-LF data. The upper right panel shows the redshift evolution of the free-electron fraction $X_e$, modeled using {\tt NNERO}. The inset zooms in on redshift $z \sim 6$, highlighting reionization constraints at the 1$\sigma$ level, marked by purple arrows. For clarity, posteriors on $\log_{10} L_X$ and $E_0$ are omitted as these parameters remain largely unconstrained. Median values and 68\% confidence intervals are displayed at the top of each column.}
    \label{fig:mcmc_wdm}
\end{figure*}

\subsubsection{Cold dark matter}

In the case of cold dark matter, we compare four different scenarios. The results are shown in Fig.~\ref{fig:mcmc_cdm}, where triangle plots from the three MCMC runs and one Fisher analysis are overlaid. The Planck18-only run, which treats $\boldsymbol\theta_{\rm \Lambda CDM}$ and $\tau$ as free parameters, is depicted in dark shades. The second run, which incorporates all likelihoods, astrophysical parameters, and calculates $\tau$, is shown in orange shades. In the third MCMC run, shown in blue shades, $\boldsymbol\theta_{\rm cosmo}$ is fixed and $\boldsymbol\theta_{\rm astro}$ is constrained from Reionization data, UV-LF and from the posterior on $\tau$ obtained in the first Planck18-only run. Finally, for the taupe contour, we combine the covariance matrix of the orange run to that of a Fisher forecast for the 21cm power spectrum as will be measured with HERA.

The orange contours provide tighter constraints on $\tau$, shifting it to slightly higher values from $\tau = 0.0557^{+0.0067}_{-0.0075}$ to $\tau = 0.059^{+0.0057}_{-0.0054}$. This agrees with Ref.\cite{Qin:2020xrg}, though our constraints are slightly tighter, maybe in part due to differences in the halo mass function and window filter functions used for the matter power spectrum variance. The higher $\tau$ values align with the $X_e(z)$ evolution shown in the top-right panel, where the distribution from the hyperbolic tangent in the Planck18-only run (dark) is compared to the {\tt NNERO}-derived distribution (orange). The reionization constraints from Ref.~\cite{McGreer:2014qwa}, marked with purple arrows, combined with the smoother variations of $X_e$ over an extended redshift range, naturally result in a higher $\tau$. Since $\tau$ is degenerate with the amplitude of the matter power spectrum, $\ln(10^{10}A_{\rm s})$ is also restricted to higher values. Other cosmological parameters show consistent results across the two runs. The constraints on astrophysical parameters align with Refs.~\cite{HERA:2021noe, Qin:2020xrg}. In addition, we observe $\alpha_{\rm esc}$ and $\log_{10}f_{\rm esc, 10}$ small correlations with $\ln(10^{10}A_{\rm s})$ due to their correlation with $\tau$.

The blue contours are obtained by fixing $\boldsymbol\theta_{\rm \Lambda CDM}$ to the best-fit values from Planck18, using {\tt emcee} and {\tt NNERO}’s built-in tools. Given the weak or negligible degeneracies between astrophysical and cosmological parameters, the blue contours closely follow the orange ones. Only small biases are observed for $\log_{10}f_{\star, 10}$ and $f_{\rm esc, 10}$ towards lower and larger values respectively. Nevertheless, the overall agreement indicates that reionization and UV-LFs data are not yet precise enough to impose significant shifts or constraints on cosmological parameters, meaning they do not strongly backreact on the astrophysical posteriors. Fixing the cosmology, therefore, provides a reasonable approximation for constraining astrophysical parameters. This is particularly interesting as the blue contours were generated within a few hours on a personal computer, whereas the orange contours required a few days of parallel computation on a cluster.

Finally, the taupe contours are the HERA Fisher forecast combined with all other likelihoods. It shows that HERA will provide a much better accuracy on the optical depth to reionization, reducing the 68\% CL limit by a factor five, from $\tau \sim 0.059\pm 0.0055 $ to $\tau \sim 0.059 \pm 0.001$. Moreover, it will also give tighter constraints on $\ln(10^{10}A_{\rm s})$ with 68\% CL errors reduced by half compared to Planck18 alone. It can also help slightly reduce the errors on $h$ and $n_{\rm s}$. In addition, it will provide much better accuracy on the astrophysical parameters that are not well constrained from the other likelihoods like $M_{\rm turn}$ and $t_{\star}$.

\subsubsection{Massive neutrinos}

The results for the massive neutrinos scenario are shown in Fig.~\ref{fig:mcmc_mnu}. Here again, we compare the results from Planck18 alone (which includes the lensing likelihood and treats $\tau$ as a free parameter), in dark shades, with those from the full MCMC that incorporates all likelihoods, in orange shades, and from the addition of a Fisher forecast for HERA, in taupe.

The main degeneracies of the neutrino mass $m_{\nu, 1}$ are with $h$ and, to a lesser extent, with $\log_{10} f_{\star, 10}$ and $\alpha_{\rm esc}$. The degeneracy with $h$ already appears in the Planck18-only run and is discussed in Ref.~\cite{Planck:2018vyg}. This discussion is particularly relevant in the context of the Hubble tension. The other two show that by increasing $\log_{10} f_{\star, 10}$ and $\alpha_{\rm esc}$ it is possible to compensate for the decrease in the matter power spectrum due to a large neutrino mass to fit the UV-LFs. Therefore, including all likelihoods (in orange), the data very slightly favor a non-zero mass for the lightest neutrino. This is associated with a slight increase of the optical depth to reionization. The best fit is obtained for $m_{\nu, 1} \sim 0.011$ eV, which is equivalent to $\sum_{\nu} m_\nu \sim 0.08$ eV (at the edge of the current 95\% CL from Planck+DESI-Y1 \cite{DESI:2024mwx}, $\sum_\nu m_\nu < 0.073$ eV). The Fisher forecast discussed below is thus performed assuming that $m_{\nu,1}$ is equal to the best-fit value (rather than the median value which has little sense in that case). 

The combination with the 21cm power spectrum measured with HERA forecasts a sensitivity slightly higher than $1\sigma$ to $m_{\nu, 1} = 0.011$ eV (with $\sigma_{m_{\nu, 1}} \sim 0.009$ eV). Although this will not be enough to claim a detection of such a small mass, it may still give the first hints towards the determination of a non-zero mass of the lightest neutrino and, at least, a complementary approach to the future constraints or discoveries from upcoming DESI data releases and from CMB stage IV experiments. This result is in agreement with Ref.~\cite{Shmueli:2023box} which argued that the 21cm power spectrum data will help constrain the mass of the neutrinos.

\subsubsection{Warm dark matter}

\begin{figure}[htbp]
    \centering
    \includegraphics[width=0.95\linewidth]{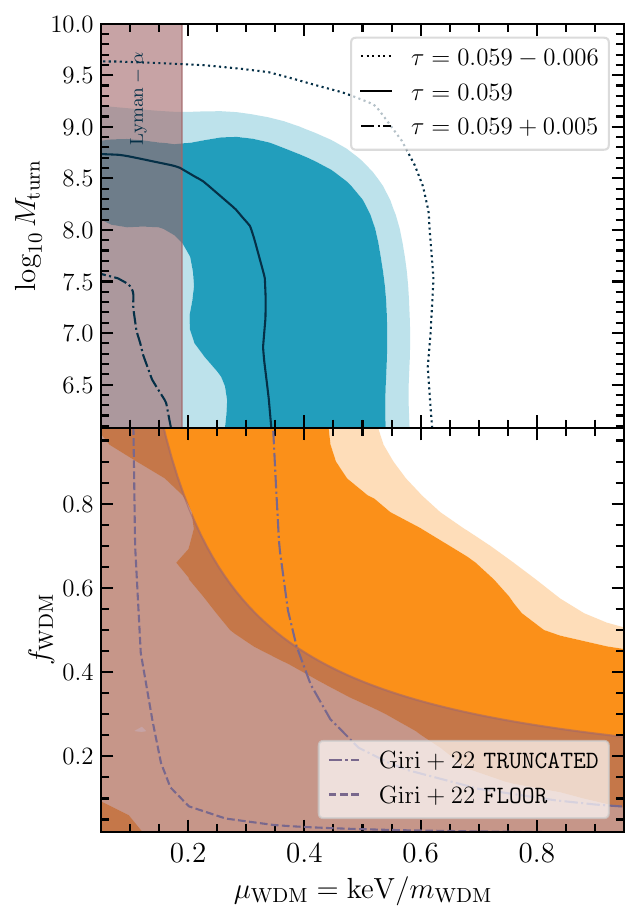}
    \caption{{\bf Upper panel.} Zoom onto the 2D-marginalised posterior in the plane $(\mu_{\rm WDM}, \log_{10} M_{\rm turn})$, assuming $f_{\rm WDM}=1$. The blue shaded area show the result of the full MCMC while the contour lines represent a quick estimate from the Planck18 constraint on the optical depth to reionization. {\bf Lower panel.} Zoom onto the 2D-marginalised posterior in the plane $(\mu_{\rm WDM}, f_{\rm WDM})$. The orange shaded area show the result of the full MCMC. Forecasts obtained in Ref.~\cite{Giri:2022nxq} in two scenarios of galaxy formations using the SKA sensitivity to the 21cm power spectrum are overlaid as purple lines. The taupe areas show the available range of masses from the best current Lyman-$\alpha$ constraints \cite{Palanque-Delabrouille:2019iyz, Hooper:2022byl}.  }
    \label{fig:zoom_mcmc_wdm}
\end{figure}

For the warm dark matter scenario, we run MCMC analyses incorporating the Planck18, Reionization, and UV-LFs likelihoods. The resulting posteriors are shown in Fig.~\ref{fig:mcmc_wdm}, while Fig.~\ref{fig:zoom_mcmc_wdm} provides a zoomed-in view of the 2D marginal posterior in the $(\mu_{\rm WDM}, \log_{10} M_{\rm turn})$ plane (upper panel) and the $(\mu_{\rm WDM}, f_{\rm WDM})$ plane (lower panel). In these figures, the blue contours correspond to the case where $f_{\rm WDM} = 1$, while the orange contours allow $f_{\rm WDM}$ to vary as a free parameter.

In the fixed $f_{\rm WDM} = 1$ case, we obtain a lower bound on the WDM mass: $m_{\rm WDM} > 1.8$ keV at the 95\% CL. This is weaker than the current best constraint from Lyman-$\alpha$ forest data ($m_{\rm WDM} > 5.3$ keV at 95\% CL \cite{Palanque-Delabrouille:2019iyz}, with the excluded region shown in taupe in Fig.~\ref{fig:zoom_mcmc_wdm}). However, as the Lyman-$\alpha$ bound is prone to modeling uncertainites, our result provides an interesting complementary approach. We find that $\mu_{\rm WDM}$ is primarily degenerate with $\log_{10} M_{\rm turn}$, $\alpha_\star$, $\log_{10} f_{\rm esc, 10}$, and $\alpha_{\rm esc}$. A nonzero $\mu_{\rm WDM}$ allows for smaller values of $\alpha_\star$ and $\alpha_{\rm esc}$ and larger values of $\log_{10} f_{\rm esc, 10}$, compensating for the reduced ionization efficiency due to the WDM-induced suppression in the matter power spectrum. This degeneracy is even more pronounced when $f_{\rm WDM} < 1$, as larger $\mu_{\rm WDM}$ values can be accommodated without violating CMB constraints on the optical depth to reionization.

In the upper panel of Fig.\ref{fig:zoom_mcmc_wdm}, we observe that small values of $M_{\rm turn}$—which would otherwise lead to early reionization and an excessively large optical depth—can be compensated by a nonzero $\mu_{\rm WDM}$. Conversely, for large $M_{\rm turn}$, ionization suppression is dominated by $M_{\rm turn}$ itself, leaving $\mu_{\rm WDM}$ unbounded from below. However, $\mu_{\rm WDM}$ cannot be arbitrarily large, as excessive suppression would delay reionization too much, setting the aforementioned lower bound of 1.8 keV. As discussed in Section\ref{sec:reio_history}, our model assumes a single population of stars forming in atomic cooling galaxies. The presence of molecular cooling galaxies could be approximately accounted for by an effectively lower $M_{\rm turn}$, which slightly favor a nonzero $\mu_{\rm WDM}$. Nevertheless, our contours remain consistent with $\mu_{\rm WDM} = 0$ at the 95\% CL, in agreement with Lyman-$\alpha$ constraints (shown in taupe).

Additionally, we overlay a simple estimate obtained by fixing all parameters to their median values, except for $\mu_{\rm WDM}$ and $\log_{10} M_{\rm turn}$, while enforcing the optical depth to reionization to be consistent with Planck18 within a given range. Imposing $0.059 - 0.006 < \tau < 0.059 + 0.005$ yields the dashed and dash-dotted contours. Interestingly, while these contours are slightly broader than the true 68\% CL region (shown in dark blue), they closely trace the shape of the full posterior, suggesting that this quick estimate can serve as a useful first approximation.

In the lower panel of Fig.~\ref{fig:zoom_mcmc_wdm}, we illustrate how allowing $f_{\rm WDM}$ to vary relaxes the constraints on $\mu_{\rm WDM}$. We compare our results to Lyman-$\alpha$ forest constraints from Ref.~\cite{Hooper:2022byl} (shaded taupe area) and to SKA forecasts from Ref.~\cite{Giri:2022nxq} (referred to as Giri+22), which consider two different star formation models in molecular-cooling galaxies: an aggressive model ({\tt FLOOR}) and a more conservative model ({\tt TRUNCATED}). This comparison highlights both the promising constraining power of future 21cm power spectrum measurements and the fact that existing reionization data already provide meaningful bounds on warm dark matter.

Finally, we note that the sensitivity of the 21cm power spectrum to warm dark matter has been recently investigated in depth in Ref.~\cite{Decant:2024bpg}. While it significantly outperforms current reionization and UV-LF constraints, it remains affected by the same degeneracy with $M_{\rm turn}$. See also Ref.~\cite{Chatterjee:2023mlh} for a complementary study combining constraints from the global 21cm signal.

\section{Summary and conclusions}
\label{sec:conclusion}

We have presented {\tt NNERO}, a neural network-based emulator for the free-electron fraction from cosmic dawn to the epoch of reionization. Trained to achieve percent-level precision on the optical depth, {\tt NNERO} efficiently models scenarios involving both nonzero neutrino masses and a component of warm dark matter. Such a level of precision is more than enough to be used in a CMB analysis that can only constrain the optical depth with a 10\% precision at 68\% CL. In addition, by generating training data in parallel within a few days and training the neural network in mere minutes, we have dramatically reduced computational costs. We then integrated this emulator into a cosmological MCMC solver, incorporating two new likelihoods: one for the free-electron fraction at redshift $z \sim 6$, and another for UV luminosity functions. The resulting MCMCs, involving $\sim 15$ parameters, converged in just a few days—a significant improvement over direct {\tt 21cmFAST}-based MCMCs, which typically require several months. This fast and flexible approach enables the exploration of a wider range of cosmological and astrophysical scenarios, including models beyond standard physics. Furthermore, we assessed the impact of upcoming HERA measurements of the 21cm power spectrum by combining our MCMC results with Fisher forecasts. Both {\tt NNERO} and the trained models are publicly available on GitHub.

Our results show that performing a full MCMC analysis -- including Planck18, reionization, and UV-LF data -- slightly tightens constraints on the optical depth compared to using Planck18 data alone in $\Lambda$CDM. Specifically, we refine the measurement from $\tau = 0.056 \pm 0.007$ to $\tau = 0.059 \pm 0.0055$ in the cold dark matter scenario. When incorporating HERA Fisher forecasts, assuming 1000 hours of observations, we find that 21cm data could significantly improve the constraint by reducing the error by a factor five to reach $\tau = 0.059 \pm 0.001$. Moreover, it will further constrain both astrophysical parameters and key cosmological quantities, including the amplitude of the primordial power spectrum $A_{\rm s}$ and the sum of the neutrino masses. The latter, which oscillation experiments confirm to be nonzero, is already being probed with great precision by DESI. Our results suggest that an independent constraint from the 21cm power spectrum should also be achievable (if not by HERA, most probably by SKA), providing an important cross-validation. In the warm dark matter scenario, the combined constraints (without 21cm data) place a lower bound on $m_{\rm WDM}$ when $f_{\rm WDM} \gtrsim 0.5$, in particular, $m_{\rm WDM} \ge 1.8$ keV for $f_{\rm WDM} = 1$. This offers a complementary limit to existing Lyman-$\alpha$ forest constraints -- though it remains weaker than the strongest bounds currently available.

Our approach relies on two key assumptions: (i) using a halo mass function model optimized for scenarios with suppressed small-scale matter power spectra, and (ii) computing the UV-LF and reionization likelihoods using a cosmology slightly different from the one they were originally derived from. However, we do not expect either of these factors to significantly affect our conclusions, particularly regarding the efficiency of our methodology.

Looking ahead, our next step will be to apply {\tt NNERO} to scenarios that either enhance the free-electron fraction compared to the standard $\Lambda$CDM prediction or change the matter power spectrum in different ways (e.g., by introducing a tilt in the primordial power spectrum). Additionally, integrating {\tt NNERO} with other datasets -- particularly the Lyman-$\alpha$ flux power spectrum -- will further improve constraints. Most importantly, once real 21cm power spectrum data becomes available, a self-consistent combination of CMB and 21cm data will be essential to perform inferences in this new era of late-time precision cosmology. To that end, we plan to extend {\tt NNERO} to emulate the 21cm power spectrum. This will allow us to refine constraints on astrophysical and $\Lambda$CDM parameters while exploring exotic dark matter models with unprecedented accuracy.

\begin{acknowledgments}
    I thank Q. Decant, V. Dandoy, and C. Döring for valuable discussions. I am especially grateful to L. Lopez-Honorez for insightful discussions and for her crucial role in shaping the ideas that led to this work. I am a Postdoctoral Researcher of the Fonds de la Recherche Scientifique – FNRS. Computational resources have been provided by the Consortium des Equipements de Calcul Intensif (CECI), funded by the Fonds de la Recherche Scientifique de Belgique (F.R.S.-FNRS) under Grant No. 2.5020.11 and by the Walloon Region of Belgium.
\end{acknowledgments}

\appendix

\section{A short {\tt NNERO} manual}
\label{app:NNERO_manual}

{\tt NNERO} is devided into four main components. First, {\tt constants.py} and {\tt data.py} handle raw data and partition it in training, testing and validation datasets. Then, {\tt network.py, classifier.py} and {\tt regressor.py} define Networks objects that contain all the neural networks structure information and weights. Afterwards, {\tt predictor.py} uses {\tt cosmology.py} as well as trained objects Regressor and Classifier to predict $X_e$ and $\tau$. Finally, {\tt NNERO} comes with build in tools to compute UV-LFs with {\tt astrophysics.py}, run likelihoods with {\tt mcmc.py} and analyse chains and produce triangle plots with {\tt analysis.py}.

The typical use case is to get $X_e$ or $\tau$ for a given set of parameters. A simple example is shown in the lines of code given below.

\begin{lstlisting}[language=Python,numbers=none]
import nnero
from nnero import predict_Xe
from nnero import predict_tau

# load classifier and regressor at <path_*>
# if no <path> given, the defaults are loaded
classifier=nnero.Classifier.load(<path_c>)
regressor=nnero.Regressor.load(<path_r>)

# print general information
# - structure of the network
# - input parameters name and training range
regressor.info()

# get Xe from loaded classifier and regressor
# **kwargs can be any parameter that is 
# printed calling the info() function above
Xe=predict_Xe(classifier, regressor, **kwargs)
z=regressor.z

# get tau similarly
tau=predict_tau(classifier, regressor, **kwargs)
\end{lstlisting}

More specialized function exists to accelerate the computation if one provides an {\tt numpy} array of input parameters for instance instead of a {\tt kwargs} dictionary. We refer to the github repository homepage for more information. \\

\section{UV-luminosity function}
\label{app:UVLF}

In this appendix, we detail the evaluation of the UV-LFs likelihood. Following \cite{Park:2018ljd}, they can be written, in terms of the UV-magnitude $M_{\rm UV}$ as
\begin{equation}
    \phi_{\rm UV}(M_{\rm UV}) = f_{\rm duty} \frac{\partial n(M, z)}{\partial M} \left| \frac{{\rm d}M}{{\rm d}M_{\rm UV}}\right| 
\end{equation}
where $\partial n(M, z) / \partial M$ is the halo mass function at redshift $z$ of the entire Universe. In order to evaluate this expression, we need to associate an halo mass $M$ to a given UV magnitude. This can be done using the following relation between the star formation rate and the UV magnitude \cite{Sun_2016, 1983ApJ...266..713O},
\begin{equation}
    \frac{\dot M_\star}{10^{10} ~{\rm M_\odot}} = \Gamma_{\rm UV}10^{-aM_{\rm UV}}\, ,
\end{equation}
with $\Gamma_{\rm UV} = 5.16 \times 10^{-18} ~ {\rm yr}^{-1}$ and $a = 0.4$. In the Park model \cite{Park:2018ljd}, the star formation rate is given by
\begin{equation}
    \dot M_\star = M_\star \frac{H(z)}{t_\star}
\end{equation}
where $t_\star$ is a characteristic star-formation timescale between 0 and 1. Using Eq.~(\ref{eq:mstar}) for the expression of $M_\star$, we obtain
\begin{equation}
    \frac{M}{10^{10}\, {\rm M_\odot }} = 10^{-aM_{\rm UV}}
    \begin{cases}
        c_0(z) ~ {\rm if} ~ f_{\star, 10} \left(\frac{M}{10^{10}\, {\rm M_\odot}}\right) > 1\\
        c_1(z) \quad {\rm else}\, ,
    \end{cases}
\end{equation}
with $c_0$ and $c_1$ are two functions of the redshift,
\begin{equation}
    c_0(z) \equiv  t_\star\frac{\omega_{\rm m}}{\omega_{\rm b}} \frac{\Gamma_{\rm UV}}{H(z)} \quad {\rm and} \quad c_1(z) \equiv \left[\frac{c_0(z)}{f_{\star, 10}}\right]^\frac{1}{\alpha_\star+1} \, 
\end{equation}
Therefore, the derivative in the expression of the UV-LFs is simply given by
\begin{equation}
    \frac{{\rm d}M}{{\rm d}M_{\rm UV}} = -a M \ln 10
    \begin{cases}
        1 ~ {\rm if} ~ f_{\star, 10} \left(\frac{M}{10^{10}\, M_\odot} \right) > 1\\
        \frac{1}{\alpha_\star+1} \quad {\rm else}\, .
    \end{cases}
\end{equation}
In practice, the observed UV-LFs are reproduced for $f_{\star, 10} M/ (10^{10}\, {\rm M_\odot}) \le 1$. Therefore, from the expression of $c_1$ we see that in the UV-LF likelihood $f_{\star, 10}$ and $t_\star$ are two degenerate parameters, but in the MCMC, the prior of $f_{\star, 10}$ is log-uniform, while that of $t_\star$ is uniform.

\bibliography{main}

\begin{thebibliography}{87}%
\makeatletter
\providecommand \@ifxundefined [1]{%
 \@ifx{#1\undefined}
}%
\providecommand \@ifnum [1]{%
 \ifnum #1\expandafter \@firstoftwo
 \else \expandafter \@secondoftwo
 \fi
}%
\providecommand \@ifx [1]{%
 \ifx #1\expandafter \@firstoftwo
 \else \expandafter \@secondoftwo
 \fi
}%
\providecommand \natexlab [1]{#1}%
\providecommand \enquote  [1]{``#1''}%
\providecommand \bibnamefont  [1]{#1}%
\providecommand \bibfnamefont [1]{#1}%
\providecommand \citenamefont [1]{#1}%
\providecommand \href@noop [0]{\@secondoftwo}%
\providecommand \href [0]{\begingroup \@sanitize@url \@href}%
\providecommand \@href[1]{\@@startlink{#1}\@@href}%
\providecommand \@@href[1]{\endgroup#1\@@endlink}%
\providecommand \@sanitize@url [0]{\catcode `\\12\catcode `\$12\catcode
  `\&12\catcode `\#12\catcode `\^12\catcode `\_12\catcode `\%12\relax}%
\providecommand \@@startlink[1]{}%
\providecommand \@@endlink[0]{}%
\providecommand \url  [0]{\begingroup\@sanitize@url \@url }%
\providecommand \@url [1]{\endgroup\@href {#1}{\urlprefix }}%
\providecommand \urlprefix  [0]{URL }%
\providecommand \Eprint [0]{\href }%
\providecommand \doibase [0]{https://doi.org/}%
\providecommand \selectlanguage [0]{\@gobble}%
\providecommand \bibinfo  [0]{\@secondoftwo}%
\providecommand \bibfield  [0]{\@secondoftwo}%
\providecommand \translation [1]{[#1]}%
\providecommand \BibitemOpen [0]{}%
\providecommand \bibitemStop [0]{}%
\providecommand \bibitemNoStop [0]{.\EOS\space}%
\providecommand \EOS [0]{\spacefactor3000\relax}%
\providecommand \BibitemShut  [1]{\csname bibitem#1\endcsname}%
\let\auto@bib@innerbib\@empty
\bibitem [{\citenamefont {Komatsu}\ \emph {et~al.}(2011)\citenamefont {Komatsu}
  \emph {et~al.}}]{WMAP:2010qai}%
  \BibitemOpen
  \bibfield  {author} {\bibinfo {author} {\bibfnamefont {E.}~\bibnamefont
  {Komatsu}} \emph {et~al.} (\bibinfo {collaboration} {WMAP}),\ }\bibfield
  {title} {\bibinfo {title} {{Seven-Year Wilkinson Microwave Anisotropy Probe
  (WMAP) Observations: Cosmological Interpretation}},\ }\href
  {https://doi.org/10.1088/0067-0049/192/2/18} {\bibfield  {journal} {\bibinfo
  {journal} {Astrophys. J. Suppl.}\ }\textbf {\bibinfo {volume} {192}},\
  \bibinfo {pages} {18} (\bibinfo {year} {2011})},\ \Eprint
  {https://arxiv.org/abs/1001.4538} {arXiv:1001.4538 [astro-ph.CO]}
  \BibitemShut {NoStop}%
\bibitem [{\citenamefont {Aghanim}\ \emph {et~al.}(2020)\citenamefont {Aghanim}
  \emph {et~al.}}]{Planck:2018vyg}%
  \BibitemOpen
  \bibfield  {author} {\bibinfo {author} {\bibfnamefont {N.}~\bibnamefont
  {Aghanim}} \emph {et~al.} (\bibinfo {collaboration} {Planck}),\ }\bibfield
  {title} {\bibinfo {title} {{Planck 2018 results. VI. Cosmological
  parameters}},\ }\href {https://doi.org/10.1051/0004-6361/201833910}
  {\bibfield  {journal} {\bibinfo  {journal} {Astron. Astrophys.}\ }\textbf
  {\bibinfo {volume} {641}},\ \bibinfo {pages} {A6} (\bibinfo {year} {2020})},\
  \bibinfo {note} {[Erratum: Astron.Astrophys. 652, C4 (2021)]},\ \Eprint
  {https://arxiv.org/abs/1807.06209} {arXiv:1807.06209 [astro-ph.CO]}
  \BibitemShut {NoStop}%
\bibitem [{\citenamefont {Zhang}\ \emph {et~al.}(2006)\citenamefont {Zhang},
  \citenamefont {Chen}, \citenamefont {Lei},\ and\ \citenamefont
  {Si}}]{Zhang:2006fr}%
  \BibitemOpen
  \bibfield  {author} {\bibinfo {author} {\bibfnamefont {L.}~\bibnamefont
  {Zhang}}, \bibinfo {author} {\bibfnamefont {X.-L.}\ \bibnamefont {Chen}},
  \bibinfo {author} {\bibfnamefont {Y.-A.}\ \bibnamefont {Lei}},\ and\ \bibinfo
  {author} {\bibfnamefont {Z.-G.}\ \bibnamefont {Si}},\ }\bibfield  {title}
  {\bibinfo {title} {{The impacts of dark matter particle annihilation on
  recombination and the anisotropies of the cosmic microwave background}},\
  }\href {https://doi.org/10.1103/PhysRevD.74.103519} {\bibfield  {journal}
  {\bibinfo  {journal} {Phys. Rev. D}\ }\textbf {\bibinfo {volume} {74}},\
  \bibinfo {pages} {103519} (\bibinfo {year} {2006})},\ \Eprint
  {https://arxiv.org/abs/astro-ph/0603425} {arXiv:astro-ph/0603425}
  \BibitemShut {NoStop}%
\bibitem [{\citenamefont {Zhang}\ \emph {et~al.}(2007)\citenamefont {Zhang},
  \citenamefont {Chen}, \citenamefont {Kamionkowski}, \citenamefont {Si},\ and\
  \citenamefont {Zheng}}]{Zhang:2007zzh}%
  \BibitemOpen
  \bibfield  {author} {\bibinfo {author} {\bibfnamefont {L.}~\bibnamefont
  {Zhang}}, \bibinfo {author} {\bibfnamefont {X.}~\bibnamefont {Chen}},
  \bibinfo {author} {\bibfnamefont {M.}~\bibnamefont {Kamionkowski}}, \bibinfo
  {author} {\bibfnamefont {Z.-g.}\ \bibnamefont {Si}},\ and\ \bibinfo {author}
  {\bibfnamefont {Z.}~\bibnamefont {Zheng}},\ }\bibfield  {title} {\bibinfo
  {title} {{Constraints on radiative dark-matter decay from the cosmic
  microwave background}},\ }\href {https://doi.org/10.1103/PhysRevD.76.061301}
  {\bibfield  {journal} {\bibinfo  {journal} {Phys. Rev. D}\ }\textbf {\bibinfo
  {volume} {76}},\ \bibinfo {pages} {061301} (\bibinfo {year} {2007})},\
  \Eprint {https://arxiv.org/abs/0704.2444} {arXiv:0704.2444 [astro-ph]}
  \BibitemShut {NoStop}%
\bibitem [{\citenamefont {Mapelli}\ \emph {et~al.}(2006)\citenamefont
  {Mapelli}, \citenamefont {Ferrara},\ and\ \citenamefont
  {Pierpaoli}}]{Mapelli:2006ej}%
  \BibitemOpen
  \bibfield  {author} {\bibinfo {author} {\bibfnamefont {M.}~\bibnamefont
  {Mapelli}}, \bibinfo {author} {\bibfnamefont {A.}~\bibnamefont {Ferrara}},\
  and\ \bibinfo {author} {\bibfnamefont {E.}~\bibnamefont {Pierpaoli}},\
  }\bibfield  {title} {\bibinfo {title} {{Impact of dark matter decays and
  annihilations on reionzation}},\ }\href
  {https://doi.org/10.1111/j.1365-2966.2006.10408.x} {\bibfield  {journal}
  {\bibinfo  {journal} {Mon. Not. Roy. Astron. Soc.}\ }\textbf {\bibinfo
  {volume} {369}},\ \bibinfo {pages} {1719} (\bibinfo {year} {2006})},\ \Eprint
  {https://arxiv.org/abs/astro-ph/0603237} {arXiv:astro-ph/0603237}
  \BibitemShut {NoStop}%
\bibitem [{\citenamefont {Padmanabhan}\ and\ \citenamefont
  {Finkbeiner}(2005)}]{Padmanabhan:2005es}%
  \BibitemOpen
  \bibfield  {author} {\bibinfo {author} {\bibfnamefont {N.}~\bibnamefont
  {Padmanabhan}}\ and\ \bibinfo {author} {\bibfnamefont {D.~P.}\ \bibnamefont
  {Finkbeiner}},\ }\bibfield  {title} {\bibinfo {title} {{Detecting dark matter
  annihilation with CMB polarization: Signatures and experimental prospects}},\
  }\href {https://doi.org/10.1103/PhysRevD.72.023508} {\bibfield  {journal}
  {\bibinfo  {journal} {Phys. Rev. D}\ }\textbf {\bibinfo {volume} {72}},\
  \bibinfo {pages} {023508} (\bibinfo {year} {2005})},\ \Eprint
  {https://arxiv.org/abs/astro-ph/0503486} {arXiv:astro-ph/0503486}
  \BibitemShut {NoStop}%
\bibitem [{\citenamefont {Galli}\ \emph {et~al.}(2009)\citenamefont {Galli},
  \citenamefont {Iocco}, \citenamefont {Bertone},\ and\ \citenamefont
  {Melchiorri}}]{Galli:2009zc}%
  \BibitemOpen
  \bibfield  {author} {\bibinfo {author} {\bibfnamefont {S.}~\bibnamefont
  {Galli}}, \bibinfo {author} {\bibfnamefont {F.}~\bibnamefont {Iocco}},
  \bibinfo {author} {\bibfnamefont {G.}~\bibnamefont {Bertone}},\ and\ \bibinfo
  {author} {\bibfnamefont {A.}~\bibnamefont {Melchiorri}},\ }\bibfield  {title}
  {\bibinfo {title} {{CMB constraints on Dark Matter models with large
  annihilation cross-section}},\ }\href
  {https://doi.org/10.1103/PhysRevD.80.023505} {\bibfield  {journal} {\bibinfo
  {journal} {Phys. Rev. D}\ }\textbf {\bibinfo {volume} {80}},\ \bibinfo
  {pages} {023505} (\bibinfo {year} {2009})},\ \Eprint
  {https://arxiv.org/abs/0905.0003} {arXiv:0905.0003 [astro-ph.CO]}
  \BibitemShut {NoStop}%
\bibitem [{\citenamefont {Galli}\ \emph {et~al.}(2011)\citenamefont {Galli},
  \citenamefont {Iocco}, \citenamefont {Bertone},\ and\ \citenamefont
  {Melchiorri}}]{Galli:2011rz}%
  \BibitemOpen
  \bibfield  {author} {\bibinfo {author} {\bibfnamefont {S.}~\bibnamefont
  {Galli}}, \bibinfo {author} {\bibfnamefont {F.}~\bibnamefont {Iocco}},
  \bibinfo {author} {\bibfnamefont {G.}~\bibnamefont {Bertone}},\ and\ \bibinfo
  {author} {\bibfnamefont {A.}~\bibnamefont {Melchiorri}},\ }\bibfield  {title}
  {\bibinfo {title} {{Updated CMB constraints on Dark Matter annihilation
  cross-sections}},\ }\href {https://doi.org/10.1103/PhysRevD.84.027302}
  {\bibfield  {journal} {\bibinfo  {journal} {Phys. Rev. D}\ }\textbf {\bibinfo
  {volume} {84}},\ \bibinfo {pages} {027302} (\bibinfo {year} {2011})},\
  \Eprint {https://arxiv.org/abs/1106.1528} {arXiv:1106.1528 [astro-ph.CO]}
  \BibitemShut {NoStop}%
\bibitem [{\citenamefont {Finkbeiner}\ \emph {et~al.}(2012)\citenamefont
  {Finkbeiner}, \citenamefont {Galli}, \citenamefont {Lin},\ and\ \citenamefont
  {Slatyer}}]{Finkbeiner:2011dx}%
  \BibitemOpen
  \bibfield  {author} {\bibinfo {author} {\bibfnamefont {D.~P.}\ \bibnamefont
  {Finkbeiner}}, \bibinfo {author} {\bibfnamefont {S.}~\bibnamefont {Galli}},
  \bibinfo {author} {\bibfnamefont {T.}~\bibnamefont {Lin}},\ and\ \bibinfo
  {author} {\bibfnamefont {T.~R.}\ \bibnamefont {Slatyer}},\ }\bibfield
  {title} {\bibinfo {title} {{Searching for Dark Matter in the CMB: A Compact
  Parameterization of Energy Injection from New Physics}},\ }\href
  {https://doi.org/10.1103/PhysRevD.85.043522} {\bibfield  {journal} {\bibinfo
  {journal} {Phys. Rev. D}\ }\textbf {\bibinfo {volume} {85}},\ \bibinfo
  {pages} {043522} (\bibinfo {year} {2012})},\ \Eprint
  {https://arxiv.org/abs/1109.6322} {arXiv:1109.6322 [astro-ph.CO]}
  \BibitemShut {NoStop}%
\bibitem [{\citenamefont {Galli}\ \emph {et~al.}(2013)\citenamefont {Galli},
  \citenamefont {Slatyer}, \citenamefont {Valdes},\ and\ \citenamefont
  {Iocco}}]{Galli:2013dna}%
  \BibitemOpen
  \bibfield  {author} {\bibinfo {author} {\bibfnamefont {S.}~\bibnamefont
  {Galli}}, \bibinfo {author} {\bibfnamefont {T.~R.}\ \bibnamefont {Slatyer}},
  \bibinfo {author} {\bibfnamefont {M.}~\bibnamefont {Valdes}},\ and\ \bibinfo
  {author} {\bibfnamefont {F.}~\bibnamefont {Iocco}},\ }\bibfield  {title}
  {\bibinfo {title} {{Systematic Uncertainties In Constraining Dark Matter
  Annihilation From The Cosmic Microwave Background}},\ }\href
  {https://doi.org/10.1103/PhysRevD.88.063502} {\bibfield  {journal} {\bibinfo
  {journal} {Phys. Rev. D}\ }\textbf {\bibinfo {volume} {88}},\ \bibinfo
  {pages} {063502} (\bibinfo {year} {2013})},\ \Eprint
  {https://arxiv.org/abs/1306.0563} {arXiv:1306.0563 [astro-ph.CO]}
  \BibitemShut {NoStop}%
\bibitem [{\citenamefont {Slatyer}(2016{\natexlab{a}})}]{Slatyer:2015jla}%
  \BibitemOpen
  \bibfield  {author} {\bibinfo {author} {\bibfnamefont {T.~R.}\ \bibnamefont
  {Slatyer}},\ }\bibfield  {title} {\bibinfo {title} {{Indirect dark matter
  signatures in the cosmic dark ages. I. Generalizing the bound on s-wave dark
  matter annihilation from Planck results}},\ }\href
  {https://doi.org/10.1103/PhysRevD.93.023527} {\bibfield  {journal} {\bibinfo
  {journal} {Phys. Rev. D}\ }\textbf {\bibinfo {volume} {93}},\ \bibinfo
  {pages} {023527} (\bibinfo {year} {2016}{\natexlab{a}})},\ \Eprint
  {https://arxiv.org/abs/1506.03811} {arXiv:1506.03811 [hep-ph]} \BibitemShut
  {NoStop}%
\bibitem [{\citenamefont {Slatyer}(2016{\natexlab{b}})}]{Slatyer:2015kla}%
  \BibitemOpen
  \bibfield  {author} {\bibinfo {author} {\bibfnamefont {T.~R.}\ \bibnamefont
  {Slatyer}},\ }\bibfield  {title} {\bibinfo {title} {{Indirect Dark Matter
  Signatures in the Cosmic Dark Ages II. Ionization, Heating and Photon
  Production from Arbitrary Energy Injections}},\ }\href
  {https://doi.org/10.1103/PhysRevD.93.023521} {\bibfield  {journal} {\bibinfo
  {journal} {Phys. Rev. D}\ }\textbf {\bibinfo {volume} {93}},\ \bibinfo
  {pages} {023521} (\bibinfo {year} {2016}{\natexlab{b}})},\ \Eprint
  {https://arxiv.org/abs/1506.03812} {arXiv:1506.03812 [astro-ph.CO]}
  \BibitemShut {NoStop}%
\bibitem [{\citenamefont {Poulin}\ \emph {et~al.}(2015)\citenamefont {Poulin},
  \citenamefont {Serpico},\ and\ \citenamefont {Lesgourgues}}]{Poulin:2015pna}%
  \BibitemOpen
  \bibfield  {author} {\bibinfo {author} {\bibfnamefont {V.}~\bibnamefont
  {Poulin}}, \bibinfo {author} {\bibfnamefont {P.~D.}\ \bibnamefont
  {Serpico}},\ and\ \bibinfo {author} {\bibfnamefont {J.}~\bibnamefont
  {Lesgourgues}},\ }\bibfield  {title} {\bibinfo {title} {{Dark Matter
  annihilations in halos and high-redshift sources of reionization of the
  universe}},\ }\href {https://doi.org/10.1088/1475-7516/2015/12/041}
  {\bibfield  {journal} {\bibinfo  {journal} {JCAP}\ }\textbf {\bibinfo
  {volume} {12}},\ \bibinfo {pages} {041}},\ \Eprint
  {https://arxiv.org/abs/1508.01370} {arXiv:1508.01370 [astro-ph.CO]}
  \BibitemShut {NoStop}%
\bibitem [{\citenamefont {Slatyer}\ and\ \citenamefont
  {Wu}(2017)}]{Slatyer:2016qyl}%
  \BibitemOpen
  \bibfield  {author} {\bibinfo {author} {\bibfnamefont {T.~R.}\ \bibnamefont
  {Slatyer}}\ and\ \bibinfo {author} {\bibfnamefont {C.-L.}\ \bibnamefont
  {Wu}},\ }\bibfield  {title} {\bibinfo {title} {{General Constraints on Dark
  Matter Decay from the Cosmic Microwave Background}},\ }\href
  {https://doi.org/10.1103/PhysRevD.95.023010} {\bibfield  {journal} {\bibinfo
  {journal} {Phys. Rev. D}\ }\textbf {\bibinfo {volume} {95}},\ \bibinfo
  {pages} {023010} (\bibinfo {year} {2017})},\ \Eprint
  {https://arxiv.org/abs/1610.06933} {arXiv:1610.06933 [astro-ph.CO]}
  \BibitemShut {NoStop}%
\bibitem [{\citenamefont {Liu}\ \emph {et~al.}(2016{\natexlab{a}})\citenamefont
  {Liu}, \citenamefont {Slatyer},\ and\ \citenamefont {Zavala}}]{Liu:2016cnk}%
  \BibitemOpen
  \bibfield  {author} {\bibinfo {author} {\bibfnamefont {H.}~\bibnamefont
  {Liu}}, \bibinfo {author} {\bibfnamefont {T.~R.}\ \bibnamefont {Slatyer}},\
  and\ \bibinfo {author} {\bibfnamefont {J.}~\bibnamefont {Zavala}},\
  }\bibfield  {title} {\bibinfo {title} {{Contributions to cosmic reionization
  from dark matter annihilation and decay}},\ }\href
  {https://doi.org/10.1103/PhysRevD.94.063507} {\bibfield  {journal} {\bibinfo
  {journal} {Phys. Rev. D}\ }\textbf {\bibinfo {volume} {94}},\ \bibinfo
  {pages} {063507} (\bibinfo {year} {2016}{\natexlab{a}})},\ \Eprint
  {https://arxiv.org/abs/1604.02457} {arXiv:1604.02457 [astro-ph.CO]}
  \BibitemShut {NoStop}%
\bibitem [{\citenamefont {Slatyer}(2013)}]{Slatyer:2012yq}%
  \BibitemOpen
  \bibfield  {author} {\bibinfo {author} {\bibfnamefont {T.~R.}\ \bibnamefont
  {Slatyer}},\ }\bibfield  {title} {\bibinfo {title} {{Energy Injection And
  Absorption In The Cosmic Dark Ages}},\ }\href
  {https://doi.org/10.1103/PhysRevD.87.123513} {\bibfield  {journal} {\bibinfo
  {journal} {Phys. Rev. D}\ }\textbf {\bibinfo {volume} {87}},\ \bibinfo
  {pages} {123513} (\bibinfo {year} {2013})},\ \Eprint
  {https://arxiv.org/abs/1211.0283} {arXiv:1211.0283 [astro-ph.CO]}
  \BibitemShut {NoStop}%
\bibitem [{\citenamefont {Lopez-Honorez}\ \emph {et~al.}(2013)\citenamefont
  {Lopez-Honorez}, \citenamefont {Mena}, \citenamefont {Palomares-Ruiz},\ and\
  \citenamefont {Vincent}}]{Lopez-Honorez:2013cua}%
  \BibitemOpen
  \bibfield  {author} {\bibinfo {author} {\bibfnamefont {L.}~\bibnamefont
  {Lopez-Honorez}}, \bibinfo {author} {\bibfnamefont {O.}~\bibnamefont {Mena}},
  \bibinfo {author} {\bibfnamefont {S.}~\bibnamefont {Palomares-Ruiz}},\ and\
  \bibinfo {author} {\bibfnamefont {A.~C.}\ \bibnamefont {Vincent}},\
  }\bibfield  {title} {\bibinfo {title} {{Constraints on dark matter
  annihilation from CMB observationsbefore Planck}},\ }\href
  {https://doi.org/10.1088/1475-7516/2013/07/046} {\bibfield  {journal}
  {\bibinfo  {journal} {JCAP}\ }\textbf {\bibinfo {volume} {07}},\ \bibinfo
  {pages} {046}},\ \Eprint {https://arxiv.org/abs/1303.5094} {arXiv:1303.5094
  [astro-ph.CO]} \BibitemShut {NoStop}%
\bibitem [{\citenamefont {Capozzi}\ \emph {et~al.}(2023)\citenamefont
  {Capozzi}, \citenamefont {Ferreira}, \citenamefont {Lopez-Honorez},\ and\
  \citenamefont {Mena}}]{Capozzi:2023xie}%
  \BibitemOpen
  \bibfield  {author} {\bibinfo {author} {\bibfnamefont {F.}~\bibnamefont
  {Capozzi}}, \bibinfo {author} {\bibfnamefont {R.~Z.}\ \bibnamefont
  {Ferreira}}, \bibinfo {author} {\bibfnamefont {L.}~\bibnamefont
  {Lopez-Honorez}},\ and\ \bibinfo {author} {\bibfnamefont {O.}~\bibnamefont
  {Mena}},\ }\bibfield  {title} {\bibinfo {title} {{CMB and
  Lyman-\ensuremath{\alpha} constraints on dark matter decays to photons}},\
  }\href {https://doi.org/10.1088/1475-7516/2023/06/060} {\bibfield  {journal}
  {\bibinfo  {journal} {JCAP}\ }\textbf {\bibinfo {volume} {06}},\ \bibinfo
  {pages} {060}},\ \Eprint {https://arxiv.org/abs/2303.07426} {arXiv:2303.07426
  [astro-ph.CO]} \BibitemShut {NoStop}%
\bibitem [{\citenamefont {Narayanan}\ \emph {et~al.}(2000)\citenamefont
  {Narayanan}, \citenamefont {Spergel}, \citenamefont {Dave},\ and\
  \citenamefont {Ma}}]{Narayanan:2000tp}%
  \BibitemOpen
  \bibfield  {author} {\bibinfo {author} {\bibfnamefont {V.~K.}\ \bibnamefont
  {Narayanan}}, \bibinfo {author} {\bibfnamefont {D.~N.}\ \bibnamefont
  {Spergel}}, \bibinfo {author} {\bibfnamefont {R.}~\bibnamefont {Dave}},\ and\
  \bibinfo {author} {\bibfnamefont {C.-P.}\ \bibnamefont {Ma}},\ }\bibfield
  {title} {\bibinfo {title} {{Constraints on the mass of warm dark matter
  particles and the shape of the linear power spectrum from the Ly$\alpha$
  forest}},\ }\href {https://doi.org/10.1086/317269} {\bibfield  {journal}
  {\bibinfo  {journal} {Astrophys. J. Lett.}\ }\textbf {\bibinfo {volume}
  {543}},\ \bibinfo {pages} {L103} (\bibinfo {year} {2000})},\ \Eprint
  {https://arxiv.org/abs/astro-ph/0005095} {arXiv:astro-ph/0005095}
  \BibitemShut {NoStop}%
\bibitem [{\citenamefont {Viel}\ \emph {et~al.}(2005)\citenamefont {Viel},
  \citenamefont {Lesgourgues}, \citenamefont {Haehnelt}, \citenamefont
  {Matarrese},\ and\ \citenamefont {Riotto}}]{Viel:2005qj}%
  \BibitemOpen
  \bibfield  {author} {\bibinfo {author} {\bibfnamefont {M.}~\bibnamefont
  {Viel}}, \bibinfo {author} {\bibfnamefont {J.}~\bibnamefont {Lesgourgues}},
  \bibinfo {author} {\bibfnamefont {M.~G.}\ \bibnamefont {Haehnelt}}, \bibinfo
  {author} {\bibfnamefont {S.}~\bibnamefont {Matarrese}},\ and\ \bibinfo
  {author} {\bibfnamefont {A.}~\bibnamefont {Riotto}},\ }\bibfield  {title}
  {\bibinfo {title} {{Constraining warm dark matter candidates including
  sterile neutrinos and light gravitinos with WMAP and the Lyman-alpha
  forest}},\ }\href {https://doi.org/10.1103/PhysRevD.71.063534} {\bibfield
  {journal} {\bibinfo  {journal} {Phys. Rev. D}\ }\textbf {\bibinfo {volume}
  {71}},\ \bibinfo {pages} {063534} (\bibinfo {year} {2005})},\ \Eprint
  {https://arxiv.org/abs/astro-ph/0501562} {arXiv:astro-ph/0501562}
  \BibitemShut {NoStop}%
\bibitem [{\citenamefont {Boyarsky}\ \emph {et~al.}(2009)\citenamefont
  {Boyarsky}, \citenamefont {Lesgourgues}, \citenamefont {Ruchayskiy},\ and\
  \citenamefont {Viel}}]{Boyarsky:2008xj}%
  \BibitemOpen
  \bibfield  {author} {\bibinfo {author} {\bibfnamefont {A.}~\bibnamefont
  {Boyarsky}}, \bibinfo {author} {\bibfnamefont {J.}~\bibnamefont
  {Lesgourgues}}, \bibinfo {author} {\bibfnamefont {O.}~\bibnamefont
  {Ruchayskiy}},\ and\ \bibinfo {author} {\bibfnamefont {M.}~\bibnamefont
  {Viel}},\ }\bibfield  {title} {\bibinfo {title} {{Lyman-alpha constraints on
  warm and on warm-plus-cold dark matter models}},\ }\href
  {https://doi.org/10.1088/1475-7516/2009/05/012} {\bibfield  {journal}
  {\bibinfo  {journal} {JCAP}\ }\textbf {\bibinfo {volume} {05}},\ \bibinfo
  {pages} {012}},\ \Eprint {https://arxiv.org/abs/0812.0010} {arXiv:0812.0010
  [astro-ph]} \BibitemShut {NoStop}%
\bibitem [{\citenamefont {Ballesteros}\ \emph {et~al.}(2021)\citenamefont
  {Ballesteros}, \citenamefont {Garcia},\ and\ \citenamefont
  {Pierre}}]{Ballesteros:2020adh}%
  \BibitemOpen
  \bibfield  {author} {\bibinfo {author} {\bibfnamefont {G.}~\bibnamefont
  {Ballesteros}}, \bibinfo {author} {\bibfnamefont {M.~A.~G.}\ \bibnamefont
  {Garcia}},\ and\ \bibinfo {author} {\bibfnamefont {M.}~\bibnamefont
  {Pierre}},\ }\bibfield  {title} {\bibinfo {title} {{How warm are non-thermal
  relics? Lyman-$\alpha$ bounds on out-of-equilibrium dark matter}},\ }\href
  {https://doi.org/10.1088/1475-7516/2021/03/101} {\bibfield  {journal}
  {\bibinfo  {journal} {JCAP}\ }\textbf {\bibinfo {volume} {03}},\ \bibinfo
  {pages} {101}},\ \Eprint {https://arxiv.org/abs/2011.13458} {arXiv:2011.13458
  [hep-ph]} \BibitemShut {NoStop}%
\bibitem [{\citenamefont {Viel}\ \emph {et~al.}(2013)\citenamefont {Viel},
  \citenamefont {Becker}, \citenamefont {Bolton},\ and\ \citenamefont
  {Haehnelt}}]{Viel:2013fqw}%
  \BibitemOpen
  \bibfield  {author} {\bibinfo {author} {\bibfnamefont {M.}~\bibnamefont
  {Viel}}, \bibinfo {author} {\bibfnamefont {G.~D.}\ \bibnamefont {Becker}},
  \bibinfo {author} {\bibfnamefont {J.~S.}\ \bibnamefont {Bolton}},\ and\
  \bibinfo {author} {\bibfnamefont {M.~G.}\ \bibnamefont {Haehnelt}},\
  }\bibfield  {title} {\bibinfo {title} {{Warm dark matter as a solution to the
  small scale crisis: New constraints from high redshift
  Lyman-\ensuremath{\alpha} forest data}},\ }\href
  {https://doi.org/10.1103/PhysRevD.88.043502} {\bibfield  {journal} {\bibinfo
  {journal} {Phys. Rev. D}\ }\textbf {\bibinfo {volume} {88}},\ \bibinfo
  {pages} {043502} (\bibinfo {year} {2013})},\ \Eprint
  {https://arxiv.org/abs/1306.2314} {arXiv:1306.2314 [astro-ph.CO]}
  \BibitemShut {NoStop}%
\bibitem [{\citenamefont {Garzilli}\ \emph {et~al.}(2017)\citenamefont
  {Garzilli}, \citenamefont {Boyarsky},\ and\ \citenamefont
  {Ruchayskiy}}]{Garzilli:2015iwa}%
  \BibitemOpen
  \bibfield  {author} {\bibinfo {author} {\bibfnamefont {A.}~\bibnamefont
  {Garzilli}}, \bibinfo {author} {\bibfnamefont {A.}~\bibnamefont {Boyarsky}},\
  and\ \bibinfo {author} {\bibfnamefont {O.}~\bibnamefont {Ruchayskiy}},\
  }\bibfield  {title} {\bibinfo {title} {{Cutoff in the Lyman
  {\textbackslash{}alpha} forest power spectrum: warm IGM or warm dark
  matter?}},\ }\href {https://doi.org/10.1016/j.physletb.2017.08.022}
  {\bibfield  {journal} {\bibinfo  {journal} {Phys. Lett. B}\ }\textbf
  {\bibinfo {volume} {773}},\ \bibinfo {pages} {258} (\bibinfo {year}
  {2017})},\ \Eprint {https://arxiv.org/abs/1510.07006} {arXiv:1510.07006
  [astro-ph.CO]} \BibitemShut {NoStop}%
\bibitem [{\citenamefont {Ir\v{s}i\v{c}}\ \emph {et~al.}(2017)\citenamefont
  {Ir\v{s}i\v{c}} \emph {et~al.}}]{Irsic:2017ixq}%
  \BibitemOpen
  \bibfield  {author} {\bibinfo {author} {\bibfnamefont {V.}~\bibnamefont
  {Ir\v{s}i\v{c}}} \emph {et~al.},\ }\bibfield  {title} {\bibinfo {title} {{New
  Constraints on the free-streaming of warm dark matter from intermediate and
  small scale Lyman-$\alpha$ forest data}},\ }\href
  {https://doi.org/10.1103/PhysRevD.96.023522} {\bibfield  {journal} {\bibinfo
  {journal} {Phys. Rev. D}\ }\textbf {\bibinfo {volume} {96}},\ \bibinfo
  {pages} {023522} (\bibinfo {year} {2017})},\ \Eprint
  {https://arxiv.org/abs/1702.01764} {arXiv:1702.01764 [astro-ph.CO]}
  \BibitemShut {NoStop}%
\bibitem [{\citenamefont {Garzilli}\ \emph {et~al.}(2021)\citenamefont
  {Garzilli}, \citenamefont {Magalich}, \citenamefont {Ruchayskiy},\ and\
  \citenamefont {Boyarsky}}]{Garzilli:2019qki}%
  \BibitemOpen
  \bibfield  {author} {\bibinfo {author} {\bibfnamefont {A.}~\bibnamefont
  {Garzilli}}, \bibinfo {author} {\bibfnamefont {A.}~\bibnamefont {Magalich}},
  \bibinfo {author} {\bibfnamefont {O.}~\bibnamefont {Ruchayskiy}},\ and\
  \bibinfo {author} {\bibfnamefont {A.}~\bibnamefont {Boyarsky}},\ }\bibfield
  {title} {\bibinfo {title} {{How to constrain warm dark matter with the
  Lyman-$\alpha$ forest}},\ }\href {https://doi.org/10.1093/mnras/stab192}
  {\bibfield  {journal} {\bibinfo  {journal} {Mon. Not. Roy. Astron. Soc.}\
  }\textbf {\bibinfo {volume} {502}},\ \bibinfo {pages} {2356} (\bibinfo {year}
  {2021})},\ \Eprint {https://arxiv.org/abs/1912.09397} {arXiv:1912.09397
  [astro-ph.CO]} \BibitemShut {NoStop}%
\bibitem [{\citenamefont {Hooper}\ \emph {et~al.}(2022)\citenamefont {Hooper},
  \citenamefont {Sch\"oneberg}, \citenamefont {Murgia}, \citenamefont
  {Archidiacono}, \citenamefont {Lesgourgues},\ and\ \citenamefont
  {Viel}}]{Hooper:2022byl}%
  \BibitemOpen
  \bibfield  {author} {\bibinfo {author} {\bibfnamefont {D.~C.}\ \bibnamefont
  {Hooper}}, \bibinfo {author} {\bibfnamefont {N.}~\bibnamefont
  {Sch\"oneberg}}, \bibinfo {author} {\bibfnamefont {R.}~\bibnamefont
  {Murgia}}, \bibinfo {author} {\bibfnamefont {M.}~\bibnamefont
  {Archidiacono}}, \bibinfo {author} {\bibfnamefont {J.}~\bibnamefont
  {Lesgourgues}},\ and\ \bibinfo {author} {\bibfnamefont {M.}~\bibnamefont
  {Viel}},\ }\bibfield  {title} {\bibinfo {title} {{One likelihood to bind them
  all: Lyman-\ensuremath{\alpha} constraints on non-standard dark matter}},\
  }\href {https://doi.org/10.1088/1475-7516/2022/10/032} {\bibfield  {journal}
  {\bibinfo  {journal} {JCAP}\ }\textbf {\bibinfo {volume} {10}},\ \bibinfo
  {pages} {032}},\ \Eprint {https://arxiv.org/abs/2206.08188} {arXiv:2206.08188
  [astro-ph.CO]} \BibitemShut {NoStop}%
\bibitem [{\citenamefont {Villasenor}\ \emph {et~al.}(2023)\citenamefont
  {Villasenor}, \citenamefont {Robertson}, \citenamefont {Madau},\ and\
  \citenamefont {Schneider}}]{Villasenor:2022aiy}%
  \BibitemOpen
  \bibfield  {author} {\bibinfo {author} {\bibfnamefont {B.}~\bibnamefont
  {Villasenor}}, \bibinfo {author} {\bibfnamefont {B.}~\bibnamefont
  {Robertson}}, \bibinfo {author} {\bibfnamefont {P.}~\bibnamefont {Madau}},\
  and\ \bibinfo {author} {\bibfnamefont {E.}~\bibnamefont {Schneider}},\
  }\bibfield  {title} {\bibinfo {title} {{New constraints on warm dark matter
  from the Lyman-\ensuremath{\alpha} forest power spectrum}},\ }\href
  {https://doi.org/10.1103/PhysRevD.108.023502} {\bibfield  {journal} {\bibinfo
   {journal} {Phys. Rev. D}\ }\textbf {\bibinfo {volume} {108}},\ \bibinfo
  {pages} {023502} (\bibinfo {year} {2023})},\ \Eprint
  {https://arxiv.org/abs/2209.14220} {arXiv:2209.14220 [astro-ph.CO]}
  \BibitemShut {NoStop}%
\bibitem [{\citenamefont {Rossi}\ \emph {et~al.}(2014)\citenamefont {Rossi},
  \citenamefont {Palanque-Delabrouille}, \citenamefont {Borde}, \citenamefont
  {Viel}, \citenamefont {Yeche}, \citenamefont {Bolton}, \citenamefont {Rich},\
  and\ \citenamefont {Le~Goff}}]{Rossi:2014wsa}%
  \BibitemOpen
  \bibfield  {author} {\bibinfo {author} {\bibfnamefont {G.}~\bibnamefont
  {Rossi}}, \bibinfo {author} {\bibfnamefont {N.}~\bibnamefont
  {Palanque-Delabrouille}}, \bibinfo {author} {\bibfnamefont {A.}~\bibnamefont
  {Borde}}, \bibinfo {author} {\bibfnamefont {M.}~\bibnamefont {Viel}},
  \bibinfo {author} {\bibfnamefont {C.}~\bibnamefont {Yeche}}, \bibinfo
  {author} {\bibfnamefont {J.~S.}\ \bibnamefont {Bolton}}, \bibinfo {author}
  {\bibfnamefont {J.}~\bibnamefont {Rich}},\ and\ \bibinfo {author}
  {\bibfnamefont {J.-M.}\ \bibnamefont {Le~Goff}},\ }\bibfield  {title}
  {\bibinfo {title} {{Suite of hydrodynamical simulations for the
  Lyman-\ensuremath{\alpha} forest with massive neutrinos}},\ }\href
  {https://doi.org/10.1051/0004-6361/201423507} {\bibfield  {journal} {\bibinfo
   {journal} {Astron. Astrophys.}\ }\textbf {\bibinfo {volume} {567}},\
  \bibinfo {pages} {A79} (\bibinfo {year} {2014})},\ \Eprint
  {https://arxiv.org/abs/1401.6464} {arXiv:1401.6464 [astro-ph.CO]}
  \BibitemShut {NoStop}%
\bibitem [{\citenamefont {Palanque-Delabrouille}\ \emph
  {et~al.}(2015)\citenamefont {Palanque-Delabrouille} \emph
  {et~al.}}]{Palanque-Delabrouille:2015pga}%
  \BibitemOpen
  \bibfield  {author} {\bibinfo {author} {\bibfnamefont {N.}~\bibnamefont
  {Palanque-Delabrouille}} \emph {et~al.},\ }\bibfield  {title} {\bibinfo
  {title} {{Neutrino masses and cosmology with Lyman-alpha forest power
  spectrum}},\ }\href {https://doi.org/10.1088/1475-7516/2015/11/011}
  {\bibfield  {journal} {\bibinfo  {journal} {JCAP}\ }\textbf {\bibinfo
  {volume} {11}},\ \bibinfo {pages} {011}},\ \Eprint
  {https://arxiv.org/abs/1506.05976} {arXiv:1506.05976 [astro-ph.CO]}
  \BibitemShut {NoStop}%
\bibitem [{\citenamefont {Palanque-Delabrouille}\ \emph
  {et~al.}(2020)\citenamefont {Palanque-Delabrouille}, \citenamefont {Y\`eche},
  \citenamefont {Sch\"oneberg}, \citenamefont {Lesgourgues}, \citenamefont
  {Walther}, \citenamefont {Chabanier},\ and\ \citenamefont
  {Armengaud}}]{Palanque-Delabrouille:2019iyz}%
  \BibitemOpen
  \bibfield  {author} {\bibinfo {author} {\bibfnamefont {N.}~\bibnamefont
  {Palanque-Delabrouille}}, \bibinfo {author} {\bibfnamefont {C.}~\bibnamefont
  {Y\`eche}}, \bibinfo {author} {\bibfnamefont {N.}~\bibnamefont
  {Sch\"oneberg}}, \bibinfo {author} {\bibfnamefont {J.}~\bibnamefont
  {Lesgourgues}}, \bibinfo {author} {\bibfnamefont {M.}~\bibnamefont
  {Walther}}, \bibinfo {author} {\bibfnamefont {S.}~\bibnamefont {Chabanier}},\
  and\ \bibinfo {author} {\bibfnamefont {E.}~\bibnamefont {Armengaud}},\
  }\bibfield  {title} {\bibinfo {title} {{Hints, neutrino bounds and WDM
  constraints from SDSS DR14 Lyman-$\alpha$ and Planck full-survey data}},\
  }\href {https://doi.org/10.1088/1475-7516/2020/04/038} {\bibfield  {journal}
  {\bibinfo  {journal} {JCAP}\ }\textbf {\bibinfo {volume} {04}},\ \bibinfo
  {pages} {038}},\ \Eprint {https://arxiv.org/abs/1911.09073} {arXiv:1911.09073
  [astro-ph.CO]} \BibitemShut {NoStop}%
\bibitem [{\citenamefont {DeBoer}\ \emph {et~al.}(2017)\citenamefont {DeBoer}
  \emph {et~al.}}]{DeBoer:2016tnn}%
  \BibitemOpen
  \bibfield  {author} {\bibinfo {author} {\bibfnamefont {D.~R.}\ \bibnamefont
  {DeBoer}} \emph {et~al.},\ }\bibfield  {title} {\bibinfo {title} {{Hydrogen
  Epoch of Reionization Array (HERA)}},\ }\href
  {https://doi.org/10.1088/1538-3873/129/974/045001} {\bibfield  {journal}
  {\bibinfo  {journal} {Publ. Astron. Soc. Pac.}\ }\textbf {\bibinfo {volume}
  {129}},\ \bibinfo {pages} {045001} (\bibinfo {year} {2017})},\ \Eprint
  {https://arxiv.org/abs/1606.07473} {arXiv:1606.07473 [astro-ph.IM]}
  \BibitemShut {NoStop}%
\bibitem [{\citenamefont {Abdurashidova}\ \emph {et~al.}(2022)\citenamefont
  {Abdurashidova} \emph {et~al.}}]{HERA:2021noe}%
  \BibitemOpen
  \bibfield  {author} {\bibinfo {author} {\bibfnamefont {Z.}~\bibnamefont
  {Abdurashidova}} \emph {et~al.} (\bibinfo {collaboration} {HERA}),\
  }\bibfield  {title} {\bibinfo {title} {{HERA Phase I Limits on the Cosmic 21
  cm Signal: Constraints on Astrophysics and Cosmology during the Epoch of
  Reionization}},\ }\href {https://doi.org/10.3847/1538-4357/ac2ffc} {\bibfield
   {journal} {\bibinfo  {journal} {Astrophys. J.}\ }\textbf {\bibinfo {volume}
  {924}},\ \bibinfo {pages} {51} (\bibinfo {year} {2022})},\ \Eprint
  {https://arxiv.org/abs/2108.07282} {arXiv:2108.07282 [astro-ph.CO]}
  \BibitemShut {NoStop}%
\bibitem [{\citenamefont {Carilli}\ and\ \citenamefont
  {Rawlings}(2004)}]{Carilli:2004nx}%
  \BibitemOpen
  \bibfield  {author} {\bibinfo {author} {\bibfnamefont {C.~L.}\ \bibnamefont
  {Carilli}}\ and\ \bibinfo {author} {\bibfnamefont {S.}~\bibnamefont
  {Rawlings}},\ }\bibfield  {title} {\bibinfo {title} {{Science with the Square
  Kilometer Array: Motivation, key science projects, standards and
  assumptions}},\ }\href {https://doi.org/10.1016/j.newar.2004.09.001}
  {\bibfield  {journal} {\bibinfo  {journal} {New Astron. Rev.}\ }\textbf
  {\bibinfo {volume} {48}},\ \bibinfo {pages} {979} (\bibinfo {year} {2004})},\
  \Eprint {https://arxiv.org/abs/astro-ph/0409274} {arXiv:astro-ph/0409274}
  \BibitemShut {NoStop}%
\bibitem [{\citenamefont {Mu\~noz}\ \emph {et~al.}(2020)\citenamefont
  {Mu\~noz}, \citenamefont {Dvorkin},\ and\ \citenamefont
  {Cyr-Racine}}]{Munoz:2019hjh}%
  \BibitemOpen
  \bibfield  {author} {\bibinfo {author} {\bibfnamefont {J.~B.}\ \bibnamefont
  {Mu\~noz}}, \bibinfo {author} {\bibfnamefont {C.}~\bibnamefont {Dvorkin}},\
  and\ \bibinfo {author} {\bibfnamefont {F.-Y.}\ \bibnamefont {Cyr-Racine}},\
  }\bibfield  {title} {\bibinfo {title} {{Probing the Small-Scale Matter Power
  Spectrum with Large-Scale 21-cm Data}},\ }\href
  {https://doi.org/10.1103/PhysRevD.101.063526} {\bibfield  {journal} {\bibinfo
   {journal} {Phys. Rev. D}\ }\textbf {\bibinfo {volume} {101}},\ \bibinfo
  {pages} {063526} (\bibinfo {year} {2020})},\ \Eprint
  {https://arxiv.org/abs/1911.11144} {arXiv:1911.11144 [astro-ph.CO]}
  \BibitemShut {NoStop}%
\bibitem [{\citenamefont {Cole}\ and\ \citenamefont
  {Silk}(2021)}]{Cole:2019zhu}%
  \BibitemOpen
  \bibfield  {author} {\bibinfo {author} {\bibfnamefont {P.~S.}\ \bibnamefont
  {Cole}}\ and\ \bibinfo {author} {\bibfnamefont {J.}~\bibnamefont {Silk}},\
  }\bibfield  {title} {\bibinfo {title} {{Small-scale primordial fluctuations
  in the 21 cm Dark Ages signal}},\ }\href
  {https://doi.org/10.1093/mnras/staa3638} {\bibfield  {journal} {\bibinfo
  {journal} {Mon. Not. Roy. Astron. Soc.}\ }\textbf {\bibinfo {volume} {501}},\
  \bibinfo {pages} {2627} (\bibinfo {year} {2021})},\ \Eprint
  {https://arxiv.org/abs/1912.02171} {arXiv:1912.02171 [astro-ph.CO]}
  \BibitemShut {NoStop}%
\bibitem [{\citenamefont {Hotinli}\ \emph {et~al.}(2022)\citenamefont
  {Hotinli}, \citenamefont {Marsh},\ and\ \citenamefont
  {Kamionkowski}}]{Hotinli:2021vxg}%
  \BibitemOpen
  \bibfield  {author} {\bibinfo {author} {\bibfnamefont {S.~C.}\ \bibnamefont
  {Hotinli}}, \bibinfo {author} {\bibfnamefont {D.~J.~E.}\ \bibnamefont
  {Marsh}},\ and\ \bibinfo {author} {\bibfnamefont {M.}~\bibnamefont
  {Kamionkowski}},\ }\bibfield  {title} {\bibinfo {title} {{Probing ultralight
  axions with the 21-cm signal during cosmic dawn}},\ }\href
  {https://doi.org/10.1103/PhysRevD.106.043529} {\bibfield  {journal} {\bibinfo
   {journal} {Phys. Rev. D}\ }\textbf {\bibinfo {volume} {106}},\ \bibinfo
  {pages} {043529} (\bibinfo {year} {2022})},\ \Eprint
  {https://arxiv.org/abs/2112.06943} {arXiv:2112.06943 [astro-ph.CO]}
  \BibitemShut {NoStop}%
\bibitem [{\citenamefont {Flitter}\ and\ \citenamefont
  {Kovetz}(2022)}]{Flitter:2022pzf}%
  \BibitemOpen
  \bibfield  {author} {\bibinfo {author} {\bibfnamefont {J.}~\bibnamefont
  {Flitter}}\ and\ \bibinfo {author} {\bibfnamefont {E.~D.}\ \bibnamefont
  {Kovetz}},\ }\bibfield  {title} {\bibinfo {title} {{Closing the window on
  fuzzy dark matter with the 21-cm signal}},\ }\href
  {https://doi.org/10.1103/PhysRevD.106.063504} {\bibfield  {journal} {\bibinfo
   {journal} {Phys. Rev. D}\ }\textbf {\bibinfo {volume} {106}},\ \bibinfo
  {pages} {063504} (\bibinfo {year} {2022})},\ \Eprint
  {https://arxiv.org/abs/2207.05083} {arXiv:2207.05083 [astro-ph.CO]}
  \BibitemShut {NoStop}%
\bibitem [{\citenamefont {Facchinetti}\ \emph {et~al.}(2024)\citenamefont
  {Facchinetti}, \citenamefont {Lopez-Honorez}, \citenamefont {Qin},\ and\
  \citenamefont {Mesinger}}]{Facchinetti:2023slb}%
  \BibitemOpen
  \bibfield  {author} {\bibinfo {author} {\bibfnamefont {G.}~\bibnamefont
  {Facchinetti}}, \bibinfo {author} {\bibfnamefont {L.}~\bibnamefont
  {Lopez-Honorez}}, \bibinfo {author} {\bibfnamefont {Y.}~\bibnamefont {Qin}},\
  and\ \bibinfo {author} {\bibfnamefont {A.}~\bibnamefont {Mesinger}},\
  }\bibfield  {title} {\bibinfo {title} {{21cm signal sensitivity to dark
  matter decay}},\ }\href {https://doi.org/10.1088/1475-7516/2024/01/005}
  {\bibfield  {journal} {\bibinfo  {journal} {JCAP}\ }\textbf {\bibinfo
  {volume} {01}},\ \bibinfo {pages} {005}},\ \Eprint
  {https://arxiv.org/abs/2308.16656} {arXiv:2308.16656 [astro-ph.CO]}
  \BibitemShut {NoStop}%
\bibitem [{\citenamefont {Sun}\ \emph {et~al.}(2023)\citenamefont {Sun},
  \citenamefont {Foster}, \citenamefont {Liu}, \citenamefont {Mu\~noz},\ and\
  \citenamefont {Slatyer}}]{Sun:2023acy}%
  \BibitemOpen
  \bibfield  {author} {\bibinfo {author} {\bibfnamefont {Y.}~\bibnamefont
  {Sun}}, \bibinfo {author} {\bibfnamefont {J.~W.}\ \bibnamefont {Foster}},
  \bibinfo {author} {\bibfnamefont {H.}~\bibnamefont {Liu}}, \bibinfo {author}
  {\bibfnamefont {J.~B.}\ \bibnamefont {Mu\~noz}},\ and\ \bibinfo {author}
  {\bibfnamefont {T.~R.}\ \bibnamefont {Slatyer}},\ }\bibfield  {title}
  {\bibinfo {title} {{Inhomogeneous Energy Injection in the 21-cm Power
  Spectrum: Sensitivity to Dark Matter Decay}},\ }\href@noop {} {\  (\bibinfo
  {year} {2023})},\ \Eprint {https://arxiv.org/abs/2312.11608}
  {arXiv:2312.11608 [hep-ph]} \BibitemShut {NoStop}%
\bibitem [{\citenamefont {Gessey-Jones}\ \emph {et~al.}(2023)\citenamefont
  {Gessey-Jones}, \citenamefont {Fialkov}, \citenamefont {Acedo}, \citenamefont
  {Handley},\ and\ \citenamefont {Barkana}}]{Gessey-Jones:2023amq}%
  \BibitemOpen
  \bibfield  {author} {\bibinfo {author} {\bibfnamefont {T.}~\bibnamefont
  {Gessey-Jones}}, \bibinfo {author} {\bibfnamefont {A.}~\bibnamefont
  {Fialkov}}, \bibinfo {author} {\bibfnamefont {E.~d.~L.}\ \bibnamefont
  {Acedo}}, \bibinfo {author} {\bibfnamefont {W.~J.}\ \bibnamefont {Handley}},\
  and\ \bibinfo {author} {\bibfnamefont {R.}~\bibnamefont {Barkana}},\
  }\bibfield  {title} {\bibinfo {title} {{Signatures of cosmic ray heating in
  21-cm observables}},\ }\href {https://doi.org/10.1093/mnras/stad3014}
  {\bibfield  {journal} {\bibinfo  {journal} {Mon. Not. Roy. Astron. Soc.}\
  }\textbf {\bibinfo {volume} {526}},\ \bibinfo {pages} {4262} (\bibinfo {year}
  {2023})},\ \Eprint {https://arxiv.org/abs/2304.07201} {arXiv:2304.07201
  [astro-ph.CO]} \BibitemShut {NoStop}%
\bibitem [{\citenamefont {Cruz}\ \emph {et~al.}(2024)\citenamefont {Cruz},
  \citenamefont {Adi}, \citenamefont {Flitter}, \citenamefont {Kamionkowski},\
  and\ \citenamefont {Kovetz}}]{Cruz:2023rmo}%
  \BibitemOpen
  \bibfield  {author} {\bibinfo {author} {\bibfnamefont {H.~A.~G.}\
  \bibnamefont {Cruz}}, \bibinfo {author} {\bibfnamefont {T.}~\bibnamefont
  {Adi}}, \bibinfo {author} {\bibfnamefont {J.}~\bibnamefont {Flitter}},
  \bibinfo {author} {\bibfnamefont {M.}~\bibnamefont {Kamionkowski}},\ and\
  \bibinfo {author} {\bibfnamefont {E.~D.}\ \bibnamefont {Kovetz}},\ }\bibfield
   {title} {\bibinfo {title} {{21-cm fluctuations from primordial magnetic
  fields}},\ }\href {https://doi.org/10.1103/PhysRevD.109.023518} {\bibfield
  {journal} {\bibinfo  {journal} {Phys. Rev. D}\ }\textbf {\bibinfo {volume}
  {109}},\ \bibinfo {pages} {023518} (\bibinfo {year} {2024})},\ \Eprint
  {https://arxiv.org/abs/2308.04483} {arXiv:2308.04483 [astro-ph.CO]}
  \BibitemShut {NoStop}%
\bibitem [{\citenamefont {Plombat}\ \emph {et~al.}(2024)\citenamefont
  {Plombat}, \citenamefont {Simon}, \citenamefont {Flitter},\ and\
  \citenamefont {Poulin}}]{Plombat:2024kla}%
  \BibitemOpen
  \bibfield  {author} {\bibinfo {author} {\bibfnamefont {H.}~\bibnamefont
  {Plombat}}, \bibinfo {author} {\bibfnamefont {T.}~\bibnamefont {Simon}},
  \bibinfo {author} {\bibfnamefont {J.}~\bibnamefont {Flitter}},\ and\ \bibinfo
  {author} {\bibfnamefont {V.}~\bibnamefont {Poulin}},\ }\bibfield  {title}
  {\bibinfo {title} {{Probing Dark Relativistic Species and Their Interactions
  with Dark Matter through CMB and 21cm surveys}},\ }\href@noop {} {\
  (\bibinfo {year} {2024})},\ \Eprint {https://arxiv.org/abs/2410.01486}
  {arXiv:2410.01486 [astro-ph.CO]} \BibitemShut {NoStop}%
\bibitem [{\citenamefont {Decant}\ \emph {et~al.}(2024)\citenamefont {Decant},
  \citenamefont {Dimitriou}, \citenamefont {Honorez},\ and\ \citenamefont
  {Zaldivar}}]{Decant:2024bpg}%
  \BibitemOpen
  \bibfield  {author} {\bibinfo {author} {\bibfnamefont {Q.}~\bibnamefont
  {Decant}}, \bibinfo {author} {\bibfnamefont {A.}~\bibnamefont {Dimitriou}},
  \bibinfo {author} {\bibfnamefont {L.~L.}\ \bibnamefont {Honorez}},\ and\
  \bibinfo {author} {\bibfnamefont {B.}~\bibnamefont {Zaldivar}},\ }\bibfield
  {title} {\bibinfo {title} {{Simulation-based inference on warm dark matter
  from HERA forecasts}},\ }\href@noop {} {\  (\bibinfo {year} {2024})},\
  \Eprint {https://arxiv.org/abs/2412.10310} {arXiv:2412.10310 [astro-ph.CO]}
  \BibitemShut {NoStop}%
\bibitem [{\citenamefont {Mesinger}\ \emph {et~al.}(2011)\citenamefont
  {Mesinger}, \citenamefont {Furlanetto},\ and\ \citenamefont
  {Cen}}]{Mesinger:2010ne}%
  \BibitemOpen
  \bibfield  {author} {\bibinfo {author} {\bibfnamefont {A.}~\bibnamefont
  {Mesinger}}, \bibinfo {author} {\bibfnamefont {S.}~\bibnamefont
  {Furlanetto}},\ and\ \bibinfo {author} {\bibfnamefont {R.}~\bibnamefont
  {Cen}},\ }\bibfield  {title} {\bibinfo {title} {{21cmFAST: A Fast,
  Semi-Numerical Simulation of the High-Redshift 21-cm Signal}},\ }\href
  {https://doi.org/10.1111/j.1365-2966.2010.17731.x} {\bibfield  {journal}
  {\bibinfo  {journal} {Mon. Not. Roy. Astron. Soc.}\ }\textbf {\bibinfo
  {volume} {411}},\ \bibinfo {pages} {955} (\bibinfo {year} {2011})},\ \Eprint
  {https://arxiv.org/abs/1003.3878} {arXiv:1003.3878 [astro-ph.CO]}
  \BibitemShut {NoStop}%
\bibitem [{\citenamefont {Murray}\ \emph {et~al.}(2020)\citenamefont {Murray},
  \citenamefont {Greig}, \citenamefont {Mesinger}, \citenamefont {Mu\~noz},
  \citenamefont {Qin}, \citenamefont {Park},\ and\ \citenamefont
  {Watkinson}}]{Murray:2020trn}%
  \BibitemOpen
  \bibfield  {author} {\bibinfo {author} {\bibfnamefont {S.~G.}\ \bibnamefont
  {Murray}}, \bibinfo {author} {\bibfnamefont {B.}~\bibnamefont {Greig}},
  \bibinfo {author} {\bibfnamefont {A.}~\bibnamefont {Mesinger}}, \bibinfo
  {author} {\bibfnamefont {J.~B.}\ \bibnamefont {Mu\~noz}}, \bibinfo {author}
  {\bibfnamefont {Y.}~\bibnamefont {Qin}}, \bibinfo {author} {\bibfnamefont
  {J.}~\bibnamefont {Park}},\ and\ \bibinfo {author} {\bibfnamefont {C.~A.}\
  \bibnamefont {Watkinson}},\ }\bibfield  {title} {\bibinfo {title} {{21cmFAST
  v3: A Python-integrated C code for generating 3D realizations of the cosmic
  21cm signal}},\ }\href {https://doi.org/10.21105/joss.02582} {\bibfield
  {journal} {\bibinfo  {journal} {J. Open Source Softw.}\ }\textbf {\bibinfo
  {volume} {5}},\ \bibinfo {pages} {2582} (\bibinfo {year} {2020})},\ \Eprint
  {https://arxiv.org/abs/2010.15121} {arXiv:2010.15121 [astro-ph.IM]}
  \BibitemShut {NoStop}%
\bibitem [{\citenamefont {Liu}\ \emph {et~al.}(2016{\natexlab{b}})\citenamefont
  {Liu}, \citenamefont {Pritchard}, \citenamefont {Allison}, \citenamefont
  {Parsons}, \citenamefont {Seljak},\ and\ \citenamefont
  {Sherwin}}]{Liu:2015txa}%
  \BibitemOpen
  \bibfield  {author} {\bibinfo {author} {\bibfnamefont {A.}~\bibnamefont
  {Liu}}, \bibinfo {author} {\bibfnamefont {J.~R.}\ \bibnamefont {Pritchard}},
  \bibinfo {author} {\bibfnamefont {R.}~\bibnamefont {Allison}}, \bibinfo
  {author} {\bibfnamefont {A.~R.}\ \bibnamefont {Parsons}}, \bibinfo {author}
  {\bibfnamefont {U.}~\bibnamefont {Seljak}},\ and\ \bibinfo {author}
  {\bibfnamefont {B.~D.}\ \bibnamefont {Sherwin}},\ }\bibfield  {title}
  {\bibinfo {title} {{Eliminating the optical depth nuisance from the CMB with
  21 cm cosmology}},\ }\href {https://doi.org/10.1103/PhysRevD.93.043013}
  {\bibfield  {journal} {\bibinfo  {journal} {Phys. Rev. D}\ }\textbf {\bibinfo
  {volume} {93}},\ \bibinfo {pages} {043013} (\bibinfo {year}
  {2016}{\natexlab{b}})},\ \Eprint {https://arxiv.org/abs/1509.08463}
  {arXiv:1509.08463 [astro-ph.CO]} \BibitemShut {NoStop}%
\bibitem [{\citenamefont {Shmueli}\ \emph {et~al.}(2023)\citenamefont
  {Shmueli}, \citenamefont {Sarkar},\ and\ \citenamefont
  {Kovetz}}]{Shmueli:2023box}%
  \BibitemOpen
  \bibfield  {author} {\bibinfo {author} {\bibfnamefont {G.}~\bibnamefont
  {Shmueli}}, \bibinfo {author} {\bibfnamefont {D.}~\bibnamefont {Sarkar}},\
  and\ \bibinfo {author} {\bibfnamefont {E.~D.}\ \bibnamefont {Kovetz}},\
  }\bibfield  {title} {\bibinfo {title} {{Mitigating the optical depth
  degeneracy in the cosmological measurement of neutrino masses using 21-cm
  observations}},\ }\href {https://doi.org/10.1103/PhysRevD.108.083531}
  {\bibfield  {journal} {\bibinfo  {journal} {Phys. Rev. D}\ }\textbf {\bibinfo
  {volume} {108}},\ \bibinfo {pages} {083531} (\bibinfo {year} {2023})},\
  \Eprint {https://arxiv.org/abs/2305.07056} {arXiv:2305.07056 [astro-ph.CO]}
  \BibitemShut {NoStop}%
\bibitem [{\citenamefont {Qin}\ \emph {et~al.}(2020{\natexlab{a}})\citenamefont
  {Qin}, \citenamefont {Poulin}, \citenamefont {Mesinger}, \citenamefont
  {Greig}, \citenamefont {Murray},\ and\ \citenamefont {Park}}]{Qin:2020xrg}%
  \BibitemOpen
  \bibfield  {author} {\bibinfo {author} {\bibfnamefont {Y.}~\bibnamefont
  {Qin}}, \bibinfo {author} {\bibfnamefont {V.}~\bibnamefont {Poulin}},
  \bibinfo {author} {\bibfnamefont {A.}~\bibnamefont {Mesinger}}, \bibinfo
  {author} {\bibfnamefont {B.}~\bibnamefont {Greig}}, \bibinfo {author}
  {\bibfnamefont {S.}~\bibnamefont {Murray}},\ and\ \bibinfo {author}
  {\bibfnamefont {J.}~\bibnamefont {Park}},\ }\bibfield  {title} {\bibinfo
  {title} {{Reionization inference from the CMB optical depth and E-mode
  polarization power spectra}},\ }\href
  {https://doi.org/10.1093/mnras/staa2797} {\bibfield  {journal} {\bibinfo
  {journal} {Mon. Not. Roy. Astron. Soc.}\ }\textbf {\bibinfo {volume} {499}},\
  \bibinfo {pages} {550} (\bibinfo {year} {2020}{\natexlab{a}})},\ \Eprint
  {https://arxiv.org/abs/2006.16828} {arXiv:2006.16828 [astro-ph.CO]}
  \BibitemShut {NoStop}%
\bibitem [{\citenamefont {Jennings}\ \emph {et~al.}(2019)\citenamefont
  {Jennings}, \citenamefont {Watkinson}, \citenamefont {Abdalla},\ and\
  \citenamefont {McEwen}}]{Jennings:2018eko}%
  \BibitemOpen
  \bibfield  {author} {\bibinfo {author} {\bibfnamefont {W.~D.}\ \bibnamefont
  {Jennings}}, \bibinfo {author} {\bibfnamefont {C.~A.}\ \bibnamefont
  {Watkinson}}, \bibinfo {author} {\bibfnamefont {F.~B.}\ \bibnamefont
  {Abdalla}},\ and\ \bibinfo {author} {\bibfnamefont {J.~D.}\ \bibnamefont
  {McEwen}},\ }\bibfield  {title} {\bibinfo {title} {{Evaluating machine
  learning techniques for predicting power spectra from reionization
  simulations}},\ }\href {https://doi.org/10.1093/mnras/sty3168} {\bibfield
  {journal} {\bibinfo  {journal} {Mon. Not. Roy. Astron. Soc.}\ }\textbf
  {\bibinfo {volume} {483}},\ \bibinfo {pages} {2907} (\bibinfo {year}
  {2019})},\ \Eprint {https://arxiv.org/abs/1811.09141} {arXiv:1811.09141
  [astro-ph.CO]} \BibitemShut {NoStop}%
\bibitem [{\citenamefont {Kern}\ \emph {et~al.}(2017)\citenamefont {Kern},
  \citenamefont {Liu}, \citenamefont {Parsons}, \citenamefont {Mesinger},\ and\
  \citenamefont {Greig}}]{Kern:2017ccn}%
  \BibitemOpen
  \bibfield  {author} {\bibinfo {author} {\bibfnamefont {N.~S.}\ \bibnamefont
  {Kern}}, \bibinfo {author} {\bibfnamefont {A.}~\bibnamefont {Liu}}, \bibinfo
  {author} {\bibfnamefont {A.~R.}\ \bibnamefont {Parsons}}, \bibinfo {author}
  {\bibfnamefont {A.}~\bibnamefont {Mesinger}},\ and\ \bibinfo {author}
  {\bibfnamefont {B.}~\bibnamefont {Greig}},\ }\bibfield  {title} {\bibinfo
  {title} {{Emulating Simulations of Cosmic Dawn for 21 cm Power Spectrum
  Constraints on Cosmology, Reionization, and X-Ray Heating}},\ }\href
  {https://doi.org/10.3847/1538-4357/aa8bb4} {\bibfield  {journal} {\bibinfo
  {journal} {Astrophys. J.}\ }\textbf {\bibinfo {volume} {848}},\ \bibinfo
  {pages} {23} (\bibinfo {year} {2017})},\ \Eprint
  {https://arxiv.org/abs/1705.04688} {arXiv:1705.04688 [astro-ph.CO]}
  \BibitemShut {NoStop}%
\bibitem [{\citenamefont {Schmit}\ and\ \citenamefont
  {Pritchard}(2018)}]{Schmit:2017pho}%
  \BibitemOpen
  \bibfield  {author} {\bibinfo {author} {\bibfnamefont {C.~J.}\ \bibnamefont
  {Schmit}}\ and\ \bibinfo {author} {\bibfnamefont {J.~R.}\ \bibnamefont
  {Pritchard}},\ }\bibfield  {title} {\bibinfo {title} {{Emulation of
  reionization simulations for Bayesian inference of astrophysics parameters
  using neural networks}},\ }\href {https://doi.org/10.1093/mnras/stx3292}
  {\bibfield  {journal} {\bibinfo  {journal} {Mon. Not. Roy. Astron. Soc.}\
  }\textbf {\bibinfo {volume} {475}},\ \bibinfo {pages} {1213} (\bibinfo {year}
  {2018})},\ \Eprint {https://arxiv.org/abs/1708.00011} {arXiv:1708.00011
  [astro-ph.CO]} \BibitemShut {NoStop}%
\bibitem [{\citenamefont {La~Plante}\ and\ \citenamefont
  {Ntampaka}(2018)}]{LaPlante:2018pst}%
  \BibitemOpen
  \bibfield  {author} {\bibinfo {author} {\bibfnamefont {P.}~\bibnamefont
  {La~Plante}}\ and\ \bibinfo {author} {\bibfnamefont {M.}~\bibnamefont
  {Ntampaka}},\ }\bibfield  {title} {\bibinfo {title} {{Machine Learning
  Applied to the Reionization History of the Universe in the 21 cm Signal}},\
  }\href {https://doi.org/10.3847/1538-4357/ab2983} {\bibfield  {journal}
  {\bibinfo  {journal} {Astrophys. J.}\ }\textbf {\bibinfo {volume} {810}},\
  \bibinfo {pages} {110} (\bibinfo {year} {2018})},\ \Eprint
  {https://arxiv.org/abs/1810.08211} {arXiv:1810.08211 [astro-ph.CO]}
  \BibitemShut {NoStop}%
\bibitem [{\citenamefont {Breitman}\ \emph {et~al.}(2023)\citenamefont
  {Breitman}, \citenamefont {Mesinger}, \citenamefont {Murray}, \citenamefont
  {Prelogovic}, \citenamefont {Qin},\ and\ \citenamefont
  {Trotta}}]{Breitman:2023pcj}%
  \BibitemOpen
  \bibfield  {author} {\bibinfo {author} {\bibfnamefont {D.}~\bibnamefont
  {Breitman}}, \bibinfo {author} {\bibfnamefont {A.}~\bibnamefont {Mesinger}},
  \bibinfo {author} {\bibfnamefont {S.~G.}\ \bibnamefont {Murray}}, \bibinfo
  {author} {\bibfnamefont {D.}~\bibnamefont {Prelogovic}}, \bibinfo {author}
  {\bibfnamefont {Y.}~\bibnamefont {Qin}},\ and\ \bibinfo {author}
  {\bibfnamefont {R.}~\bibnamefont {Trotta}},\ }\bibfield  {title} {\bibinfo
  {title} {{21cmemu: an emulator of 21cmfast summary observables}},\ }\href
  {https://doi.org/10.1093/mnras/stad3849} {\bibfield  {journal} {\bibinfo
  {journal} {Mon. Not. Roy. Astron. Soc.}\ }\textbf {\bibinfo {volume} {527}},\
  \bibinfo {pages} {9833} (\bibinfo {year} {2023})},\ \bibinfo {note}
  {[Erratum: Mon.Not.Roy.Astron.Soc. 533, 1045--1047 (2024)]},\ \Eprint
  {https://arxiv.org/abs/2309.05697} {arXiv:2309.05697 [astro-ph.CO]}
  \BibitemShut {NoStop}%
\bibitem [{\citenamefont {Montero-Camacho}\ \emph {et~al.}(2024)\citenamefont
  {Montero-Camacho}, \citenamefont {Li},\ and\ \citenamefont
  {Cranmer}}]{Montero-Camacho:2024dzs}%
  \BibitemOpen
  \bibfield  {author} {\bibinfo {author} {\bibfnamefont {P.}~\bibnamefont
  {Montero-Camacho}}, \bibinfo {author} {\bibfnamefont {Y.}~\bibnamefont
  {Li}},\ and\ \bibinfo {author} {\bibfnamefont {M.}~\bibnamefont {Cranmer}},\
  }\bibfield  {title} {\bibinfo {title} {{Five parameters are all you need (in
  $\Lambda$CDM)}}\ }\href@noop {} {} (\bibinfo {year} {2024}),\ \Eprint
  {https://arxiv.org/abs/2405.13680} {arXiv:2405.13680 [astro-ph.CO]}
  \BibitemShut {NoStop}%
\bibitem [{\citenamefont {Blas}\ \emph {et~al.}(2011)\citenamefont {Blas},
  \citenamefont {Lesgourgues},\ and\ \citenamefont {Tram}}]{Blas:2011rf}%
  \BibitemOpen
  \bibfield  {author} {\bibinfo {author} {\bibfnamefont {D.}~\bibnamefont
  {Blas}}, \bibinfo {author} {\bibfnamefont {J.}~\bibnamefont {Lesgourgues}},\
  and\ \bibinfo {author} {\bibfnamefont {T.}~\bibnamefont {Tram}},\ }\bibfield
  {title} {\bibinfo {title} {{The Cosmic Linear Anisotropy Solving System
  (CLASS) II: Approximation schemes}},\ }\href
  {https://doi.org/10.1088/1475-7516/2011/07/034} {\bibfield  {journal}
  {\bibinfo  {journal} {JCAP}\ }\textbf {\bibinfo {volume} {07}},\ \bibinfo
  {pages} {034}},\ \Eprint {https://arxiv.org/abs/1104.2933} {arXiv:1104.2933
  [astro-ph.CO]} \BibitemShut {NoStop}%
\bibitem [{\citenamefont {Lesgourgues}\ and\ \citenamefont
  {Tram}(2011)}]{Lesgourgues:2011rh}%
  \BibitemOpen
  \bibfield  {author} {\bibinfo {author} {\bibfnamefont {J.}~\bibnamefont
  {Lesgourgues}}\ and\ \bibinfo {author} {\bibfnamefont {T.}~\bibnamefont
  {Tram}},\ }\bibfield  {title} {\bibinfo {title} {{The Cosmic Linear
  Anisotropy Solving System (CLASS) IV: efficient implementation of non-cold
  relics}},\ }\href {https://doi.org/10.1088/1475-7516/2011/09/032} {\bibfield
  {journal} {\bibinfo  {journal} {JCAP}\ }\textbf {\bibinfo {volume} {09}},\
  \bibinfo {pages} {032}},\ \Eprint {https://arxiv.org/abs/1104.2935}
  {arXiv:1104.2935 [astro-ph.CO]} \BibitemShut {NoStop}%
\bibitem [{\citenamefont {Dandoy}\ \emph {et~al.}(2025)\citenamefont {Dandoy},
  \citenamefont {Doering}, \citenamefont {Facchinetti}, \citenamefont
  {Lopez-Honorez},\ and\ \citenamefont {Schwagereit}}]{Dandoy:2025}%
  \BibitemOpen
  \bibfield  {author} {\bibinfo {author} {\bibfnamefont {V.}~\bibnamefont
  {Dandoy}}, \bibinfo {author} {\bibfnamefont {C.}~\bibnamefont {Doering}},
  \bibinfo {author} {\bibfnamefont {G.}~\bibnamefont {Facchinetti}}, \bibinfo
  {author} {\bibfnamefont {L.}~\bibnamefont {Lopez-Honorez}},\ and\ \bibinfo
  {author} {\bibfnamefont {J.}~\bibnamefont {Schwagereit}},\ }\bibfield
  {title} {\bibinfo {title} {{HERA sensitivity to non-cold dark matter}},\
  }\href@noop {} {\bibfield  {journal} {\bibinfo  {journal} {in prep.}\ }
  (\bibinfo {year} {2025})}\BibitemShut {NoStop}%
\bibitem [{\citenamefont {Trac}\ and\ \citenamefont {Cen}(2007)}]{Trac:2006vr}%
  \BibitemOpen
  \bibfield  {author} {\bibinfo {author} {\bibfnamefont {H.}~\bibnamefont
  {Trac}}\ and\ \bibinfo {author} {\bibfnamefont {R.}~\bibnamefont {Cen}},\
  }\bibfield  {title} {\bibinfo {title} {{Radiative transfer simulations of
  cosmic reionization. 1. Methodology and initial results}},\ }\href
  {https://doi.org/10.1086/522566} {\bibfield  {journal} {\bibinfo  {journal}
  {Astrophys. J.}\ }\textbf {\bibinfo {volume} {671}},\ \bibinfo {pages} {1}
  (\bibinfo {year} {2007})},\ \Eprint {https://arxiv.org/abs/astro-ph/0612406}
  {arXiv:astro-ph/0612406} \BibitemShut {NoStop}%
\bibitem [{\citenamefont {Puchwein}\ \emph {et~al.}(2019)\citenamefont
  {Puchwein}, \citenamefont {Haardt}, \citenamefont {Haehnelt},\ and\
  \citenamefont {Madau}}]{Puchwein:2018arm}%
  \BibitemOpen
  \bibfield  {author} {\bibinfo {author} {\bibfnamefont {E.}~\bibnamefont
  {Puchwein}}, \bibinfo {author} {\bibfnamefont {F.}~\bibnamefont {Haardt}},
  \bibinfo {author} {\bibfnamefont {M.~G.}\ \bibnamefont {Haehnelt}},\ and\
  \bibinfo {author} {\bibfnamefont {P.}~\bibnamefont {Madau}},\ }\bibfield
  {title} {\bibinfo {title} {{Consistent modelling of the meta-galactic UV
  background and the thermal/ionization history of the intergalactic medium}},\
  }\href {https://doi.org/10.1093/mnras/stz222} {\bibfield  {journal} {\bibinfo
   {journal} {Mon. Not. Roy. Astron. Soc.}\ }\textbf {\bibinfo {volume}
  {485}},\ \bibinfo {pages} {47} (\bibinfo {year} {2019})},\ \Eprint
  {https://arxiv.org/abs/1801.04931} {arXiv:1801.04931 [astro-ph.GA]}
  \BibitemShut {NoStop}%
\bibitem [{\citenamefont {Park}\ \emph {et~al.}(2022)\citenamefont {Park},
  \citenamefont {Greig},\ and\ \citenamefont {Mesinger}}]{Park:2021eux}%
  \BibitemOpen
  \bibfield  {author} {\bibinfo {author} {\bibfnamefont {J.}~\bibnamefont
  {Park}}, \bibinfo {author} {\bibfnamefont {B.}~\bibnamefont {Greig}},\ and\
  \bibinfo {author} {\bibfnamefont {A.}~\bibnamefont {Mesinger}},\ }\bibfield
  {title} {\bibinfo {title} {{Calibrating excursion set reionization models to
  approximately conserve ionizing photons}},\ }\href
  {https://doi.org/10.1093/mnras/stac2756} {\bibfield  {journal} {\bibinfo
  {journal} {Mon. Not. Roy. Astron. Soc.}\ }\textbf {\bibinfo {volume} {517}},\
  \bibinfo {pages} {192} (\bibinfo {year} {2022})},\ \Eprint
  {https://arxiv.org/abs/2112.05184} {arXiv:2112.05184 [astro-ph.CO]}
  \BibitemShut {NoStop}%
\bibitem [{\citenamefont {Park}\ \emph {et~al.}(2019)\citenamefont {Park},
  \citenamefont {Mesinger}, \citenamefont {Greig},\ and\ \citenamefont
  {Gillet}}]{Park:2018ljd}%
  \BibitemOpen
  \bibfield  {author} {\bibinfo {author} {\bibfnamefont {J.}~\bibnamefont
  {Park}}, \bibinfo {author} {\bibfnamefont {A.}~\bibnamefont {Mesinger}},
  \bibinfo {author} {\bibfnamefont {B.}~\bibnamefont {Greig}},\ and\ \bibinfo
  {author} {\bibfnamefont {N.}~\bibnamefont {Gillet}},\ }\bibfield  {title}
  {\bibinfo {title} {{Inferring the astrophysics of reionization and cosmic
  dawn from galaxy luminosity functions and the 21-cm signal}},\ }\href
  {https://doi.org/10.1093/mnras/stz032} {\bibfield  {journal} {\bibinfo
  {journal} {Mon. Not. Roy. Astron. Soc.}\ }\textbf {\bibinfo {volume} {484}},\
  \bibinfo {pages} {933} (\bibinfo {year} {2019})},\ \Eprint
  {https://arxiv.org/abs/1809.08995} {arXiv:1809.08995 [astro-ph.GA]}
  \BibitemShut {NoStop}%
\bibitem [{\citenamefont {Qin}\ \emph {et~al.}(2020{\natexlab{b}})\citenamefont
  {Qin}, \citenamefont {Mesinger}, \citenamefont {Park}, \citenamefont
  {Greig},\ and\ \citenamefont {Mu\~noz}}]{Qin:2020xyh}%
  \BibitemOpen
  \bibfield  {author} {\bibinfo {author} {\bibfnamefont {Y.}~\bibnamefont
  {Qin}}, \bibinfo {author} {\bibfnamefont {A.}~\bibnamefont {Mesinger}},
  \bibinfo {author} {\bibfnamefont {J.}~\bibnamefont {Park}}, \bibinfo {author}
  {\bibfnamefont {B.}~\bibnamefont {Greig}},\ and\ \bibinfo {author}
  {\bibfnamefont {J.~B.}\ \bibnamefont {Mu\~noz}},\ }\bibfield  {title}
  {\bibinfo {title} {{A tale of two sites \textendash{} I. Inferring the
  properties of minihalo-hosted galaxies from current observations}},\ }\href
  {https://doi.org/10.1093/mnras/staa1131} {\bibfield  {journal} {\bibinfo
  {journal} {Mon. Not. Roy. Astron. Soc.}\ }\textbf {\bibinfo {volume} {495}},\
  \bibinfo {pages} {123} (\bibinfo {year} {2020}{\natexlab{b}})},\ \Eprint
  {https://arxiv.org/abs/2003.04442} {arXiv:2003.04442 [astro-ph.CO]}
  \BibitemShut {NoStop}%
\bibitem [{\citenamefont {Stefanon}\ \emph {et~al.}(2021)\citenamefont
  {Stefanon}, \citenamefont {Bouwens}, \citenamefont {Labbé}, \citenamefont
  {Illingworth}, \citenamefont {Gonzalez},\ and\ \citenamefont
  {Oesch}}]{Stefanon_2021}%
  \BibitemOpen
  \bibfield  {author} {\bibinfo {author} {\bibfnamefont {M.}~\bibnamefont
  {Stefanon}}, \bibinfo {author} {\bibfnamefont {R.~J.}\ \bibnamefont
  {Bouwens}}, \bibinfo {author} {\bibfnamefont {I.}~\bibnamefont {Labbé}},
  \bibinfo {author} {\bibfnamefont {G.~D.}\ \bibnamefont {Illingworth}},
  \bibinfo {author} {\bibfnamefont {V.}~\bibnamefont {Gonzalez}},\ and\
  \bibinfo {author} {\bibfnamefont {P.~A.}\ \bibnamefont {Oesch}},\ }\bibfield
  {title} {\bibinfo {title} {Galaxy stellar mass functions from z $\sim$ 10 to
  z $\sim$ 6 using the deepest spitzer/infrared array camera data: No
  significant evolution in the stellar-to-halo mass ratio of galaxies in the
  first gigayear of cosmic time},\ }\href
  {https://doi.org/10.3847/1538-4357/ac1bb6} {\bibfield  {journal} {\bibinfo
  {journal} {The Astrophysical Journal}\ }\textbf {\bibinfo {volume} {922}},\
  \bibinfo {pages} {29} (\bibinfo {year} {2021})}\BibitemShut {NoStop}%
\bibitem [{\citenamefont {Shuntov}\ \emph {et~al.}(2022)\citenamefont {Shuntov}
  \emph {et~al.}}]{Shuntov:2022qwu}%
  \BibitemOpen
  \bibfield  {author} {\bibinfo {author} {\bibfnamefont {M.}~\bibnamefont
  {Shuntov}} \emph {et~al.},\ }\bibfield  {title} {\bibinfo {title}
  {{COSMOS2020: Cosmic evolution of the stellar-to-halo mass relation for
  central and satellite galaxies up to z \ensuremath{\sim} 5}},\ }\href
  {https://doi.org/10.1051/0004-6361/202243136} {\bibfield  {journal} {\bibinfo
   {journal} {Astron. Astrophys.}\ }\textbf {\bibinfo {volume} {664}},\
  \bibinfo {pages} {A61} (\bibinfo {year} {2022})},\ \Eprint
  {https://arxiv.org/abs/2203.10895} {arXiv:2203.10895 [astro-ph.GA]}
  \BibitemShut {NoStop}%
\bibitem [{\citenamefont {Zackrisson}\ \emph {et~al.}(2011)\citenamefont
  {Zackrisson}, \citenamefont {Rydberg}, \citenamefont {Schaerer},
  \citenamefont {Ostlin},\ and\ \citenamefont {Tuli}}]{Zackrisson:2011ct}%
  \BibitemOpen
  \bibfield  {author} {\bibinfo {author} {\bibfnamefont {E.}~\bibnamefont
  {Zackrisson}}, \bibinfo {author} {\bibfnamefont {C.~E.}\ \bibnamefont
  {Rydberg}}, \bibinfo {author} {\bibfnamefont {D.}~\bibnamefont {Schaerer}},
  \bibinfo {author} {\bibfnamefont {G.}~\bibnamefont {Ostlin}},\ and\ \bibinfo
  {author} {\bibfnamefont {M.}~\bibnamefont {Tuli}},\ }\bibfield  {title}
  {\bibinfo {title} {{The spectral evolution of the first galaxies. I. James
  Webb Space Telescope detection limits and color criteria for population III
  galaxies}},\ }\href {https://doi.org/10.1088/0004-637X/740/1/13} {\bibfield
  {journal} {\bibinfo  {journal} {Astrophys. J.}\ }\textbf {\bibinfo {volume}
  {740}},\ \bibinfo {pages} {13} (\bibinfo {year} {2011})},\ \Eprint
  {https://arxiv.org/abs/1105.0921} {arXiv:1105.0921 [astro-ph.CO]}
  \BibitemShut {NoStop}%
\bibitem [{\citenamefont {Rydberg}\ \emph {et~al.}(2013)\citenamefont
  {Rydberg}, \citenamefont {Zackrisson}, \citenamefont {Lundqvist},\ and\
  \citenamefont {Scott}}]{Rydberg:2012ez}%
  \BibitemOpen
  \bibfield  {author} {\bibinfo {author} {\bibfnamefont {C.-E.}\ \bibnamefont
  {Rydberg}}, \bibinfo {author} {\bibfnamefont {E.}~\bibnamefont {Zackrisson}},
  \bibinfo {author} {\bibfnamefont {P.}~\bibnamefont {Lundqvist}},\ and\
  \bibinfo {author} {\bibfnamefont {P.}~\bibnamefont {Scott}},\ }\bibfield
  {title} {\bibinfo {title} {{Detection of isolated population III stars with
  the James Webb Space Telescope}},\ }\href
  {https://doi.org/10.1093/mnras/sts653} {\bibfield  {journal} {\bibinfo
  {journal} {Mon. Not. Roy. Astron. Soc.}\ }\textbf {\bibinfo {volume} {429}},\
  \bibinfo {pages} {3658} (\bibinfo {year} {2013})},\ \Eprint
  {https://arxiv.org/abs/1206.0007} {arXiv:1206.0007 [astro-ph.CO]}
  \BibitemShut {NoStop}%
\bibitem [{\citenamefont {Riaz}\ \emph {et~al.}(2022)\citenamefont {Riaz},
  \citenamefont {Hartwig},\ and\ \citenamefont {Latif}}]{Riaz:2022prd}%
  \BibitemOpen
  \bibfield  {author} {\bibinfo {author} {\bibfnamefont {S.}~\bibnamefont
  {Riaz}}, \bibinfo {author} {\bibfnamefont {T.}~\bibnamefont {Hartwig}},\ and\
  \bibinfo {author} {\bibfnamefont {M.~A.}\ \bibnamefont {Latif}},\ }\bibfield
  {title} {\bibinfo {title} {{Unveiling the Contribution of Population III
  Stars in Primeval Galaxies at Redshift \ensuremath{\geq}6}},\ }\href
  {https://doi.org/10.3847/2041-8213/ac8ea6} {\bibfield  {journal} {\bibinfo
  {journal} {Astrophys. J. Lett.}\ }\textbf {\bibinfo {volume} {937}},\
  \bibinfo {pages} {L6} (\bibinfo {year} {2022})},\ \Eprint
  {https://arxiv.org/abs/2208.01673} {arXiv:2208.01673 [astro-ph.GA]}
  \BibitemShut {NoStop}%
\bibitem [{\citenamefont {Larkin}\ \emph {et~al.}(2023)\citenamefont {Larkin},
  \citenamefont {Gerasimov},\ and\ \citenamefont {Burgasser}}]{Larkin:2022asx}%
  \BibitemOpen
  \bibfield  {author} {\bibinfo {author} {\bibfnamefont {M.~M.}\ \bibnamefont
  {Larkin}}, \bibinfo {author} {\bibfnamefont {R.}~\bibnamefont {Gerasimov}},\
  and\ \bibinfo {author} {\bibfnamefont {A.~J.}\ \bibnamefont {Burgasser}},\
  }\bibfield  {title} {\bibinfo {title} {{Characterization of Population III
  Stars with Stellar Atmosphere and Evolutionary Modeling and Predictions of
  their Observability with the JWST}},\ }\href
  {https://doi.org/10.3847/1538-3881/ac9b43} {\bibfield  {journal} {\bibinfo
  {journal} {Astron. J.}\ }\textbf {\bibinfo {volume} {165}},\ \bibinfo {pages}
  {2} (\bibinfo {year} {2023})},\ \Eprint {https://arxiv.org/abs/2210.09185}
  {arXiv:2210.09185 [astro-ph.SR]} \BibitemShut {NoStop}%
\bibitem [{\citenamefont {Das}\ \emph {et~al.}(2017)\citenamefont {Das},
  \citenamefont {Mesinger}, \citenamefont {Pallottini}, \citenamefont
  {Ferrara},\ and\ \citenamefont {Wise}}]{Das:2017fys}%
  \BibitemOpen
  \bibfield  {author} {\bibinfo {author} {\bibfnamefont {A.}~\bibnamefont
  {Das}}, \bibinfo {author} {\bibfnamefont {A.}~\bibnamefont {Mesinger}},
  \bibinfo {author} {\bibfnamefont {A.}~\bibnamefont {Pallottini}}, \bibinfo
  {author} {\bibfnamefont {A.}~\bibnamefont {Ferrara}},\ and\ \bibinfo {author}
  {\bibfnamefont {J.~H.}\ \bibnamefont {Wise}},\ }\bibfield  {title} {\bibinfo
  {title} {{High Mass X-ray Binaries and the Cosmic 21-cm Signal: Impact of
  Host Galaxy Absorption}},\ }\href {https://doi.org/10.1093/mnras/stx943}
  {\bibfield  {journal} {\bibinfo  {journal} {Mon. Not. Roy. Astron. Soc.}\
  }\textbf {\bibinfo {volume} {469}},\ \bibinfo {pages} {1166} (\bibinfo {year}
  {2017})},\ \Eprint {https://arxiv.org/abs/1702.00409} {arXiv:1702.00409
  [astro-ph.CO]} \BibitemShut {NoStop}%
\bibitem [{\citenamefont {de~Salas}\ \emph {et~al.}(2021)\citenamefont
  {de~Salas}, \citenamefont {Forero}, \citenamefont {Gariazzo}, \citenamefont
  {Mart\'\i{}nez-Mirav\'e}, \citenamefont {Mena}, \citenamefont {Ternes},
  \citenamefont {T\'ortola},\ and\ \citenamefont {Valle}}]{deSalas:2020pgw}%
  \BibitemOpen
  \bibfield  {author} {\bibinfo {author} {\bibfnamefont {P.~F.}\ \bibnamefont
  {de~Salas}}, \bibinfo {author} {\bibfnamefont {D.~V.}\ \bibnamefont
  {Forero}}, \bibinfo {author} {\bibfnamefont {S.}~\bibnamefont {Gariazzo}},
  \bibinfo {author} {\bibfnamefont {P.}~\bibnamefont {Mart\'\i{}nez-Mirav\'e}},
  \bibinfo {author} {\bibfnamefont {O.}~\bibnamefont {Mena}}, \bibinfo {author}
  {\bibfnamefont {C.~A.}\ \bibnamefont {Ternes}}, \bibinfo {author}
  {\bibfnamefont {M.}~\bibnamefont {T\'ortola}},\ and\ \bibinfo {author}
  {\bibfnamefont {J.~W.~F.}\ \bibnamefont {Valle}},\ }\bibfield  {title}
  {\bibinfo {title} {{2020 global reassessment of the neutrino oscillation
  picture}},\ }\href {https://doi.org/10.1007/JHEP02(2021)071} {\bibfield
  {journal} {\bibinfo  {journal} {JHEP}\ }\textbf {\bibinfo {volume} {02}},\
  \bibinfo {pages} {071}},\ \Eprint {https://arxiv.org/abs/2006.11237}
  {arXiv:2006.11237 [hep-ph]} \BibitemShut {NoStop}%
\bibitem [{\citenamefont {Schneider}(2015)}]{Schneider:2014rda}%
  \BibitemOpen
  \bibfield  {author} {\bibinfo {author} {\bibfnamefont {A.}~\bibnamefont
  {Schneider}},\ }\bibfield  {title} {\bibinfo {title} {{Structure formation
  with suppressed small-scale perturbations}},\ }\href
  {https://doi.org/10.1093/mnras/stv1169} {\bibfield  {journal} {\bibinfo
  {journal} {Mon. Not. Roy. Astron. Soc.}\ }\textbf {\bibinfo {volume} {451}},\
  \bibinfo {pages} {3117} (\bibinfo {year} {2015})},\ \Eprint
  {https://arxiv.org/abs/1412.2133} {arXiv:1412.2133 [astro-ph.CO]}
  \BibitemShut {NoStop}%
\bibitem [{\citenamefont {Greig}\ \emph {et~al.}(2024)\citenamefont {Greig},
  \citenamefont {Prelogovi\'c}, \citenamefont {Mirocha}, \citenamefont {Qin},
  \citenamefont {Ting},\ and\ \citenamefont {Mesinger}}]{Greig:2024zso}%
  \BibitemOpen
  \bibfield  {author} {\bibinfo {author} {\bibfnamefont {B.}~\bibnamefont
  {Greig}}, \bibinfo {author} {\bibfnamefont {D.}~\bibnamefont {Prelogovi\'c}},
  \bibinfo {author} {\bibfnamefont {J.}~\bibnamefont {Mirocha}}, \bibinfo
  {author} {\bibfnamefont {Y.}~\bibnamefont {Qin}}, \bibinfo {author}
  {\bibfnamefont {Y.-S.}\ \bibnamefont {Ting}},\ and\ \bibinfo {author}
  {\bibfnamefont {A.}~\bibnamefont {Mesinger}},\ }\bibfield  {title} {\bibinfo
  {title} {{Exploring the role of the halo-mass function for inferring
  astrophysical parameters during reionization}},\ }\href
  {https://doi.org/10.1093/mnras/stae1983} {\bibfield  {journal} {\bibinfo
  {journal} {Mon. Not. Roy. Astron. Soc.}\ }\textbf {\bibinfo {volume} {533}},\
  \bibinfo {pages} {2502} (\bibinfo {year} {2024})},\ \Eprint
  {https://arxiv.org/abs/2403.14061} {arXiv:2403.14061 [astro-ph.CO]}
  \BibitemShut {NoStop}%
\bibitem [{\citenamefont {McGreer}\ \emph {et~al.}(2015)\citenamefont
  {McGreer}, \citenamefont {Mesinger},\ and\ \citenamefont
  {D'Odorico}}]{McGreer:2014qwa}%
  \BibitemOpen
  \bibfield  {author} {\bibinfo {author} {\bibfnamefont {I.}~\bibnamefont
  {McGreer}}, \bibinfo {author} {\bibfnamefont {A.}~\bibnamefont {Mesinger}},\
  and\ \bibinfo {author} {\bibfnamefont {V.}~\bibnamefont {D'Odorico}},\
  }\bibfield  {title} {\bibinfo {title} {{Model-independent evidence in favour
  of an end to reionization by $z \approx$ 6}},\ }\href
  {https://doi.org/10.1093/mnras/stu2449} {\bibfield  {journal} {\bibinfo
  {journal} {Mon. Not. Roy. Astron. Soc.}\ }\textbf {\bibinfo {volume} {447}},\
  \bibinfo {pages} {499} (\bibinfo {year} {2015})},\ \Eprint
  {https://arxiv.org/abs/1411.5375} {arXiv:1411.5375 [astro-ph.CO]}
  \BibitemShut {NoStop}%
\bibitem [{\citenamefont {Habib}\ \emph {et~al.}(2007)\citenamefont {Habib},
  \citenamefont {Heitmann}, \citenamefont {Higdon}, \citenamefont {Nakhleh},\
  and\ \citenamefont {Williams}}]{Habib:2007ca}%
  \BibitemOpen
  \bibfield  {author} {\bibinfo {author} {\bibfnamefont {S.}~\bibnamefont
  {Habib}}, \bibinfo {author} {\bibfnamefont {K.}~\bibnamefont {Heitmann}},
  \bibinfo {author} {\bibfnamefont {D.}~\bibnamefont {Higdon}}, \bibinfo
  {author} {\bibfnamefont {C.}~\bibnamefont {Nakhleh}},\ and\ \bibinfo {author}
  {\bibfnamefont {B.}~\bibnamefont {Williams}},\ }\bibfield  {title} {\bibinfo
  {title} {{Cosmic Calibration: Constraints from the Matter Power Spectrum and
  the Cosmic Microwave Background}},\ }\href
  {https://doi.org/10.1103/PhysRevD.76.083503} {\bibfield  {journal} {\bibinfo
  {journal} {Phys. Rev. D}\ }\textbf {\bibinfo {volume} {76}},\ \bibinfo
  {pages} {083503} (\bibinfo {year} {2007})},\ \Eprint
  {https://arxiv.org/abs/astro-ph/0702348} {arXiv:astro-ph/0702348}
  \BibitemShut {NoStop}%
\bibitem [{\citenamefont {Higdon}\ \emph {et~al.}(2008)\citenamefont {Higdon},
  \citenamefont {Nakhleh}, \citenamefont {Gattiker},\ and\ \citenamefont
  {Williams}}]{HIGDON20082431}%
  \BibitemOpen
  \bibfield  {author} {\bibinfo {author} {\bibfnamefont {D.}~\bibnamefont
  {Higdon}}, \bibinfo {author} {\bibfnamefont {C.}~\bibnamefont {Nakhleh}},
  \bibinfo {author} {\bibfnamefont {J.}~\bibnamefont {Gattiker}},\ and\
  \bibinfo {author} {\bibfnamefont {B.}~\bibnamefont {Williams}},\ }\bibfield
  {title} {\bibinfo {title} {A bayesian calibration approach to the thermal
  problem},\ }\href {https://doi.org/https://doi.org/10.1016/j.cma.2007.05.031}
  {\bibfield  {journal} {\bibinfo  {journal} {Computer Methods in Applied
  Mechanics and Engineering}\ }\textbf {\bibinfo {volume} {197}},\ \bibinfo
  {pages} {2431} (\bibinfo {year} {2008})},\ \bibinfo {note} {validation
  Challenge Workshop}\BibitemShut {NoStop}%
\bibitem [{\citenamefont {Audren}\ \emph {et~al.}(2013)\citenamefont {Audren},
  \citenamefont {Lesgourgues}, \citenamefont {Benabed},\ and\ \citenamefont
  {Prunet}}]{Audren:2012wb}%
  \BibitemOpen
  \bibfield  {author} {\bibinfo {author} {\bibfnamefont {B.}~\bibnamefont
  {Audren}}, \bibinfo {author} {\bibfnamefont {J.}~\bibnamefont {Lesgourgues}},
  \bibinfo {author} {\bibfnamefont {K.}~\bibnamefont {Benabed}},\ and\ \bibinfo
  {author} {\bibfnamefont {S.}~\bibnamefont {Prunet}},\ }\bibfield  {title}
  {\bibinfo {title} {{Conservative Constraints on Early Cosmology: an
  illustration of the Monte Python cosmological parameter inference code}},\
  }\href {https://doi.org/10.1088/1475-7516/2013/02/001} {\bibfield  {journal}
  {\bibinfo  {journal} {JCAP}\ }\textbf {\bibinfo {volume} {1302}},\ \bibinfo
  {pages} {001}},\ \Eprint {https://arxiv.org/abs/1210.7183} {arXiv:1210.7183
  [astro-ph.CO]} \BibitemShut {NoStop}%
\bibitem [{\citenamefont {Alam}\ \emph {et~al.}(2017)\citenamefont {Alam} \emph
  {et~al.}}]{BOSS:2016wmc}%
  \BibitemOpen
  \bibfield  {author} {\bibinfo {author} {\bibfnamefont {S.}~\bibnamefont
  {Alam}} \emph {et~al.} (\bibinfo {collaboration} {BOSS}),\ }\bibfield
  {title} {\bibinfo {title} {{The clustering of galaxies in the completed
  SDSS-III Baryon Oscillation Spectroscopic Survey: cosmological analysis of
  the DR12 galaxy sample}},\ }\href {https://doi.org/10.1093/mnras/stx721}
  {\bibfield  {journal} {\bibinfo  {journal} {Mon. Not. Roy. Astron. Soc.}\
  }\textbf {\bibinfo {volume} {470}},\ \bibinfo {pages} {2617} (\bibinfo {year}
  {2017})},\ \Eprint {https://arxiv.org/abs/1607.03155} {arXiv:1607.03155
  [astro-ph.CO]} \BibitemShut {NoStop}%
\bibitem [{\citenamefont {Beutler}\ \emph {et~al.}(2011)\citenamefont
  {Beutler}, \citenamefont {Blake}, \citenamefont {Colless}, \citenamefont
  {Jones}, \citenamefont {Staveley-Smith}, \citenamefont {Campbell},
  \citenamefont {Parker}, \citenamefont {Saunders},\ and\ \citenamefont
  {Watson}}]{Beutler:2011hx}%
  \BibitemOpen
  \bibfield  {author} {\bibinfo {author} {\bibfnamefont {F.}~\bibnamefont
  {Beutler}}, \bibinfo {author} {\bibfnamefont {C.}~\bibnamefont {Blake}},
  \bibinfo {author} {\bibfnamefont {M.}~\bibnamefont {Colless}}, \bibinfo
  {author} {\bibfnamefont {D.~H.}\ \bibnamefont {Jones}}, \bibinfo {author}
  {\bibfnamefont {L.}~\bibnamefont {Staveley-Smith}}, \bibinfo {author}
  {\bibfnamefont {L.}~\bibnamefont {Campbell}}, \bibinfo {author}
  {\bibfnamefont {Q.}~\bibnamefont {Parker}}, \bibinfo {author} {\bibfnamefont
  {W.}~\bibnamefont {Saunders}},\ and\ \bibinfo {author} {\bibfnamefont
  {F.}~\bibnamefont {Watson}},\ }\bibfield  {title} {\bibinfo {title} {{The 6dF
  Galaxy Survey: Baryon Acoustic Oscillations and the Local Hubble Constant}},\
  }\href {https://doi.org/10.1111/j.1365-2966.2011.19250.x} {\bibfield
  {journal} {\bibinfo  {journal} {Mon. Not. Roy. Astron. Soc.}\ }\textbf
  {\bibinfo {volume} {416}},\ \bibinfo {pages} {3017} (\bibinfo {year}
  {2011})},\ \Eprint {https://arxiv.org/abs/1106.3366} {arXiv:1106.3366
  [astro-ph.CO]} \BibitemShut {NoStop}%
\bibitem [{\citenamefont {Ross}\ \emph {et~al.}(2015)\citenamefont {Ross},
  \citenamefont {Samushia}, \citenamefont {Howlett}, \citenamefont {Percival},
  \citenamefont {Burden},\ and\ \citenamefont {Manera}}]{Ross:2014qpa}%
  \BibitemOpen
  \bibfield  {author} {\bibinfo {author} {\bibfnamefont {A.~J.}\ \bibnamefont
  {Ross}}, \bibinfo {author} {\bibfnamefont {L.}~\bibnamefont {Samushia}},
  \bibinfo {author} {\bibfnamefont {C.}~\bibnamefont {Howlett}}, \bibinfo
  {author} {\bibfnamefont {W.~J.}\ \bibnamefont {Percival}}, \bibinfo {author}
  {\bibfnamefont {A.}~\bibnamefont {Burden}},\ and\ \bibinfo {author}
  {\bibfnamefont {M.}~\bibnamefont {Manera}},\ }\bibfield  {title} {\bibinfo
  {title} {{The clustering of the SDSS DR7 main Galaxy sample \textendash{} I.
  A 4 per cent distance measure at $z = 0.15$}},\ }\href
  {https://doi.org/10.1093/mnras/stv154} {\bibfield  {journal} {\bibinfo
  {journal} {Mon. Not. Roy. Astron. Soc.}\ }\textbf {\bibinfo {volume} {449}},\
  \bibinfo {pages} {835} (\bibinfo {year} {2015})},\ \Eprint
  {https://arxiv.org/abs/1409.3242} {arXiv:1409.3242 [astro-ph.CO]}
  \BibitemShut {NoStop}%
\bibitem [{\citenamefont {Bouwens}\ \emph {et~al.}(2017)\citenamefont
  {Bouwens}, \citenamefont {Oesch}, \citenamefont {Illingworth}, \citenamefont
  {Ellis},\ and\ \citenamefont {Stefanon}}]{Bouwens_2017}%
  \BibitemOpen
  \bibfield  {author} {\bibinfo {author} {\bibfnamefont {R.~J.}\ \bibnamefont
  {Bouwens}}, \bibinfo {author} {\bibfnamefont {P.~A.}\ \bibnamefont {Oesch}},
  \bibinfo {author} {\bibfnamefont {G.~D.}\ \bibnamefont {Illingworth}},
  \bibinfo {author} {\bibfnamefont {R.~S.}\ \bibnamefont {Ellis}},\ and\
  \bibinfo {author} {\bibfnamefont {M.}~\bibnamefont {Stefanon}},\ }\bibfield
  {title} {\bibinfo {title} {The z$\sim$6 luminosity function fainter than -15
  mag from the hubble frontier fields: The impact of magnification
  uncertainties},\ }\href {https://doi.org/10.3847/1538-4357/aa70a4} {\bibfield
   {journal} {\bibinfo  {journal} {The Astrophysical Journal}\ }\textbf
  {\bibinfo {volume} {843}},\ \bibinfo {pages} {129} (\bibinfo {year}
  {2017})}\BibitemShut {NoStop}%
\bibitem [{\citenamefont {Bouwens}\ \emph {et~al.}(2021)\citenamefont
  {Bouwens}, \citenamefont {Oesch}, \citenamefont {Stefanon}, \citenamefont
  {Illingworth}, \citenamefont {Labbé}, \citenamefont {Reddy}, \citenamefont
  {Atek}, \citenamefont {Montes}, \citenamefont {Naidu}, \citenamefont
  {Nanayakkara}, \citenamefont {Nelson},\ and\ \citenamefont
  {Wilkins}}]{Bouwens_2021}%
  \BibitemOpen
  \bibfield  {author} {\bibinfo {author} {\bibfnamefont {R.~J.}\ \bibnamefont
  {Bouwens}}, \bibinfo {author} {\bibfnamefont {P.~A.}\ \bibnamefont {Oesch}},
  \bibinfo {author} {\bibfnamefont {M.}~\bibnamefont {Stefanon}}, \bibinfo
  {author} {\bibfnamefont {G.}~\bibnamefont {Illingworth}}, \bibinfo {author}
  {\bibfnamefont {I.}~\bibnamefont {Labbé}}, \bibinfo {author} {\bibfnamefont
  {N.}~\bibnamefont {Reddy}}, \bibinfo {author} {\bibfnamefont
  {H.}~\bibnamefont {Atek}}, \bibinfo {author} {\bibfnamefont {M.}~\bibnamefont
  {Montes}}, \bibinfo {author} {\bibfnamefont {R.}~\bibnamefont {Naidu}},
  \bibinfo {author} {\bibfnamefont {T.}~\bibnamefont {Nanayakkara}}, \bibinfo
  {author} {\bibfnamefont {E.}~\bibnamefont {Nelson}},\ and\ \bibinfo {author}
  {\bibfnamefont {S.}~\bibnamefont {Wilkins}},\ }\bibfield  {title} {\bibinfo
  {title} {New determinations of the uv luminosity functions from z $\sim$ 9 to
  2 show a remarkable consistency with halo growth and a constant star
  formation efficiency},\ }\href {https://doi.org/10.3847/1538-3881/abf83e}
  {\bibfield  {journal} {\bibinfo  {journal} {The Astronomical Journal}\
  }\textbf {\bibinfo {volume} {162}},\ \bibinfo {pages} {47} (\bibinfo {year}
  {2021})}\BibitemShut {NoStop}%
\bibitem [{\citenamefont {Adame}\ \emph {et~al.}(2025)\citenamefont {Adame}
  \emph {et~al.}}]{DESI:2024mwx}%
  \BibitemOpen
  \bibfield  {author} {\bibinfo {author} {\bibfnamefont {A.~G.}\ \bibnamefont
  {Adame}} \emph {et~al.} (\bibinfo {collaboration} {DESI}),\ }\bibfield
  {title} {\bibinfo {title} {{DESI 2024 VI: cosmological constraints from the
  measurements of baryon acoustic oscillations}},\ }\href
  {https://doi.org/10.1088/1475-7516/2025/02/021} {\bibfield  {journal}
  {\bibinfo  {journal} {JCAP}\ }\textbf {\bibinfo {volume} {02}},\ \bibinfo
  {pages} {021}},\ \Eprint {https://arxiv.org/abs/2404.03002} {arXiv:2404.03002
  [astro-ph.CO]} \BibitemShut {NoStop}%
\bibitem [{\citenamefont {Giri}\ and\ \citenamefont
  {Schneider}(2022)}]{Giri:2022nxq}%
  \BibitemOpen
  \bibfield  {author} {\bibinfo {author} {\bibfnamefont {S.~K.}\ \bibnamefont
  {Giri}}\ and\ \bibinfo {author} {\bibfnamefont {A.}~\bibnamefont
  {Schneider}},\ }\bibfield  {title} {\bibinfo {title} {{Imprints of fermionic
  and bosonic mixed dark matter on the 21-cm signal at cosmic dawn}},\ }\href
  {https://doi.org/10.1103/PhysRevD.105.083011} {\bibfield  {journal} {\bibinfo
   {journal} {Phys. Rev. D}\ }\textbf {\bibinfo {volume} {105}},\ \bibinfo
  {pages} {083011} (\bibinfo {year} {2022})},\ \Eprint
  {https://arxiv.org/abs/2201.02210} {arXiv:2201.02210 [astro-ph.CO]}
  \BibitemShut {NoStop}%
\bibitem [{\citenamefont {Chatterjee}\ and\ \citenamefont
  {Choudhury}(2023)}]{Chatterjee:2023mlh}%
  \BibitemOpen
  \bibfield  {author} {\bibinfo {author} {\bibfnamefont {A.}~\bibnamefont
  {Chatterjee}}\ and\ \bibinfo {author} {\bibfnamefont {T.~R.}\ \bibnamefont
  {Choudhury}},\ }\bibfield  {title} {\bibinfo {title} {{Warm Dark matter
  constraints from the joint analysis of CMB, Ly~\ensuremath{\alpha}, and
  global 21~cm data}},\ }\href {https://doi.org/10.1093/mnras/stad3930}
  {\bibfield  {journal} {\bibinfo  {journal} {Mon. Not. Roy. Astron. Soc.}\
  }\textbf {\bibinfo {volume} {527}},\ \bibinfo {pages} {10777} (\bibinfo
  {year} {2023})},\ \Eprint {https://arxiv.org/abs/2304.09810}
  {arXiv:2304.09810 [astro-ph.CO]} \BibitemShut {NoStop}%
\bibitem [{\citenamefont {Sun}\ and\ \citenamefont
  {Furlanetto}(2016)}]{Sun_2016}%
  \BibitemOpen
  \bibfield  {author} {\bibinfo {author} {\bibfnamefont {G.}~\bibnamefont
  {Sun}}\ and\ \bibinfo {author} {\bibfnamefont {S.~R.}\ \bibnamefont
  {Furlanetto}},\ }\bibfield  {title} {\bibinfo {title} {Constraints on the
  star formation efficiency of galaxies during the epoch of reionization},\
  }\href {https://doi.org/10.1093/mnras/stw980} {\bibfield  {journal} {\bibinfo
   {journal} {Monthly Notices of the Royal Astronomical Society}\ }\textbf
  {\bibinfo {volume} {460}},\ \bibinfo {pages} {417–433} (\bibinfo {year}
  {2016})}\BibitemShut {NoStop}%
\bibitem [{\citenamefont {{Oke}}\ and\ \citenamefont
  {{Gunn}}(1983)}]{1983ApJ...266..713O}%
  \BibitemOpen
  \bibfield  {author} {\bibinfo {author} {\bibfnamefont {J.~B.}\ \bibnamefont
  {{Oke}}}\ and\ \bibinfo {author} {\bibfnamefont {J.~E.}\ \bibnamefont
  {{Gunn}}},\ }\bibfield  {title} {\bibinfo {title} {{Secondary standard stars
  for absolute spectrophotometry.}},\ }\href {https://doi.org/10.1086/160817}
  {\bibfield  {journal} {\bibinfo  {journal} {\apj}\ }\textbf {\bibinfo
  {volume} {266}},\ \bibinfo {pages} {713} (\bibinfo {year}
  {1983})}\BibitemShut {NoStop}%
\end{thebibliography}%

\end{document}